\documentclass[
    aps,
    physrev,
    twocolumn,
    superscriptaddress,
    nofootinbib,
    floatfix
]{revtex4-2}

\usepackage{mathtools}
\usepackage{amssymb}
\usepackage{bm}
\usepackage{dsfont}
\usepackage{braket}
\usepackage{nicematrix}
\NiceMatrixOptions{cell-space-limits=1pt}

\usepackage{amsthm}
\newtheorem{theorem}{Theorem}

\usepackage{graphicx}
\usepackage{ragged2e}
\usepackage{caption}
\usepackage{hyperref}
\hypersetup{
    colorlinks=true,
    linkcolor=blue,
    urlcolor=blue,
    citecolor=blue
}
\usepackage{verbatim} 

\newcommand{\NiceBlock}[3][name]{\Block[name=#1,draw=gray,rounded-corners]{#2}{#3}}

\newcommand{\BlockMatrix}[3][2pt]{
    \begin{pNiceMatrix}[margin][columns-width=#1]
        \Block[name=b1]{2-2}{#2} & & & \Block{2-2}{\bm{0}}\\
        & & & & \\
        \Block{2-2}{\bm{0}} & & & \Block[name=b2]{2-2}{#3}\\
        & & & &
    \CodeAfter
        \line[radius=0.6pt,inter=0.4em,shorten=0.2em]{b1}{b2}
    \end{pNiceMatrix}
}

\DeclareMathOperator{\Tr}{Tr}
\DeclareMathOperator{\Span}{span}

\newcommand{\id}{\mathds{1}}
\newcommand{\dsC}{\mathds{C}}
\newcommand{\dsT}{\mathds{T}}
\newcommand{\dsZ}{\mathds{Z}}

\newcommand{\mA}{\mathcal{A}}
\newcommand{\mE}{\mathcal{E}}
\newcommand{\mH}{\mathcal{H}}
\newcommand{\mK}{\mathcal{K}}
\newcommand{\mL}{\mathcal{L}}

\newcommand{\mO}{\mathcal{O}}
\newcommand{\mS}{\mathcal{S}}
\newcommand{\mX}{\mathcal{X}}
\newcommand{\mZ}{\mathcal{Z}}

\newcommand{\kett}[1]{\left| #1 \middle\rangle\!\right\rangle}
\newcommand{\bbra}[1]{\left\langle\!\middle\langle #1 \right|}
\newcommand{\bigkett}[1]{\bigl| #1 \bigl\rangle\!\bigr\rangle}
\newcommand{\bigbbra}[1]{\bigl\langle\!\bigr\langle #1 \bigr|}
\newcommand{\ketbra}[2]{\Ket{#1}\!\Bra{#2}}
\newcommand{\bbrakett}[2]{\left\langle\!\middle\langle #1 \middle| #2 \middle\rangle\!\right\rangle}
\newcommand{\proj}[1]{\ketbra{#1}{#1}}
\newcommand{\projj}[1]{\kett{#1} \! \bbra{#1}}

\newcommand{\cAFL}{H_\textup{AFL}}
\newcommand{\AFL}{h_\textup{AFL}}
\newcommand{\shannon}[1]{H_\textup{S}(\{#1\})}
\newcommand{\sigAFL}[2][]{\sigma_{#1}\!\left[ #2 \right]}
\newcommand{\AFLstate}[2][]{\widetilde{\rho_{#1}}\!\left[ #2 \right]}
\newcommand{\DCTsum}{\bigoplus_{\lambda=1}^{d}}
\newcommand{\MTK}[2][]{U_{#1} X^{#2_t}_{#1} \cdots U_{#1} X^{#2_1}_{#1}}
\newcommand{\MTKd}[2][]{X^{#2_1 \dagger}_{#1} U^\dagger_{#1} \cdots X^{#2_t \dagger}_{#1} U^\dagger_{#1}}

\newcommand{\catmat}[4]{\begin{pmatrix} #1 & #2 \\ #3 & #4 \end{pmatrix}}
\newcommand{\smallcatmat}[4]{\big( \begin{smallmatrix} #1 & #2 \\ #3 & #4 \end{smallmatrix} \big)}
\newcommand{\torvec}[2]{\begin{pNiceMatrix} #1 \\ #2 \end{pNiceMatrix}}
\newcommand{\bfA}{\mathbf{A}}

\newcommand{\eqtext}[1]{\:\:\text{#1}\:\:}

\newcommand{\be}{\begin{equation}}
\newcommand{\ee}{\end{equation}}

\begin{document}

\title{Symmetry Properties of Quantum Dynamical Entropy}

\author{Eric D.~Schultz}
\email{schul211@purdue.edu}
\affiliation{Department of Physics and Astronomy, Purdue University, West Lafayette, Indiana 47906, USA}

\author{Keiichiro Furuya}
\email{k.furuya@northeastern.edu}
\affiliation{Department of Physics, Northeastern University, Boston, Massachusetts 02115, USA}

\author{Laimei Nie}
\email{nlm@purdue.edu}
\affiliation{Department of Physics and Astronomy, Purdue University, West Lafayette, Indiana 47906, USA}

\date{\today}

\begin{abstract}
As quantum analogs of the classical Kolmogorov-Sinai entropy, quantum dynamical entropies have emerged as important tools to characterize complex quantum dynamics. In particular, Alicki-Fannes-Lindblad (AFL) entropy, which quantifies the information production rate of a coherent quantum system subjected to repeated measurement, has received considerable attention as a potential diagnostic for quantum chaos.
Despite this interest, the precise behavior of quantum dynamical entropy in the presence of symmetry remains largely unexplored. In this work, we establish rigorous inequalities of the AFL entropy for arbitrary unitary dynamics (single-particle and many-body) in the presence of various types of symmetry. Our theorems encompass three cases: Abelian symmetry, an anticommuting unitary, and non-Abelian symmetries. 
In particular, we show that, while the cumulative AFL entropy generally saturates to the dimensional bound at late times for chaotic dynamics, this saturation value is distinctively lower when the measurements respect the symmetries.
We motivate our main results with numerical simulations of the perturbed quantum cat maps.
Our findings highlight the crucial role of symmetry in quantum dynamics under measurements, and our framework is readily adaptable for investigating symmetry's influence across diverse probes of quantum chaos.

\end{abstract}

\maketitle

\section{Introduction}

Complex dynamical phenomena in quantum systems, such as quantum chaos, thermalization, and information scrambling, have been the subject of extensive investigation.~\cite{dalessio-2016-QuantumChaos, abanin-2019-ColloquiumManybody, xu-2024-ScramblingDynamics}. Of the numerous operational tools proposed to characterize quantum dynamics, the ones that connect to classical chaos---for instance, out-of-time-ordered correlators (OTOC)~\cite{garcia-mata-2018-ChaosSignatures, fortes-2019-GaugingClassical, foini-2019-OTOCandETH, chan-2019-ETH, brenes-2021-OTOCandETH, xu-2024-ScramblingDynamics} and random matrix theory~\cite{bohigas-1984-CharacterizationChaotic, vikram-2023-DynamicalQuantum, Cotler-2017-chaoscomplexity, Kos-2018-analyticconnection}---have proven particularly insightful. Complementary to these are quantum information-theoretic approaches, exemplified by entanglement entropy and mutual information~\cite{hosur-2016-ChaosQuantum, nie-2019-SignatureQuantum, bianchi-2022-VolumeLawEntanglement}. A third avenue is to study quantum dynamics under measurements or dissipation, including measurement induced phase transitions (MIPT)~\cite{Li-2018-Zeno, Skinner-2019-mipt, potter-2022-MIPT, fisher-2023-MIPT, skinner-2023-MIPT} and quantum Ruelle-Policott (RP) resonances~\cite{nonnenmacher-2003-SpectralProperties, prosen-2004-RuelleResonances, garcia-mata-2005-SpectralApproach, yoshimura-2024-RobustnessQuantum, mori-2024-LiouvilliangapAnalysis, jacoby-2024-SpectralGaps, znidaric-2024-MomentumdependentQuantum, zhang-2024-ThermalizationRates, yoshimura2025irreversibility}.
This motivates the following question: is there a framework that seamlessly integrates the strengths of all three approaches? 
Indeed, quantum dynamical entropy, a concept rooted in the classical Kolmogorov-Sinai (KS) entropy, offers such a synthesis. 

In classical dynamical systems, the KS entropy serves as a key indicator of chaos~\cite{WaltersBook}. It is defined as the maximum asymptotic rate at which information about the dynamics is gained by successive measurements. These measurements refer to the partitioning of phase space into finite cells whose evolution yields a coarse-grained picture of the dynamics. If the dynamics is chaotic, the exponential sensitivity to initial conditions means further measurements may always yield more information of the underlying dynamics, leading to a nonzero KS entropy.

\begin{figure*}[t]
\includegraphics[width=0.97\linewidth]{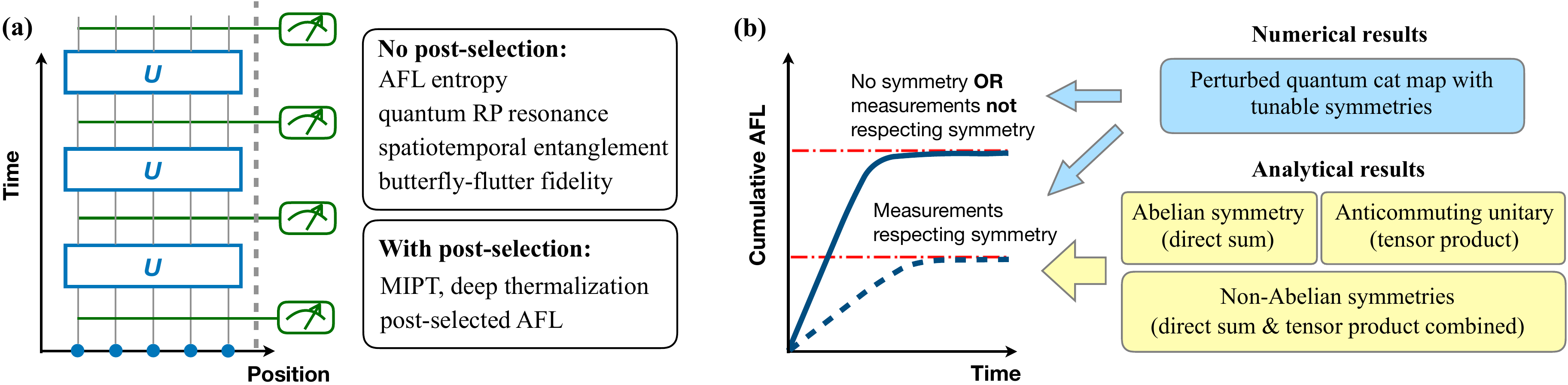}
\caption{\justifying \textbf{(a)} Overview of the general setup. 
The system undergoes repeated steps of measurement and discrete (Floquet) time evolution. Without post-selection, this generates entanglement with the auxiliary measurement/environment Hilbert space. The entanglement cut is represented by the dashed line. The AFL entropy, defined in Section~\ref{sect:dynamical-entropy}, is the maximal growth rate of this entanglement entropy over all possible measurement choices. The cumulative AFL entropy, defined in Section~\ref{sect:dynamical-entropy} as a more suitable measure for finite-dimension systems, is the entanglement entropy under a particular set of measurements. The same construction is used in the computation of quantum RP resonances~\cite{garcia-mata-2005-SpectralApproach, yoshimura-2025-TheoryIrreversibility}, spatiotemporal entanglement~\cite{cotler-2018-SuperdensityOperators, odonovan-2025-DiagnosingChaos}, and butterfly-flutter fidelity~\cite{dowling-2024-OperationalMetric}. Post-selecting on the measurement results yields a projected ensemble, as used in MIPT~\cite{Li-2018-Zeno, Skinner-2019-mipt}, deep thermalization~\cite{ippoliti-2022-DeepTherm}, and post-selected dynamical entropy~\cite{slomczynski-2017-QuantumDynamical}. This is discussed in more detail in Appendix~\ref{sect:projected-ensembles}. \textbf{(b)} A sketch of the typical behavior of cumulative AFL entropy as a function of time. For chaotic dynamics, the initial growth is linear until nearing the dimensional capacity for entanglement and saturating. In the presence of symmetry, the growth and saturation are lowered, provided the measurements are compatible with the symmetry. Our analytical results in Sections~\ref{sect:abelian},~\ref{sect:tensorproduct}, and~\ref{sect:nonabelian} (along with numerical results in Section~\ref{sect:catmapresults}) show that, the lowered value depends on how the Hilbert space decomposes under different types of symmetries. 
In contrast, our numerical results suggest the cumulative AFL entropy is not sensitive to the symmetries if the measurements do not respect them. Note that the analytical results apply to both single-particle and many-body unitary dynamics. }
\label{fig:intro}
\end{figure*}

Quantum dynamical entropies are generalizations of the classical KS entropy to noncommutative (quantum) dynamical systems. Various quantum dynamical entropies have been proposed with the goal of elucidating quantum chaos~\cite{connes-1975-EntropyAutomorphisms, connes-1987-DynamicalEntropy, slomczynski-1994-QuantumChaos, voiculescu-1995-DynamicalApproximation, alicki-1994-DefiningQuantum, slomczynski-2017-QuantumDynamical, Swingle2019LyapunovSpectrum, Goldfriend2021KSPesin, Maier2022holographicKS}. Among these, the Alicki-Fannes-Lindblad (AFL) entropy~\cite{alicki-1994-DefiningQuantum} has proven particularly useful. This entropy quantifies the entanglement generated per time step, by repeated measurement, between a quantum system and a measurement device (Fig.~\ref{fig:intro}(a)). Importantly, there is no post-selection on the measurement outcomes. AFL entropy has been used to quantify quantum chaos in semiclassical models~\cite{alicki-1996-KickedTop, alicki-2004-DecoherenceRate} and bound the communication capacity of quantum channels~\cite{alicki-1997-QuantumErgodic, alicki-2002-InformationtheoreticalMeaning}. More recently, AFL entropy and similar constructions were used to study spatiotemporal aspects of quantum information, such as entanglement cuts beyond only spatial~\cite{cotler-2018-SuperdensityOperators, odonovan-2025-DiagnosingChaos}. This setup has also been used to extract quantum RP resonances to compute the decay of correlation functions~\cite{garcia-mata-2005-SpectralApproach, yoshimura-2025-TheoryIrreversibility}, quantify relaxation to equilibrium~\cite{dowling-2023-RelaxationMultitime}, and define a quantum butterfly effect~\cite{dowling-2024-OperationalMetric}.

Alternatively, one may post-select on measurement outcomes to yield a projected ensemble of pure states. Projected ensembles see use in MIPT and deep thermalization~\cite{kaneko-2020-DeepTherm, ho-2022-DeepTherm, ippoliti-2022-DeepTherm, cotler-2023-DeepTherm, ippoliti-2023-DeepTherm, bhore-2023-DeepTherm, daniel-2024-DeepTherm}, and further define a ``post-selected'' quantum dynamical entropy~\cite{srinivas-1978-QuantumGeneralization, pechukas-1982-KolmogorovEntropy, beck-1992-SymbolicDynamics, kollar-2014-EntropyRate, slomczynski-2017-QuantumDynamical}.
Another common quantum dynamical entropy, studied primarily in mathematics, is the Connes-Narnhofer-Thirring (CNT) entropy, whose construction involves a set of completely positive maps between C$^*$-algebras~\cite{connes-1975-EntropyAutomorphisms, connes-1987-DynamicalEntropy}.

While quantum dynamical entropies have been extensively studied, their properties under quantum symmetries remain largely unexplored. In general, symmetries play a crucial role in quantum dynamics. Probes of quantum chaos are correspondingly sensitive to symmetry, as seen in level statistics~\cite{bohigas-1984-CharacterizationChaotic, Pandey-1993-symmetry-breaking, Javier-2020-symmetry-chaos}, eigenstate thermalization~\cite{murthy-2023-NonAbelianEigenstate, chang-2024-DeepThermCharges}, OTOC~\cite{Chen:2024pxm, Sharma:2024fqc}, and other measures~\cite{Bracken20, kudler-flam-2022-InformationScrambling, Balasubramanian:2024ghv}. Subsystem entanglement is also influenced by symmetries, in both single-particle~\cite{abreu-2006-EntanglingPower, santhanam-2008-EffectClassicalBifurcations} and many-body~\cite{Goldstein-2018-SymmetryResolvedEntanglement, bianchi-2024-NonAbelianSymmetryResolved} systems. Symmetries have further impact in the presence of measurement or noise. For example, recent work has shown symmetries in open systems enforce nontrivial mixed-state entanglement of the steady states~\cite{moharramipour-2024-SymmetryEnforcedEntanglement, li-2025-HighlyEntangled}. Our main contribution in this work is to establish rigorous inequalities for AFL entropy that reveal the impact of a wide variety of symmetries.

\subsection{Summary of Main Results}

In this study, we focus on finite-dimensional quantum systems. For such systems, the usual definition of AFL entropy vanishes due to the finite capacity for entanglement. Consequently, the \textit{cumulative} AFL entropy is a more meaningful measure. Cumulative AFL entropy, for chaotic dynamics, typically exhibits an initial linear growth phase with growth rate approaching the Lyapunov exponent in the classical limit, followed by saturation at a value determined by the system’s dimension (Fig.~\ref{fig:intro}(b)). This work proves inequalities for the cumulative AFL entropy in the presence of symmetry, using the saturation value as an easily-computable point of comparison with example numerics.

The presence of symmetry induces structure in the Hilbert space. 
Most familiar is that of Abelian symmetries, which imply the Hilbert space can be organized into a direct sum of charge sectors that do not mix under dynamics. We have found that, if the chosen measurements for computing cumulative AFL entropy do not couple charge sectors, then the entanglement with the measurement device is restricted to degrees of freedom \textit{within} a charge sector, lowering the entropy, as pictorially shown in Fig.~\ref{fig:intro}(b). The resulting saturation value is thus composed of contributions from each charge sector, determined by the number of sectors, their individual cumulative AFL entropies, and a weighting factor set by the initial state.
Our results phrase this phenomenon precisely and extend it to the cases of a non-Abelian symmetry algebra and the existence of a unitary anticommuting with the dynamics. For anticommuting unitary,  the Hilbert space acquires tensor product structure, whereas for non-Abelian symmetries, it decomposes into a direct sum of tensor products. Both lead to lowered values of cumulative AFL entropy at all times, given that the measurement is symmetry-compatible.
Conversely, our numerics show that when measurements do not respect a given symmetry, the cumulative AFL entropy remains unchanged, showing insensitivity to the symmetry's presence. 

Our study was motivated by numerical observations in quantum cat maps, a paradigmatic family of models for the study of quantum chaos (see~\cite{chen-2018-OperatorScrambling, garcia-mata-2018-ChaosSignatures, moudgalya-2019-OperatorSpreading, nie-2021-OperatorGrowth} for some recent work). 
They are notable for their well-understood structure, including a thoroughly characterized classical limit and tunable symmetries. In particular, we utilize their number-theoretic nature to explicitly construct the representation of a non-Abelian symmetry algebra and prove the existence of an anticommuting unitary. These features make quantum cat maps an ideal playground for exploring how symmetries influence quantum dynamics.

It is crucial to note that, while our numerical results focus on single-particle models, our analytical results are applicable to all unitary dynamics, including many-body systems. The framework used to prove our results can be readily applied for the study of symmetries in other probes of quantum dynamics.

This paper is organized as follows. Section~\ref{sect:dynamical-entropy} details the construction of AFL entropy, with an emphasis on finite dimensional systems. Section~\ref{sect:catmapresults} introduces the quantum cat map and presents numerical results of the cumulative AFL entropy with various types of symmetry. 
Our analytical results, applicable to any unitary dynamics, are presented in Sections~\ref{sect:abelian} (Abelian symmetry),~\ref{sect:tensorproduct} (ancicommuting unitaries) and~\ref{sect:nonabelian} (non-Abelian symmetries). These sections also explain our numerical results as special cases of our theorems.
Further discussions regarding (cumulative) AFL entropy's potential as a quantum chaos indicator, its relation to CNT entropy and Holevo information, as well as other future directions can be found in Section~\ref{sect:discussion}.


\section{Alicki-Fannes-Lindblad Entropy}
\label{sect:dynamical-entropy}

\subsection{Definitions}

The construction of AFL entropy is due to Alicki and Fannes~\cite{alicki-1994-DefiningQuantum} based on earlier work by Lindblad~\cite{lindblad-1979-NonMarkovianQuantum,lindblad-1991-QuantumEntropy}, and is analogous to that of KS entropy (see also~\cite{alicki-1996-KickedTop, alicki-1997-QuantumErgodic, alicki-2002-InformationtheoreticalMeaning, alicki-2004-DecoherenceRate}). The system $S$ is governed by a discrete unitary dynamics $U$ and a density matrix $\rho$ on an $N$-dimensional Hilbert space $\mH_S$. The state $\rho$ is analogous to the measure in classical dynamical systems, and so is typically taken to be stationary with $[U,\rho]=0$ (i.e.~an equilibrium state), but this is not strictly necessary. A generalized measurement on the system may be represented with Kraus operators $\mX = \{X^1,\dots,X^K\}$ with $\sum_i X^{i\dagger} X^i = \id$~\cite{nielsen-2010-QuantumComputation, wilde-2021-ClassicalQuantum}. The measurement channel (without post-selection) $\mE_\mX$ acts as $\mE_\mX(\rho) = \sum_i X^i \rho X^{i\dagger}$. In analogy to the phase-space partition of KS entropy, $\mX$ is referred to as an \textit{operational partition of unity} (or simply a \textit{partition}) of size $K$.

\subsubsection{Entropy Exchange}

The measurement channel may be written as a unitary process on an enlarged Hilbert space including the environment (measurement device). In this view, the channel generates entanglement between the system and environment known as entropy exchange~\cite{lindblad-1991-QuantumEntropy, schumacher-1996-SendingEntanglement}. To discount the entanglement entropy present in $\rho$ prior to the channel, we purify the state by introducing a purifier $P$ with Hilbert space $\mH_P$ isomorphic to $\mH_S$. Diagonalizing the density matrix as 
$\rho = \sum_\alpha r_\alpha \ketbra{\psi_\alpha}{\psi_\alpha}$,
a canonical purification is
$\kett{\sqrt{\rho}} = \sum_\alpha \sqrt{r_\alpha} \ket{\psi_\alpha}\!\ket{\psi_\alpha}$, which lives in the doubled Hilbert space $\mH_S \otimes \mH_P$.
Here, $\kett{\cdot}$ notates the Choi state of an operator with respect to the eigenbasis of $\rho$, given by
$\kett{\mO} = \sum_{\alpha\beta} \mO_{\alpha\beta} \ket{\psi_\alpha}\!\ket{\psi_\beta}$ for an operator $\mO$ on $\mH_S$~\cite{wilde-2021-ClassicalQuantum}.

\begin{figure}[ht]
    \includegraphics[width=0.75\columnwidth]{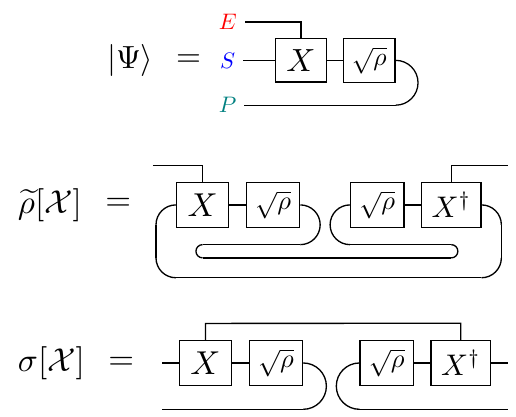}
    \caption{\justifying Tensor network representations of $\ket{\Psi}$, $\Tilde{\rho}[\mX]$, and $\sigma[\mX]$. These are the pure state on $ESP$ and the reduced states on $E$ and $SP$ respectively. The top line of $X$ is the Kraus index, which is the environment Hilbert space $\mH_E$. See~\cite{cotler-2018-SuperdensityOperators} for further diagrams and~\cite{bridgeman-2017-HandwavingInterpretive, biamonte-2020-LecturesQuantum} for details on this notation more generally.}
    \label{fig:AFL_state}
\end{figure}

Given a partition of size $K$, the measurement outcomes correspond to orthogonal states $\ket{1},\dots,\ket{K}$ in the environment $E$ with Hilbert space $\mH_E$. Dilating the channel to a unitary process on $\mH_E \otimes \mH_S \otimes \mH_P$, a measurement maps the state $\ket{1}\kett{\sqrt{\rho}}$ in this bigger Hilbert space to
\begin{equation} \label{eq:measurement-channel}
    \ket{1}\kett{\sqrt{\rho}} \mapsto \sum_{i=1}^{K} \ket{i}\kett{X^i\sqrt{\rho}} \eqqcolon \ket{\Psi}
\end{equation}
which one may check is normalized. Tracing out $SP$, the state of the environment is
\begin{equation} \label{eq:AFL-state}
    \Tilde{\rho}[\mX] = \Tr_{ SP}\ketbra{\Psi}{\Psi}
    = \sum_{ij} \Tr(X^i \rho X^{j\dagger}) \ketbra{i}{j}.
\end{equation}
The \textit{entropy exchange} is defined as the entanglement with the environment generated by the measurement. More precisely, it is the von Neumann entropy of $\Tilde{\rho}[\mX]$,
\begin{equation}
S(\Tilde{\rho}[\mX]) = -\Tr (\Tilde{\rho}[\mX] \log \Tilde{\rho}[\mX]).
\end{equation}
Equivalently, we can compute the entropy exchange by tracing out $E$ to get the state on $SP$~\cite{alicki-1996-KickedTop}
\begin{equation}
    \sigma[\mX]
    = \Tr_E \ketbra{\Psi}{\Psi}
    = \sum_i \projj{X^i \sqrt{\rho}}.
\end{equation}
Since the state $\ket{\Psi}$ on $ESP$ is pure, this has the same von Neumann entropy as $\Tilde{\rho}[\mX]$:
\begin{equation}
    S(\sigma[\mX]) = -\Tr(\sigma[\mX] \log \sigma[\mX]) = S(\Tilde{\rho}[\mX]).
\end{equation}
Both constructions are shown graphically in Fig.~\ref{fig:AFL_state}.

\subsubsection{AFL Entropy}

\begin{figure}[t]
    \includegraphics[width=0.8\columnwidth]{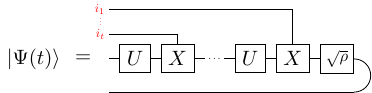}
    \caption{\justifying Tensor network representation of the pure state $\ket{\Psi}$ after $t$ time steps of measurement $X$ and unitary evolution $U$.}
    \label{fig:multitime}
\end{figure}

In analogy with KS entropy, AFL entropy is the maximum rate of entropy production from periodic measurements. This means we apply the measurement channel~\eqref{eq:measurement-channel} after each discrete evolution $U$. The graphical representation easily incorporates multiple time steps as shown in Fig.~\ref{fig:multitime}. This is equivalent to acting with a time-evolved partition, which we notate after $t$ time steps as
\begin{equation}
    (U\mX)^t \coloneqq \Set{U X^{i_t} \cdots U X^{i_2} U X^{i_1} | i_j = 1,\dots,K}.
\end{equation}
The \textit{cumulative AFL entropy} is defined as the entropy exchange of this multitime measurement channel:
\begin{equation}
    \cAFL(\rho, U,\mX,t) = S(\Tilde{\rho}[(U\mX)^t])  = S(\sigma[(U\mX)^t]).
\end{equation}
The \textit{AFL entropy}, as with the classical KS entropy, is defined as the asymptotic growth rate of $\cAFL$ maximized over all partitions (measurements):
\begin{equation} \label{eq:AFL-def}
    \AFL(\rho, U) = \sup_\mX \limsup_{t\rightarrow\infty} \frac{1}{t} \cAFL(\rho, U,\mX,t). 
\end{equation}
This definition is readily generalized to C$^*$-algebras, including infinite-dimensional quantum systems and classical dynamical systems. In the classical case, the algebra of observables is commutative, but the formalism above still holds and yields the classical KS entropy~\cite{alicki-1996-AlgebraicApproach, alicki-1998-QuantumMechanical}. 
One can further obtain a post-selected version of the AFL entropy, which involves an additional dephasing step detailed in Appendix~\ref{sect:projected-ensembles}.

One may note a curious fact that replacing $U \mapsto \id$ and $\mX \mapsto U\mX$ yields the same value for the cumulative entropy $\cAFL$. This is because quantum measurement itself may add uncertainty to the monitored dynamics via noncommutivity of the Kraus operators. AFL entropy makes no distinction between the entropy arising from the dynamics $U$ and from the measurement $\mX$~\cite{alicki-2004-DecoherenceRate}. This also motivates restricting the choice of Kraus operators to avoid this ambiguity. For instance, choosing commuting Kraus operators guarantees all entropy growth is due to the dynamics $U$ under $\mX$ and not from $\mX$ alone. Another motivated choice is to demand the measurement respect the symmetry structure of $U$, which is the focus of this work.

\subsection{AFL entropy in Finite Dimensions}
\label{sect:AFL_finiteD}

Cumulative AFL entropy, as the entanglement entropy of a subsystem of a pure state, is dimensionally bounded~\cite{alicki-2004-DecoherenceRate}. 
The first dimensional bound comes from the environment. The dimension of $\mH_E$ at time $t$ is at most $K^t$, so
\begin{equation}
    \cAFL(\rho, U,\mX,t) \leq t \log K
\end{equation}
The complementary bound comes from system and purifier. Further traces of the state $\sigma[\mX]$ in $SP$ space give
\begin{align}
    \Tr_S (\sigma[\mX]) &= \rho^\mathrm{T} = \rho^* \\
    \Tr_P (\sigma[\mX]) &= \mE_\mX(\rho)
\end{align}
as expected. By subaddivitivy of von Neumann entropy, the cumulative AFL entropy then obeys
\begin{align}
    \cAFL(\rho, U,\mX,t) &= S(\sigma[(U\mX)^t]) \nonumber \\
    &\leq S(\rho) + S(\mE_{(U\mX)^t} (\rho)) \nonumber \\
    &\leq S(\rho) + \log N \label{eq:dim-bound-state}
\end{align}
For the special case of a maximally mixed $\rho$, the dimensional bound takes the simple form
\begin{equation}
    \cAFL(\rho, U,\mX,t) \leq 2 \log N. \label{eq:dim-bound}
\end{equation}
In summary, for any initial state $\rho$, the cumulative AFL entropy is bounded above by a finite constant. This means the asymptotic growth rate must vanish: $\AFL(\rho, U) = 0$ for all dynamics $U$ on $\mH_S$.

From this property, AFL entropy $\AFL$ is fundamentally ill-suited for studying quantum chaos in finite dimensions. The cumulative AFL entropy $\cAFL$, on the other hand, has been proposed as a possible alternative, potentially offering insight beyond the vanishing limit of equation~\eqref{eq:AFL-def}. In practice, $\cAFL$ is typically evaluated numerically. This involves selecting an initial state $\rho$ and partition $\mX$, and then tracking the density matrix of the $SP$ space, $\sigma[(U\mX)^t]$. The advantage of using $\sigma[(U\mX)^t]$ instead of $\Tilde{\rho}[(U\mX)^t]$ of the environment space is that the former maintains a constant size over time.

Typically, the cumulative AFL entropy will approach its saturation value as an exponential at late times. This was argued in Ref.~\cite{cotler-2018-SuperdensityOperators} using the spectrum of the corresponding quantum channel. The particular bound~\eqref{eq:dim-bound-state} assumes $\mE_{(U\mX)^t} (\rho)$ approaches the maximally mixed state at late times, or in other words that $\id/N$ is the unique eigenoperator of the channel with unit eigenvalue. The symmetry-resolved entropy inequalities we will introduce in later sections can be seen as choosing measurements such that $\id/N$ is not the unique steady-state and choosing initial states that overlap with these new steady-states.

Several studies have examined the behavior of $\cAFL$ in various models.
In the quantum kicked top and quantum baker's map, numerics have indicated that cumulative AFL can identify the presence or absence of semiclassical chaos through, for example, early time%
\footnote{\textit{Early time} refers to times prior to the semiclassical Ehrenfest time $t_\text{E} \sim |\log(\hbar)| / h_{\text{KS}}$ where $h_{\text{KS}}$ is the Kolmogorov-Sinai entropy of the dynamics~\cite{benatti-2003-ClassicalLimit, alicki-2004-DecoherenceRate, shepelyansky-2020-EhrenfestTime}.}
$\cAFL$'s growth rate matching the corresponding classical Lyapunov exponent~\cite{alicki-1996-KickedTop, alicki-2004-DecoherenceRate}.
This matching has been proved rigorously in the infinite-dimensional limit of the quantum cat map~\cite{benatti-2003-ClassicalLimit}. The cumulative AFL entropy was also studied on free fermions and the Sachdev-Ye-Kitaev model under the name ``spacetime entropy''~\cite{cotler-2018-SuperdensityOperators}. Spin chains have also seen study using the second R\'{e}nyi entropy of $\Tilde{\rho}$ instead of von Neumann entropy~\cite{odonovan-2025-DiagnosingChaos}.

Despite these initial promising observations, the suitability of $\cAFL$ for diagnosing quantum chaos, especially in finite dimensions, remains an open question. 
We address this gap by focusing on the influence of symmetries, which impose significant constraints on the quantum dynamics.
In particular, we will show that the effectiveness of $\cAFL$ strongly depends on the choice of partitions (measurements). When partitions are chosen to be compatible with a system's symmetries, the growth of $\cAFL$ is limited, saturating at a lower value. However, as shown in the next section, in practice the challenge lies in identifying these symmetries a priori, as their absence can lead to $\cAFL$ saturation that mimics fully chaotic dynamics.

\section{Example Numerics: Cumulative AFL in Quantum Cat Map}
\label{sect:catmapresults}

Here, we present numerics showcasing the behaviors of cumulative AFL entropy in the perturbed quantum cat maps, a well-studied family of single particle models, in the presence of various types of symmetry. Our numerical results are consistent with rigorous theorems proved in Sec.~\ref{sect:abelian}, \ref{sect:tensorproduct}, and \ref{sect:nonabelian}. We stress that the presented theorems are fully general and apply to \textit{any} unitary dynamics, including many-particle models. Our choice of the cat map for numerics is motivated by its number-theoretic nature, which allows us to prove the existence of an anticommuting unitary and analytically compute the representation of a non-Abelian symmetry algebra for easy comparison with our general results. Our code is available on GitHub at~\cite{github-link}.

\subsection{The Quantum Cat Map and its Symmetries}

 The unitary dynamics of the quantum cat map is derived from the Arnol'd cat map, a discrete classical map on torus $\dsT^2$~\cite{arnold-1968-ErgodicProblems}. For more information on this map and its quantization, see Appendix~\ref{sect:cat-appendix} and references therein. In our work, we choose the perturbed cat map
\begin{multline} \label{eq:pert-cat-map}
    \torvec{q}{p} \mapsto
    \catmat{A_{11}}{A_{12}}{A_{21}}{A_{22}} \torvec{q}{p} \\
    + \frac{\kappa}{2\pi} \cos (2\pi q) \torvec{A_{12}}{A_{22}} \mod 1
\end{multline}
where $(q,p) \in \dsT^2$, $\bf{A} = \smallcatmat{A_{11}}{A_{12}}{A_{21}}{A_{22}}$ is an $\operatorname{SL}(2,\dsZ)$ matrix, and $\kappa$ is a small perturbation. The notation $(q,p)$ is meant to evoke position and conjugate momentum forming a toral phase space, which is the identification made for quantization. Demanding wavefunctions respect the periodicity of the torus (set to unity) in both $q$ and its Fourier partner $p$ yields a finite dimensional Hilbert space with rational Planck's constant $h = 1 / N$, where $N$ is the Hilbert space dimension~\cite{hannay-1980-QuantizationLinear, esposti-1993-QuantizationOrientation}. The Hilbert space is spanned by position eigenstates $\ket{q_j}$ delta-localized to $q_j = j / N$ for $j=0,\dots,N-1$ (more generally, the index is identified modulo $N$). For a given classical map defined by $\bfA$ and $\kappa$, the quantized unitary map $U$ is defined by the semiclassical (stationary phase) formula for the propagator~\cite{hannay-1980-QuantizationLinear, matos-1993-QuantizationAnosov}.

\subsubsection{Abelian Symmetry}
    The quantum cat map at a given dimension $N$ can host several number-theoretic symmetries~\cite{kurlberg-2000-HeckeTheory, keating-2000-PseudosymmetriesAnosov, esposti-2005-QuantumPerturbed}. We will focus the momentum shift $R$ and inversion $W$ from Ref.~\cite{esposti-2005-QuantumPerturbed}, given a perturbation of the form~\eqref{eq:pert-cat-map}. $R$ occurs when the dimension $N$ is an integer multiple of $s \coloneqq \gcd (A_{12}, A_{22}-1)$ for odd $A_{12}$, where $\gcd$ is the greatest common divisor. Then the momentum kick
    \begin{equation} \label{eq:cat-R}
        R \ket{q_j} = \exp \left( i \frac{2\pi j}{s} \right) \ket{q_j}
    \end{equation}
    commutes with the dynamics $U$ and generates a symmetry group $\dsZ_s$. $W$ is the quantization of the classical map
    \begin{equation}
        \torvec{q}{p} \mapsto \torvec{\frac{1}{2}-q}{\frac{1}{2}-p}
    \end{equation}
    which takes the form
    \begin{equation} \label{eq:cat-W}
        W \ket{q_j} = (-1)^j \ket{q_{\frac{N}{2}-j}}
    \end{equation}
    for even dimension $N$. The $W$ operator commutes with the dynamics $U$ when $N$ is divisible by 4 and generates a $\dsZ_2$ symmetry.

\subsubsection{Anticommuting Unitary}
    When $N$ is even but \textit{not} divisible by 4, and with particular constraints on $\bfA$, the $W$ operator anticommutes with $U$. This is proved in Appendix~\ref{sect:cat-AC-proof}. The existence of an anticommuting unitary induces a pseudospin structure in the unitary dynamics:
    \begin{equation}\label{eq:pseudospin-U}
        U = \sigma_z \otimes \overline{U}
    \end{equation}
    where $\sigma_z$ is the Pauli-z matrix acting on a pseudospin, and $\overline{U}$ a unitary. The details of this are discussed in Sec.~\ref{subsec:anticommuting}.

\subsubsection{Non-Abelian Symmetry}
    The $R$ and $W$ operators do not commute, and so generate a non-Abelian symmetry algebra when both symmetries are present. The irreducible representations of the algebra must be either one-dimensional (simultaneous eigenspaces of $R$ and $W$) or two-dimensional, since $W$ exchanges two eigenvectors of $R$. Multiplicity of the representations when embedded in the cat map Hilbert space means each representation is tensored with a vector space on which the algebra acts trivially (as a multiple of identity), sometimes known as the multiplicity space. Representing such algebras is discussed in more detail in Sec.~\ref{sect:nonabelian}, while computation of the algebra generated by $R$ and $W$ is reserved for Appendix~\ref{sect:cat-nonabelian-algebra}.

\subsubsection{Quantum Chaos in the Cat Map}
    In general, the energy spectra of the perturbed quantum cat maps possess random matrix statistics~\cite{matos-1993-QuantizationAnosov, keating-2000-PseudosymmetriesAnosov, backer-2003-NumericalAspects}. However, once $R$ or $W$ symmetry is present the spectral statistics deviate from random matrices, which can be interpreted as the superposition of spectra from independent symmetry sectors~\cite{esposti-2005-QuantumPerturbed}. In the case of $\{ W, U \} = 0$, equation~\eqref{eq:pseudospin-U} implies the quasi-energies appear in pairs $(E, E + \pi)$. Consequently, only half of the quasi-energies show meaningful spectral statistics. The perturbed cat maps also display Lyapunov growth in OTOC~\cite{chen-2018-OperatorScrambling, garcia-mata-2018-ChaosSignatures}, and a non-vanishing gap in the noisy quantum propagator in the semiclassical (large-$N$) and small noise limit~\cite{nonnenmacher-2003-SpectralProperties, garcia-mata-2003-ClassicalDecays, garcia-mata-2004-SpectralProperties, garcia-mata-2005-SpectralApproach}, reminiscent of the recently studied quantum Ruelle-Pollicott resonance in many-body systems~\cite{yoshimura-2024-RobustnessQuantum, mori-2024-LiouvilliangapAnalysis, jacoby-2024-SpectralGaps, znidaric-2024-MomentumdependentQuantum, zhang-2024-ThermalizationRates, yoshimura2025irreversibility}.


\subsection{Cumulative AFL in Quantum Cat Map}

Unless otherwise noted, the perturbation strength is fixed at $\kappa = 0.05$ throughout our numerics. We compute the cumulative AFL entropy with respect to the maximally mixed state and choose the partitions to always be commuting projectors (that is, $[X^i, X^j] = 0$ and $(X^i)^2 = X^i = X^{i\dagger}$). Cumulative AFL is nondecreasing under such a partition, as the resulting channel $\mE_\mX \otimes \id$ on the system and purifier is doubly stochastic.%
\footnote{Doubly stochastic channels are completely positive, trace preserving, and unital (sends $\id \mapsto \id$). This means that $\id = \sum_i X^{i\dagger}X^i = \sum_i X^iX^{i\dagger}$. Such channels do not decrease von Neumann entropy~\cite{nielsen-2002-IntroductionMajorization}.}

For partitions that do not respect any symmetry, we choose to have $\mO(1)$-many projectors that are diagonal in a random basis, which we call a \textbf{random partition}. For chaotic dynamics, under such a partition the cumulative AFL entropy is expected to asymptote to the dimensional bound $S(\rho) + \log N = 2 \log N$ as an exponential~\cite{alicki-1996-KickedTop, cotler-2018-SuperdensityOperators}.
For partitions respecting an Abelian symmetry $Z$, we choose a random partition independently in each charge sector, which we call a \textbf{$Z$-symmetric partition}.
For partitions respecting an anticommuting unitary, we pick a \textbf{tensor product partition} which is the product of a channel on the pseudospin space (e.g.~pseudospin-z measurement or identity channel) and a random partition on the non-pseudospin space.
For partitions respecting a non-Abelian symmetry, we notice that a $Z$-symmetric partition is built from of partitions in the multiplicity space of each one-dimensional irrep of the Abelian symmetry. Generalizing this, we construct a \textbf{commutant partition} by, for each distinct irrep of the algebra, choosing a random partition in the corresponding multiplicity space and acting trivially on the irrep itself. This implies each Kraus operator commutes with the whole algebra (in other words, they are in the commutant).
Further details, including the relaxed conditions on the partitions, and the complete analysis of the numerical results in light of the analytical results (Theorems~\ref{thm:abelian},~\ref{thm:tensor-dynamics},~\ref{thm:nonabelian}), can be found in Sections~\ref{sect:abelian}, \ref{sect:tensorproduct}, and \ref{sect:nonabelian}.


\begin{figure}[t]
    \includegraphics[width=0.93\columnwidth]{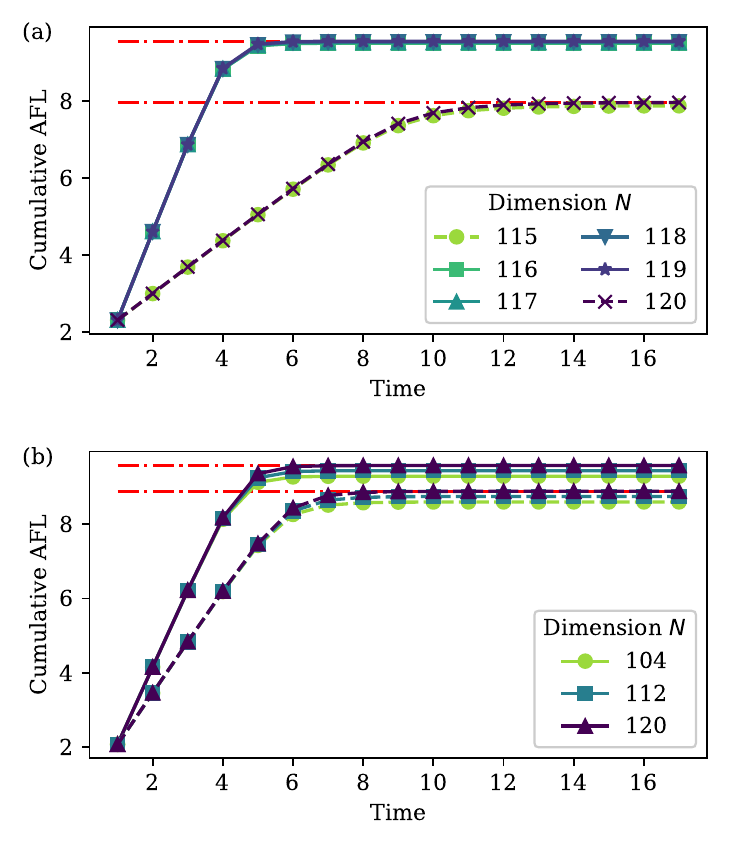}
    \caption{\justifying Cumulative AFL entropy of quantum cat map with Abelian symmetries. \textbf{(a)} $R$ symmetry. The plot shows the cumulative AFL entropy with an $R$-symmetric partition. $R$ is a symmetry of the dynamics for dimensions 115 and 120 (dashed lines) only. The horizontal dash-dotted lines are the dimensional bounds for the largest appropriate dimension plotted: $2 \log N$ for no symmetry $(N=119)$, and the lowered bound $2 \log M + \log s$ for $R$ symmetry $(N=120)$.  
    \textbf{(b)} $W$ symmetry. The cat maps for all dimensions plotted are symmetric under $W$. The plot shows the cumulative AFL entropy  with random partitions (solid lines) and $W$-symmetric partitions (dashed lines). The horizontal dash-dotted lines show the dimensional bounds for the largest appropriate dimension plotted $(N=120)$: $2\log N$ for random partition, and the lowered bound of $2\log(N/2) + \log 2$ for $W$-symmetric partition.}
    \label{fig:Abelian}
\end{figure}

\subsubsection{Abelian Symmetries}
    The cumulative AFL entropy for the Abelian $R$ and $W$ symmetries is plotted in Fig.~\ref{fig:Abelian}(a) and Fig.~\ref{fig:Abelian}(b) respectively.
    For the $R$ operator, we choose the cat map matrix
    \begin{equation} \label{eq:cat-6576}
        \bfA = \catmat{6}{5}{7}{6}
    \end{equation}
    with $s = \gcd(A_{12},A_{22}-1) = 5$. We compute the cumulative AFL entropy on $R$-symmetric partitions of size $K=10$ (2 Kraus operators per charge sector). If $N = sM$, the dynamics commutes with $R$ and has $s$ charge sectors of dimension $M$. In this case, the late-time saturation of $\cAFL$ reduces from $2\log N$ to $2\log M + \log s$, as one might expect of independent/uncoupled dynamics. We may interpret the $\log s$ as the Shannon entropy of choosing which sector to start in. We note that we still expect the channel $\mE_{U\mX}$ to have steady-states that are maximally mixed within each charge sector (possibly with different weights, yielding the $\log M$ factor) since $U\mX$ has no further symmetries to resolve and is still doubly-stochastic within each sector. In other words, to get a bound lower than $2\log M + \log s$ the channel must admit a steady state that is \textit{not} maximally mixed in each charge sector, which implies further structure beyond our construction here.
    
    For the $W$ operator, we choose the classical cat map
    \begin{equation} \label{eq:cat-2132}
        \bfA = \catmat{2}{1}{3}{2}
    \end{equation}
    which has no nontrivial $R$ symmetry, but commutes with $W$ when $4 | N$. The cumulative AFL entropy is computed on $W$-symmetric dynamics with random and $W$-symmetric partitions of total size $K=8$. The $W$ operator has two charge sectors, and correspondingly $\cAFL$ saturates to $2\log (N/2) + \log 2$ for $W$-symmetric partitions, instead of the usual $2\log N$.

\begin{figure}[t]
        \includegraphics[width=0.9\columnwidth]{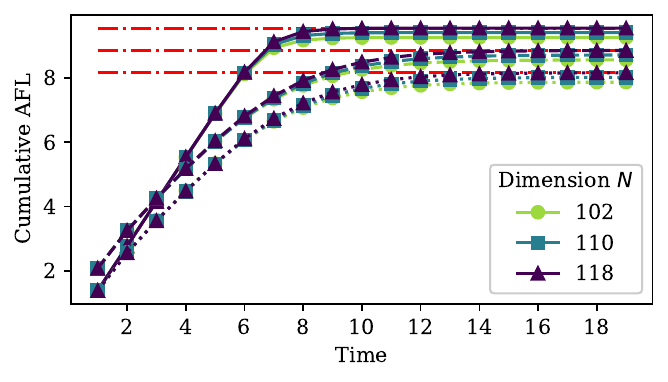}
        \caption{\justifying Cumulative AFL entropy of quantum cat map with an anticommuting unitary. The figure compares random partitions (solid lines) and two tensor product partitions:  measurement of pseudospin-z (dashed lines), and no pseudospin measurement (dotted lines). The respective bounds of $2\log N$, $2\log(N/2) + \log 2$, and $2\log(N/2)$ for $N=118$ are shown by the horizontal dash-dotted lines.}
        \label{fig:cat-W-AC}
\end{figure}

\subsubsection{Anticommuting Unitary}
    When $N$ is even but not divisible by 4, the dynamics $U$ of the map~\eqref{eq:cat-2132} obeys the anticommutation condition $\{W, U\} = 0$. In Fig.~\ref{fig:cat-W-AC}, we compare random partitions on the whole space to tensor product partitions with pseudospin-z measurement and no pseudospin measurement (identity channel). The respective dimensional bounds are $2\log N$, $2\log(N/2) + \log 2$, and $2\log(N/2)$ as shown. The pseudospin dynamics are $\sigma_z$, which contributes a constant one bit ($\log 2$) of entropy when measured. In all cases, the random partition has size $K=4$.

\subsubsection{Non-Abelian Symmetry}
\begin{figure}[t]
    \includegraphics[width=0.9\columnwidth]{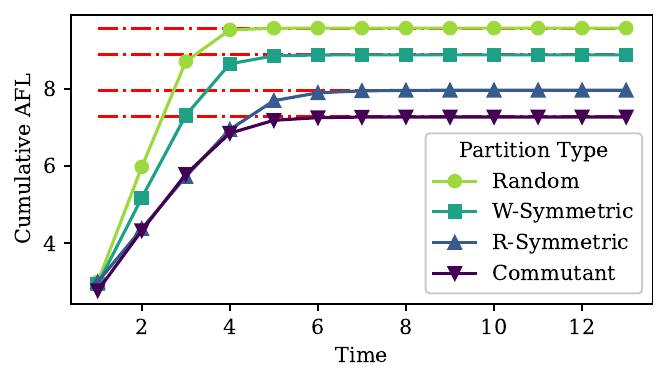}
    \caption{\justifying Cumulative AFL entropy of quantum cat map with non-Abelian symmetries. Here, both $R$ and $W$ symmetries are present, and various partition types are used. The horizontal dash-dotted lines show the expected bounds in each case as described in the text.}
    \label{fig:cat-nonabelian}
\end{figure}

    We choose the cat map matrix~\eqref{eq:cat-6576} and Hilbert space dimension $N=120$ (divisible by 4 and by $s=5$) so that both the $R$ and $W$ symmetries are present. The cumulative AFL entropy is plotted in Fig.~\ref{fig:cat-nonabelian} for random, symmetric, and commutant partitions of size 20. The random, $R$-symmetric, and $W$-symmetric partitions have respective bounds of $2\log N$, $2\log M + \log s$, and $2\log (N/2) + \log 2$ as discussed previously, but the commutant partition is even lower. The precise form of the bound is explained in Sec.~\ref{sect:nonabelian}.

\section{Analytical Result: Abelian Symmetry}
\label{sect:abelian}

Below we provide rigorous results on $\cAFL$ in the presence of Abelian symmetries for any unitary dynamics $U$ (including many-body systems)  and apply them to explain the results depicted in Fig~\ref{fig:Abelian}. To clean up notation, we drop the $S$ subscript from the system Hilbert space ($\mH = \mH_S$).


Take a Hermitian operator $Z$ on a finite-dimensional Hilbert space $\mH$ (in our numerics, we have $Z=W$ or $Z=R$, and $\mH$ is the system Hilbert space $\mH_S$). We may diagonalize the operator as $Z = \sum_{\lambda, \ell} z_\lambda \proj{\lambda, \ell}$ where $\lambda=1,\dots,d$ index the distinct eigenvalues $z_\lambda$ of $Z$, and $\ell$ runs over the degeneracy. Any such eigenbasis of $Z$ decomposes the Hilbert space into orthogonal subspaces $\mH_\lambda = \Span \{ \ket{\lambda, \ell} \}_\ell$ such that
\begin{equation} \label{eq:abelian-dirsum}
    \mH = \bigoplus_\lambda \mH_\lambda \eqtext{and} Z = \bigoplus_\lambda z_\lambda \id_{\mH_\lambda}.
\end{equation}
We will also refer to $\mH_\lambda$ as a charge sector.

Now consider some other linear operator $Q$ on $\mH$ that commutes with $Z$. From equation~\eqref{eq:abelian-dirsum}, it is clear that $[Q, Z] = 0$ if and only if $\mH_\lambda$ is stable under $Q$, meaning $Q\mH_\lambda \subseteq \mH_\lambda$ for all $\lambda$. Then $Q$ has a direct sum decomposition $Q = \bigoplus_\lambda Q_\lambda$ where $Q_\lambda = Q |_{\mH_\lambda}$ is the restriction to the $\lambda^\text{th}$ charge sector. Intuitively, in some eigenbasis grouped by charges of $Z$, the matrix forms of $Z$ and $Q$ are block diagonal:
\begin{equation*}
    Z = \BlockMatrix[6pt]{z_1 \id_{\mH_1}}{z_d \id_{\mH_d}}
    \eqtext{and}
    Q = \BlockMatrix{Q_1}{Q_d}.
\end{equation*}
We will occasionally abuse notation and also use $Q_\lambda$ to refer to the projection $\sum_\ell \braket{\lambda, \ell | Q | \lambda, \ell} \proj{\lambda, \ell}$ acting on full Hilbert space $\mH$. We will note a few final aspects and notations of operators commuting with $Z$.

Due to the block diagonal structure, there are no cross terms when composing direct-sum operators. Given any two operators, $Q$ and $Q'$, that both commute with $Z$, we have
\begin{equation}
    QQ' = \left( \bigoplus_\lambda Q_\lambda \right)
    \left( \bigoplus_\lambda Q'_\lambda \right)
    = \bigoplus_\lambda Q_\lambda Q'_\lambda
\end{equation}
which is clear from the matrix form
\begin{equation*}
    \BlockMatrix{Q_1}{Q_d}
    \BlockMatrix{Q'_1}{Q'_d}
    = \BlockMatrix[5pt]{Q_1 Q'_1}{Q_d Q'_d}.
\end{equation*}

When the operator commuting with $Z$ is a density matrix $\rho$, we will alter the notation by defining a density matrix $\rho_\lambda$ on $\mH_\lambda$ and a probability distribution $\{ p_\lambda \}$ given by
\begin{equation}
    p_\lambda = \Tr \left( \rho|_{\mH_\lambda} \right)
    \eqtext{and} \rho_\lambda = \frac{1}{p_\lambda} \rho|_{\mH_\lambda},
\end{equation}
from which we can write the state as a convex direct sum
\begin{equation}\label{eq:convex-sum}
    \rho = \bigoplus_\lambda p_\lambda \rho_\lambda.
\end{equation}

Given a partition of unity $\mX = \{X^1, \dots, X^K\}$ on $\mH$, if all the operators in $\mX$ commute with $Z$ then the set of restrictions to $\mH_\lambda$ is a partition of unity on $\mH_\lambda$, which we denote $\mX_\lambda = \{X^1_\lambda, \dots, X^K_\lambda\}$.\footnote{In open systems literature, the case given here where all the Kraus operators commute with the symmetry generator is known as a strong symmetry~\cite{buca-2012-SymmetryReductions}.} As matrices, this looks like
\begin{equation*}
    \sum_i
    \BlockMatrix{X^{i\dagger}_1}{X^{i\dagger}_d}
    \BlockMatrix{X^i_1}{X^i_d}
    = \BlockMatrix{\id_{\mH_1}}{\id_{\mH_d}}.
\end{equation*}

\subsection{Theorem 1}
For convenience, we denote the Shannon entropy of a probability distribution $\{ p_i \}$ as $\shannon{p_i} = -\sum_i p_i \log p_i$. We have the following theorem regarding the cumulative AFL entropy $\cAFL$ in the presence of a symmetry operator $Z$:

\begin{theorem} \label{thm:abelian}
    If the dynamics $U$, density matrix $\rho$, and all the operators in the partition $\mX$ commute with a Hermitian operator $Z$, then 
    \begin{multline}
        \cAFL(\rho, U,\mX,t) \leq \smash{\sum_\lambda}\, p_\lambda \cAFL(\rho_\lambda, U_\lambda, \mX_\lambda, t)
        \\ + \shannon{p_\lambda}
    \end{multline}
    with equality if the measurement channel admits a Kraus representation where $X^i \in \mX$ have support on exactly one charge sector each.
\end{theorem}

The equality condition means there is a map $\phi$ from a Kraus index to the corresponding charge index. We can then write
\begin{equation}
    X^i_\lambda = X^i |_{\mH_\lambda} = \delta(\phi(i) = \lambda) X^i_{\phi(i)} \label{eq:single-charge}
\end{equation}
which is visually represented as
\begin{equation*}
    X^i =
    \BlockMatrix{X^i_1}{X^i_d} =
    \begin{pNiceMatrix}[margin][columns-width=2pt]
        \Block[name=b1]{2-1}{\bm{0}} & & & & \Block{2-2}{\bm{0}} \\ 
        & & & & & \\
        & & \Block[name=b2]{2-2}{X^i_{\phi(i)}} & & \\ 
        & & & & & \\
        \Block{2-2}{\bm{0}} & & & & & \Block[name=b3]{2-1}{\bm{0}} \\
        & & & & &
    \CodeAfter
        \line[radius=0.6pt,inter=0.4em,shorten=0.3em]{b1}{b2}
        \line[radius=0.6pt,inter=0.4em,shorten=0.2em]{b2}{b3}
    \end{pNiceMatrix}.
\end{equation*}

Intuitively, the full dynamics is a set of independent dynamics on each charge sector. If our measurements on our chosen state do not mix charge sectors, then the resulting dynamics in the environment are also independent and combine with weights $\{ p_\lambda \}$ to form the full distribution. Thus, their entropies add (weighted by the state) with an additional Shannon entropy corresponding to the uncertainty of picking which sector to start in. Note that Theorem \ref{thm:abelian} holds whether or not the dynamics within each sector are chaotic\footnote{We appreciate a referee pointing this out. We leave the analysis of the AFL entropy for the constrained dynamics within a sector for future work.}.   

We now proceed to apply the analysis above to the quantum cat map. The cumulative AFL entropy on a maximally mixed state with an $R$- or $W$-symmetric partition falls under Theorem~\ref{thm:abelian}. The dynamics splits into $s$ charge sectors of dimension $M$ (for the $W$ symmetry, take $s=2$ and $M=N/2$) and the state is maximally mixed, so the distribution $p_\lambda$ is a uniform $1/s$. This means $\shannon{p_\lambda} = \log s$ and the dimensional upper bound~\eqref{eq:dim-bound} in each sector is $2\log M$. Thus, the cumulative AFL entropy at all times is bounded above by
\begin{equation}
    \cAFL \leq 2 \log M + \log s,
\end{equation}
as we observe in our numerics from Fig.~\ref{fig:Abelian}.

It is worth noting that the conditions of Theorem~\ref{thm:abelian} for the $R$ symmetry (which are more general than our construction of an $R$-symmetric partition) means that any partition composed of $q$-projectors would also obey this bound. Such a partition is a reasonable choice if studying semiclassics. Ref.~\cite{alicki-2004-DecoherenceRate} used partitions of $p$-projectors, for example, but could have just as well chosen $q$. The choice of initial state and measurement in monitored systems like we study here is not typically random and may be compatible with some underlying symmetry.

\subsection{Proof of Theorem 1}
\begin{proof}
We will prove Theorem~\ref{thm:abelian} in the environment Hilbert space. Another proof using states in the system and purifier spaces can be found in Appendix~\ref{sect:abelian-proof2-appendix}. Before proceeding, recall that von Neumann entropy obeys the following inequality for a convex sum $\rho = \sum_i p_i \rho_i$,
\begin{equation} \label{eq:EE-convexity}
    S(\rho) \leq \sum_i p_i S(\rho_i) + \shannon{p_i}
\end{equation}
with equality if the set of $\rho_i$ have orthogonal support~\cite{nielsen-2010-QuantumComputation}.

After $t$ time steps, the multitime Kraus operators of the measurement channel are indexed by $\bm{i} = (i_1, \dots, i_t)$. The kets $\ket{\bm{i}}$ are a basis of the environment Hilbert space, as in equation~\eqref{eq:measurement-channel}.
The state of the measurement device is
\begin{align}
    \AFLstate{(U\mX)^t}_{\bm{i}\bm{j}}
    &= \Braket{\bm{i} | \AFLstate{(U\mX)^t} | \bm{j}} \nonumber \\
    &= \Tr%
        \!\begin{multlined}[t]
            \bigl( \MTK{i} \\
            \rho \MTKd{j} \bigr)
        \end{multlined}
    \nonumber \\
    &=\smash{\sum_\lambda}\, p_\lambda \Tr%
        \!\begin{multlined}[t]
            \bigl( \MTK[\lambda]{i} \\
            \rho_\lambda \MTKd[\lambda]{j} \bigr)
            \phantom{(00)} 
        \end{multlined}
    \nonumber \\
    &= \sum_\lambda p_\lambda
        \AFLstate[\lambda]{(U_\lambda \mX_\lambda)^t}_{\bm{i}\bm{j}} \label{eq:decomp-meas}
\end{align}
where we have used the fact that $U, \rho, $ and $\mX$ all admit a block diagonal structure due to their commutation relation with $Z$, and
$\AFLstate[\lambda]{(U_\lambda \mX_\lambda)^t}$ is the state of the measurement device if the system had started in $\rho_\lambda$.%
\footnote{We mean the initial state of the system was $\bm{0} \oplus \rho_\lambda \oplus \bm{0}$ where the zeros cover all charge sectors with index differing from $\lambda$.}
It follows from~\eqref{eq:EE-convexity} that
\begin{align}\label{eq:symm-bound-meas}
\begin{split}
    S \left( \AFLstate{(U\mX)^t} \right) \leq \smash{\sum_\lambda}\, p_\lambda
    S \left( \AFLstate[\lambda]{(U_\lambda \mX_\lambda)^t} \right) \quad& \\
    + \shannon{p_\lambda}&.
\end{split}
\end{align}

Next, we prove the equality is achieved when the measurement channel admits a Kraus representation where $X^i \in \mX$ have support on exactly one charge sector each. In general, while the set of $\rho_\lambda$ have orthogonal support, the same is not necessarily true of $\widetilde{\rho_\lambda}$, preventing equality. However, if each Kraus operator in $\mX$ has support on exactly one charge sector, we will show that the   
$\widetilde{\rho_\lambda}$
do indeed have orthogonal support.

It follows from equation~\eqref{eq:single-charge} that $X^i_\lambda U_\lambda X^j_\lambda$ is nonzero only if $\phi(i) = \phi(j) = \lambda$. Thus, we have 
\begin{equation}
    \AFLstate[\lambda]{(U_\lambda \mX_\lambda)^t}_{\bm{i}\bm{j}}
    \propto \delta(\bm{i} \sim \lambda)\delta(\bm{j} \sim \lambda)
\end{equation}
where $\bm{i} \sim \lambda$ is shorthand for $\phi(i_k) = \lambda$ for all $k=1,\dots,t$.
This means the support of $\widetilde{\rho_\lambda}$ is a subspace of $\Span\{ \ket{\bm{j}} \in \mH_E \ |\ \bm{j} \sim \lambda \}$, which can be explicitly seen from the relation
\begin{equation}
    \widetilde{\rho_\lambda} \ket{\bm{j}}
    = \sum_{\bm{i}} \left[\widetilde{\rho_\lambda}\right]_{\bm{i}\bm{j}} \ket{\bm{i}}
    \propto \delta(\bm{j} \sim \lambda)
\end{equation}
Consider now $\widetilde{\rho_\mu}$ with $\mu \neq \lambda$, whose support is a subspace of $\Span\{ \ket{\bm{j}} \in \mH_E \ |\ \bm{j} \sim \mu \}$. Given a fixed vector $\ket{\bm{j}}$, since each index $j_k$ belongs to only one charge sector $\phi(j_k)$, one cannot have both $\bm{j} \sim \mu$ and $\bm{j} \sim \lambda$. In other words, $\ket{\bm{j}}$ cannot be in both supports. Thus, the supports are orthogonal as desired.

\end{proof}

\section{Analytical Result: Tensor Product Dynamics}
\label{sect:tensorproduct}

Having studied dynamics with a direct sum structure, we now turn to general dynamics with a tensor product structure. We will see in Section~\ref{subsec:anticommuting} that the anticommuting unitary is a special case. 
Let the Hilbert space be a tensor product $\mH = \mH_A \otimes \mH_B$ and $\rho$ be a separable state, meaning there is a decomposition
\begin{equation}
    \rho = \sum_\nu q^\nu \rho^\nu_A \otimes \rho^\nu_B
\end{equation}
where $\rho^\nu_A,\rho^\nu_B$ are states on $\mH_A,\mH_B$ respectively, and $\{ q^\nu \}$ is a probability distribution over Schmidt index $\nu$ (we use superscripts for later notational convenience in Section~\ref{sect:nonabelian}).
We will consider dynamics and measurements that act ``locally'' on subsystems $A$ and $B$. Explicitly, $U = U_A \otimes U_B$, and given partitions $\mX_A$ and $\mX_B$ on $\mH_A$ and $\mH_B$ respectively, the measurement channel takes the form
\begin{align}
\begin{split}
    \mX &= \mX_A \otimes \mX_B \\
    &= \Set{ X_A \otimes X_B | X_A \in \mX_A, X_B \in \mX_B}
\end{split}
\end{align}
This partition corresponds to measuring the subsystems independently.

\subsection{Theorem 2}
\begin{theorem} \label{thm:tensor-dynamics}
    With the separable state $\rho = \sum_\nu q^\nu \rho^\nu_A \otimes \rho^\nu_B$,  tensor product dynamics $U = U_A \otimes U_B$, and partition $\mX = \mX_A \otimes \mX_B$ given above, the cumulative AFL entropy obeys
    \begin{align}
    \cAFL(\rho, U, \mX, t)
        \leq \smash{\sum_\nu} q^\nu
        &\bigl[
            \cAFL(\rho^\nu_A, U_A, \mX_A, t) \nonumber \\
            &+ \cAFL(\rho^\nu_B, U_B, \mX_B, t)
        \bigr] \phantom{(00)} \nonumber \\
        &+ \shannon{q^\nu}
    \end{align}
\end{theorem}

The generalization to multipartite systems is obvious. In some special cases, we get simpler equalities as listed below.

If the partition $\mX_B$ is a unitary channel (a single unitary Kraus operator, typically $\id_B$), then we have
\begin{equation} \label{eq:B-unitary}
    \cAFL(\rho, U, \mX, t) = \cAFL(\rho_A, U_A, \mX_A, t)
\end{equation}
where $\rho_A = \sum_\nu q^\nu \rho^\nu_A = \Tr_B (\rho)$ is the reduced state of subsystem $A$. Intuitively, we are not measuring system $B$ and so the dynamical entropy reduces to that of $A$.

If $\rho$ is a product state $\rho_A \otimes \rho_B$ (no constraints on $\mX_B$), then
\begin{align}
\begin{split} \label{eq:prod-state-case}
    \cAFL(\rho, U, \mX, t) =\; &\cAFL(\rho_A, U_A, \mX_A, t) \\
    &\quad + \cAFL(\rho_B, U_B, \mX_B, t)
\end{split}
\end{align}
as we would expect of independent dynamics.

\subsection{Anticommuting Unitary}
\label{subsec:anticommuting}

One way to guarantee tensor product dynamics is the existence of an anticommuting unitary. Consider a unitary $W$ that anticommutes with the dynamics $U$. Note anticommuting with the unitary time evolution $U$ is distinct from anticommuting with a Hamiltonian $H$. Given $U\ket{\phi} = e^{i\phi}\ket{\phi}$ for $\phi \in [0, 2\pi)$, then $W\ket{\phi}$ is also an eigenstate of $U$ with eigenvalue $-e^{i\phi} = e^{i(\phi + \pi)}$. Since $-e^{i\phi} \neq e^{i\phi}$ for all $\phi$, the Hilbert space must be even dimensional and splits into two spaces of equal dimension: $\mH = \overline{\mH} \oplus W\overline{\mH}$ with $\overline{\mH}$ the eigenspace for phases $\phi \in [0, \pi)$. We can identify these two spaces as a pseudospin-$1/2$ degree of freedom and write $\mH \cong \dsC^2 \otimes \overline{\mH}$, where $\dsC^2$ notates a 2-dimensional complex vector space. In some basis respecting the pseudospin, we have
\begin{equation} \label{eq:anticomm-blocks}
    U = \sigma_z \otimes \overline{U}
    \eqtext{and}
    W = \sigma_x \otimes \overline{W}
\end{equation}
which have matrix forms
\begin{equation*}
    U =
    \begin{pNiceMatrix}[margin][columns-width=4pt]
        \Block{2-2}{\overline{U}} & & \Block{2-2}{\bm{0}} \\
        & & & \\
        \Block{2-2}{\bm{0}} & & \Block{2-2}{-\overline{U}} \\
        & & &
    \end{pNiceMatrix}
    \eqtext{and}
    W =
    \begin{pNiceMatrix}[margin][columns-width=4pt]
        \Block{2-2}{\bm{0}} & & \Block{2-2}{\overline{W}} \\
        & & & \\
        \Block{2-2}{\overline{W}} & & \Block{2-2}{\bm{0}} \\
        & & &
    \end{pNiceMatrix}
    \nonumber
\end{equation*}
in the pseudospin-z basis, where $\overline{U} = U|_{\overline{\mH}}$ and $\overline{W}$ is a unitary on $\overline{\mH}$ commuting with $\overline{U}$.

The induced tensor product structure allows us to make use of Theorem~\ref{thm:tensor-dynamics} in the presence of an anticommuting unitary, such as for the cat map in Fig.~\ref{fig:cat-W-AC}. There, we consider two tensor product partitions, one measuring pseudospin-z and one with no psuedospin measurement. Explicitly, the partitions on the $\dsC^2$ pseudospin space are $\mX_{\dsC^2} = \{ \ketbra{\uparrow}{\uparrow}, \ketbra{\downarrow}{\downarrow}\}$ and $\mX_{\dsC^2} = \{ \id_2 \}$ respectively. The pseudospin-z measurement falls under the special case~\eqref{eq:prod-state-case}, as the maximally mixed state is a product state
\begin{equation}
    \rho = \frac{1}{N} \id_N = \frac{1}{2} \id_2 \otimes \frac{1}{N/2} \id_{\frac{N}{2}}.
\end{equation}
To be explicit, $\rho_A = \id_2/2$ on $\mH_A = \dsC^2$ and $\rho_B = 2 \id_{N/2}/N$ on $\mH_B = \overline{\mH}$. The cumulative AFL entropy on $\overline{\mH}$ is dimensionally bounded by $2 \log \dim \overline{\mH} = 2 \log (N/2)$. The pseudospin space adds a constant entropy of $\log 2$ at all times. This is because the pseudospin dynamics are $\sigma_z$, so the pseudospin-z measurement simply contributes one bit of information: which pseudospin-z sector the state begins in. Thus, the dimensional bound of $\cAFL$ has lowered from $2\log N$ to
\begin{equation}
    \cAFL \leq 2 \log (N/2) + \log 2.
\end{equation}
The case of no pseudospin measurement falls under~\eqref{eq:B-unitary}, so the $\log 2$ from the pseudospin space is absent, consistent with our numerics.

\subsection{Proof of Theorem 2}
\label{sect:tensorproductproof}

\begin{proof}
The partition after $t$ time steps is indexed by two vectors $\bm{a},\bm{b}$ with
\begin{align}
\begin{split}
    (U\mX)^t_{\bm{a}\bm{b}} =\; &(\MTK[A]{a}) \\
    &\quad \otimes (\MTK[B]{b})
\end{split}
\end{align}
Indexing the rows by vectors $\bm{a}\bm{b}$ and columns by $\bm{\alpha}\bm{\beta}$, the state of the environment is

\begin{widetext}
\begin{align}
    \AFLstate{(U\mX)^t}_{\bm{a}\bm{b};\bm{\alpha}\bm{\beta}}
    &= \smash{\sum_\nu}\, q^{\nu} \Tr \Bigl[
        \begin{multlined}[t]
            (\MTK[A]{a}) \otimes (\MTK[B]{b})
            \, \rho_A^{\nu} \otimes \rho_B^{\nu} \\
            (\MTK[A]{\alpha})^{\dagger} \otimes
            (\MTK[B]{\beta})^{\dagger} \Bigr]
        \end{multlined}
    \nonumber \\
    &= \smash{\sum_\nu}\, q^\nu
        \begin{multlined}[t]
            \Tr \bigl(
                \MTK[A]{a} \rho^\nu_A \MTKd[A]{\alpha}
            \bigr) \\
            \times\Tr \bigl(
                \MTK[B]{b} \rho^\nu_B \MTKd[B]{\beta}
            \bigr)
        \end{multlined}
    \nonumber \\
    &= \smash{\sum_\nu}\, q^\nu \:
    \widetilde{\rho^\nu_A}\left[(U_A\mX_A)^t\right]_{\bm{a}\bm{\alpha}}
    \widetilde{\rho^\nu_B}\left[(U_B\mX_B)^t\right]_{\bm{b}\bm{\beta}}
    \\ &\Downarrow \nonumber \\
    \AFLstate{(U\mX)^t} &= \sum_\nu q^\nu \:
    \widetilde{\rho^\nu_A}\left[(U_A\mX_A)^t\right] \otimes
    \widetilde{\rho^\nu_B}\left[(U_B\mX_B)^t\right] \label{eq:AFL-product-state}
\end{align}
\end{widetext}

Since the von Neumann entropy of a product state is additive, $S(\rho \otimes \sigma) = S(\rho) + S(\sigma)$, applying the convex sum inequality~\eqref{eq:EE-convexity} gives the desired result.

Now we turn to the special cases. If $\mX_B$ is a unitary channel, then $(U_B\mX_B)^t$ is also a unitary channel with only one Kraus operator. Thus, following equation~\eqref{eq:AFL-state}, $\widetilde{\rho^\nu_B}$ is simply the scalar $1$ and can be dropped from~\eqref{eq:AFL-product-state}. 
By linearity of the trace in equation~\eqref{eq:AFL-state}, 
\begin{equation}
    \sum_\nu q^\nu \widetilde{\rho^\nu_A} = \widetilde{\rho_A}
\end{equation}
for $\rho_A = \sum_\nu q^\nu \rho_A^\nu$ the reduced state on $A$ and $\widetilde{\rho_A}$ the corresponding state of the environment after measurement, which gives us case~\eqref{eq:B-unitary}.

Now making no assumptions of $\mX_B$ but restricting the state $\rho$ to be a product state $\rho_A \otimes \rho_B$, the sum over $\nu$ is absent. With no need to use the inequality~\eqref{eq:EE-convexity}, the proof above immediately yields the equality~\eqref{eq:prod-state-case}.

\end{proof}

\section{Analytical Result: Non-Abelian Symmetries}
\label{sect:nonabelian}

The direct sum and tensor product structures naturally combine when we generalize from a single symmetry operator $Z$ to a (von Neumann) algebra of symmetry operators $\mA$ on $\mH$ all commuting with $U$. For simplicity, we assume $\id_\mH \in \mA$.
Let $\mZ$ be the center of $\mA$ and $d = \dim\mZ$. Since $\mZ$ is Abelian, it has a diagonal representation $\mZ = \bigoplus_\lambda \dsC\, \id_{\mH_\lambda}$ on $\mH = \bigoplus_\lambda \mH_\lambda$, where $\lambda = 1,\dots,d$ label the (one-dimensional) irreps of the center~\cite{beny-2020-AlgebraicApproach}. Each subspace $\mH_\lambda$ can be further decomposed as a tensor product of two spaces, one $\mA$ acts irreducibly on (call it $\mK_\lambda$), the other $\mA$ acts trivially on (call it $\mK'_\lambda$). One can think of $\mK'_\lambda$ as multiple copies of the irrep $\mK_\lambda$ embedded in $\mH$.
Thus, we have an overall decomposition
\begin{equation}\label{eq:algebra-reps}
    \mH = \DCTsum \mK_\lambda \otimes \mK'_\lambda.
\end{equation}
We will set $\dim \mK_\lambda = n_\lambda$ and $\dim \mK'_\lambda = n'_\lambda$. With this decomposition, the symmetry algebra takes the form
\begin{equation}
    \mA \cong \DCTsum \mL(\mH_\lambda) \otimes \id_{n'_\lambda} \\
\end{equation}
where $\mL(\mH_\lambda)$ is all linear operators on $\mH_\lambda$. Since the dynamics $U$ commutes with the entire algebra $\mA$, it must be of the form
\begin{equation} \label{eq:nonabelian-U}
    U = \DCTsum \id_{n_\lambda} \otimes U_\lambda
\end{equation}
where $U_\lambda$ is the restriction $U |_{\mK'_\lambda}$. Visually, in a basis respecting the decomposition~\eqref{eq:algebra-reps}, the dynamics has the matrix form
\begin{equation*}
    U =
    \begin{pNiceMatrix}[margin][columns-width=5pt]
        \NiceBlock[b1]{2-2}{\scriptstyle \lambda=1} & & & \Block{2-2}{\bm{0}}\\
        & & & & \\
        & & & & \\
        \Block{2-2}{\bm{0}} & & & \NiceBlock[b2]{2-2}{\scriptstyle \lambda=d}\\
        & & & &
    \CodeAfter
        \line[radius=0.6pt,inter=0.4em,shorten=0.2em]{b1}{b2}
    \end{pNiceMatrix}
\end{equation*}
where the blocks labelled by $\lambda$ are of the form
\begin{equation*}
    \begin{NiceArray}{w{c}{15pt} c c}
        \NiceBlock{2-2}{\lambda} \\
        & &
    \end{NiceArray}
    =
    \begin{pNiceMatrix}[margin][columns-width=6pt]
        \Block[name=b1]{2-2}{U_\lambda} & & & \Block{3-3}{\bm{0}}\\
        & & & & & \\
        & & & & & \\
        \Block{3-3}{\bm{0}} & & & & & \\
        & & & & \Block[name=b2]{2-2}{U_\lambda} \\
        & & & & &
    \CodeAfter
        \line[radius=0.7pt,inter=0.6em,shorten=-0.1em]{b1}{b2}^{n_\lambda \text{ times}}
    \end{pNiceMatrix}.
\end{equation*}
For further discussion and examples of this structure, see~\cite{harlow-2017-RyuTakayanagi, beny-2020-AlgebraicApproach, moudgalya-2022-HilbertSpace, moudgalya-2023-SymmetriesCommutant}. These representation theory techniques have previously been used to study how non-Abelian symmetry affects subsystem entanglement in many-body systems, including symmetric state constructions similar to what we define below~\cite{bianchi-2024-NonAbelianSymmetryResolved}.

For clarity, we can explicitly show this formalism subsumes the previous case of a single symmetry operator $Z$. Consider the case where $\mA$ is Abelian. Then $\mA = \mZ = \bigoplus_\lambda \dsC\, \id_{\mH_\lambda}$, which induces the same decomposition as a single symmetry operator $Z = \bigoplus_\lambda z_\lambda \id_{\mH_\lambda}$ for some set of distinct scalars $\{z_\lambda\}$. In fact, any such $Z$ generates an Abelian $\mA$.

Now we return to the $\cAFL$ inequalities. Given a symmetry algebra $\mA$ commuting with the dynamics, then $U$ takes the form~\eqref{eq:nonabelian-U} which is a direct sum over tensor product dynamics. The obvious extension of our prior requirements is for the state to commute with the center $\mZ$ and be separable in each charge sector:
\begin{equation}
    \rho = \bigoplus_\lambda \sum_\nu p_\lambda^\nu \:
    \rho^\nu_\lambda \otimes \rho'^\nu_\lambda
\end{equation}
where $\{ p_\lambda^\nu \}$ is a probability distribution over $(\lambda, \nu)$ and $\rho^\nu_\lambda,\rho'^\nu_\lambda$ density matrices on $\mK_\lambda,\mK'_\lambda$ respectively. Similarly, we require the partition to admit a form
\begin{align}
    \mX &= \bigoplus_\lambda \mX_\lambda \otimes \mX'_\lambda \nonumber \\
    &= \Bigl\{ \bigoplus_\lambda X_\lambda \otimes X'_\lambda \ |\ X_\lambda \in \mX_\lambda,\, X'_\lambda \in \mX'_\lambda,\, \Bigr\}
\end{align}
for partitions $\mX_\lambda,\mX'_\lambda$ on $\mK_\lambda,\mK'_\lambda$ respectively.
\begin{theorem} \label{thm:nonabelian}
    If the dynamics $U$ commute with a von Neumann algebra $\mA$, the state takes the form
    $\rho = \bigoplus_\lambda \sum_\nu p_\lambda^\nu \:
    \rho^\nu_\lambda \otimes \rho'^\nu_\lambda$
    and the partition admits a Kraus representation
    $\mX = \bigoplus_\lambda \mX_\lambda \otimes \mX'_\lambda$
    as defined above, then the cumulative AFL entropy obeys
    \begin{align}
        \cAFL(\rho,U,\mX,t) \leq \smash{\sum_{\lambda,\nu}}\, p_\lambda^\nu
        &\bigl[
            \cAFL(\rho^\nu_\lambda, \id_{n_\lambda}, \mX_\lambda,t) \nonumber \\
            &+ \cAFL(\rho'^\nu_\lambda, U_\lambda, \mX'_\lambda,t)
        \bigr] \phantom{(00)} \nonumber \\
        &+ \shannon{p_\lambda^\nu}
    \end{align}
\end{theorem}
If we choose $\mX_\lambda$ to be a partition of commuting Kraus operators (e.g. commuting projectors), then the cumulative AFL entropy on $\mK_\lambda$ (the first term) is a constant. If $\mX_\lambda$ is a unitary channel, then the term vanishes.

The commutant partition used for computing $\cAFL$ in Fig.~\ref{fig:cat-nonabelian} falls under Theorem~\ref{thm:nonabelian}. Namely, we have $\mX_\lambda = \{ \id_{n_\lambda} \}$, $\mX'_\lambda$ a random partition, and a state
\begin{equation}
    \rho = \frac{1}{N} \id_N
    = \DCTsum p_\lambda \, \frac{1}{n_\lambda} \id_{n_\lambda} \otimes \frac{1}{n'_\lambda} \id_{n'_\lambda}
\end{equation}
for Hilbert space dimension $N$ and probability distribution $p_\lambda = n_\lambda n'_\lambda / N$. Thus, Theorem~\ref{thm:nonabelian} yields a dimensional upper bound of
\begin{equation} \label{eq:commutant-partition-bound}
    \cAFL \leq 2 \sum_{\lambda=1}^d p_\lambda \log n'_\lambda + \shannon{p_\lambda}.
\end{equation}
For the algebra generated by the cat map symmetries $R$ and $W$, as computed in Appendix~\ref{sect:cat-nonabelian-algebra}, the center has dimension $d = (s+3)/2$. The dimensions of the irreps of the algebra are
$n_1 = n_2 = 1$, and $n_\lambda = 2$ otherwise. The dimensions of the trivial spaces are $n'_1 = \tfrac{M}{2}+1$, $n'_2 = \tfrac{M}{2}-1$, and $n'_\lambda = M$ otherwise. Given these values, the bound~\eqref{eq:commutant-partition-bound} is computed and shown in Fig.~\ref{fig:cat-nonabelian} as the lowest horizontal dash-dotted line. The cumulative AFL entropy under a commutant partition asymptotes to this bound as expected.

\begin{proof}
    Theorem~\ref{thm:nonabelian} follows immediately from Theorems~\ref{thm:abelian} and~\ref{thm:tensor-dynamics}, and the identity
    \begin{equation}
        \shannon{p^\nu_\lambda} = \shannon{q_\lambda} + \sum_\lambda q_\lambda \shannon{r_{\nu | \lambda}}
    \end{equation}
    for marginal distribution $q_\lambda = \sum_\nu p_\lambda^\nu$ and conditional distribution $r_{\nu | \lambda} = p_\lambda^\nu / q_\lambda$. Note $p^\nu_\lambda$ is a probability distribution over $(\lambda,\nu)$ but $r_{\nu | \lambda}$ is a distribution over $\nu$ only.
    
\end{proof}

\section{Concluding Remarks}
\label{sect:discussion}
The entropy rate $\AFL$~\eqref{eq:AFL-def} is partition(measurement)-independent by definition,%
\footnote{There can still be some subtlety in infinite-dimensional systems over which space of partitions to take the supremum over~\cite{fannes-1999-ContinuityProperty}.}
but in finite dimensions, where cumulative AFL entropy is the appropriate quantity, the choice of partition plays an important role.
Outside of comparing semiclassical partitions to random partitions~\cite{alicki-2004-DecoherenceRate}, the effects of partition choice have---to our knowledge---not been studied. In this work, we showed how the presence of symmetries in the dynamics leads to dramatic dependence of $\cAFL$ on partition choice. Specifically, the cumulative AFL entropy is insensitive to a given symmetry unless the measurements are chosen to respect the resulting symmetry structure in Hilbert space. Symmetries have important consequences for quantum chaos, and one rarely has full knowledge of the symmetries for a given dynamics. This means in practice that $\cAFL$ may fail to fully diagnose the chaoticity of a given dynamics. Beyond the study of quantum chaos, symmetry structure in open quantum systems has garnered significant interest, such as with regard to mixed-state topological order~\cite{deGroot-2022-SymmetryProtected, ramanjit-2025-NoisyApproach} and mixed-state entanglement~\cite{moharramipour-2024-SymmetryEnforcedEntanglement, li-2025-HighlyEntangled}. AFL entropy and similar quantities may still serve as an interesting playground to explore symmetric open quantum dynamics.

\subsection{Types of Quantum Chaos}

Although the perturbed quantum cat map has random matrix spectral statistics, the unperturbed map does not~\cite{matos-1993-QuantizationAnosov, keating-2000-PseudosymmetriesAnosov, backer-2003-NumericalAspects}. Despite this, the map still has a semiclassically chaotic limit (see~\cite{esposti-2003-MathematicalAspects} for a review), in apparent violation of the Bohigas-Giannoni-Schmit (BGS) conjecture~\cite{bohigas-1984-CharacterizationChaotic}. Indeed, the unperturbed cat map still displays Lyapunov growth in OTOC~\cite{chen-2018-OperatorScrambling, garcia-mata-2018-ChaosSignatures} and in cumulative AFL (see Appendix~\ref{sect:cat-perts}), and obeys eigenstate thermalization hypothesis~\cite{axenides-2018-QuantumCat}.
However, as pointed out by Magan and Wu~\cite{Magan-2024n-BGS-chaostype}, spectral chaos (random matrix statistics) and basis/semiclassical chaos (eigenstate thermalization and Lyapunov growth) are notably distinct characteristics of quantum chaos, except when the Hamiltonian is $k$-local. The unperturbed cat map is an example of dynamics displaying only basis chaos. In this language, AFL entropy is a measure of basis chaos and exhibits limited sensitivity to spectral chaos, and so may not capture all aspects of \textit{quantum} chaos. 

\subsection{Relation to CNT Entropy}

The quantum dynamical entropies, as information theoretic quantities, are often related conceptually and can even bound one another using various quantum information constructions. Of particular importance is the Holevo bound on the {\it classical} capacity of a quantum channel~\cite{nielsen-2010-QuantumComputation, wilde-2021-ClassicalQuantum}. Entropy exchange and AFL entropy upper bound instances of the Holevo quantity~\cite{schumacher-1996-SendingEntanglement, alicki-1997-QuantumErgodic, alicki-2002-InformationtheoreticalMeaning}, and can be slightly modified to match it~\cite{hudetz-1998-QuantumDynamical}. The other common quantum dynamical entropy construction by St\o rmer, Connes, Narnhofer, and Thirring (CNT)~\cite{connes-1975-EntropyAutomorphisms, connes-1987-DynamicalEntropy} can be considered a generalization of the Holevo quantity to multiple channel inputs~\cite{benatti-1998-QuantumChaos}. Like AFL entropy, CNT entropy coincides with Kolmogorov-Sinai entropy in classical systems~\cite{benatti-2023-DynamicsInformation}. However, they generally differ in quantum systems such as the noncommutative shift, with AFL conjectured to upper-bound CNT entropy~\cite{alicki-1995-ComparisonDynamical, accardi-1996-NoteQuantum, tuyls-1998-ComparingQuantum, fannes-2003-QuantumDynamical}. Further study of both quantum dynamical entropies could offer valuable insights on {\it quantum} channel capacities and the diagnostics of quantum chaos. We leave such explorations to future work.

\subsection{Other Directions}

The construction of AFL entropy as entanglement arising from measurement/interaction with an environment is a framework with broad applications for studying quantum dynamics and thermalization. 
Past work on AFL entropy has focused on strong interaction with the environment, such as projective measurements. However, other work uses the same framework of an open system in the \textit{weak} dissipation limit to compute the decay of correlators, or quantum Ruelle-Pollicott resonances~\cite{nonnenmacher-2003-SpectralProperties, prosen-2004-RuelleResonances, garcia-mata-2005-SpectralApproach, yoshimura-2024-RobustnessQuantum, mori-2024-LiouvilliangapAnalysis, jacoby-2024-SpectralGaps, znidaric-2024-MomentumdependentQuantum, zhang-2024-ThermalizationRates, yoshimura2025irreversibility}. Exploring late-time AFL entropy or similar spatiotemporal entanglement in the weak measurement limit may provide further insight in this direction. 

Another promising avenue is to post-select on the measurement outcomes to study deep thermalization and a post-selected dynamical entropy~\cite{odonovan-2025-DiagnosingChaos} distinct from AFL, as discussed in Appendix~\ref{sect:projected-ensembles}. 
In holography, the effects on the geometric bulk dual of post-selected local measurements on a subregion of boundary CFT have been explored in \cite{Numasawa-2016-localmeasurements,Antonini-2022-HMbulkteleportation,Antonini-2023-HMCFT}. It will be interesting to study the potential relations between the boundary quantities, for instance the post-selected dynamical entropy, and the corresponding bulk dual.

The notion of turning time steps into effective spatial degrees of freedom allows for entanglement cuts beyond just spatial, for example across spacetime (e.g.~spatiotemporal entanglement). AFL entropy itself is an example of a temporal cut, and some initial work has been done for other spacetime cuts~\cite{cotler-2018-SuperdensityOperators, dowling-2023-ScramblingNecessary, odonovan-2025-DiagnosingChaos}. This mapping between time steps and the spatial degrees of freedom resembles the Wick rotation. Time-like entanglement entropy, particularly in holographic quantum systems, allows for the study of the spatiotemporal entanglement structure in the Wick rotated quantum states~\cite{Doi-2023-timelikeEE,Heller-2025-GeometrictimelikeEE,Li-2023-holographicTEE,Mollabashi-2021-PseudoEntropy,Mollabashi-2021-pseudoentropyfreefields,Nishioka-2021-topologicalPE,Goto-2021-sff,Miyaji-2021-Island,Guo-2023-pseudo-Hermiticity,He:2023eap}. Despite its non-positive semi-definite nature, it would still be interesting to explore the relation between holographic time-like entanglement entropy and quantum dynamical entropy, such as AFL entropy, which inherently involves spacetime cuts.

The role of locality is indispensable in the analysis of quantum chaos, referring to, for instance,  \cite{Magan-2024n-BGS-chaostype, Shi2023localdynamics-ETH}. Our work presented, as a demonstration of Theorem \ref{thm:abelian}, \ref{thm:tensor-dynamics}, and \ref{thm:nonabelian}, the numerical studies of the AFL entropy for unitary dynamics on a single particle quantum system, where locality is not apparent. It is in our future interest to explore how the locality of a Hamiltonian and that of the measurement channel on a quantum many-body system play a role in the behaviors of the AFL entropy.


\acknowledgments
The authors thank Nima Lashkari, Hong Liu, Ayush Raj, and Carolyn Zhang for inspiring discussions.
K.F.~acknowledges support from Prof.~Ning Bao at Northeastern University.

\appendix

\section{Projected Ensembles} \label{sect:projected-ensembles}

Here, we briefly show how the framework for constructing AFL entropy relates to projected ensembles and post-selected quantum dynamical entropy. First, put
\begin{equation}
p_i = \Tr(\rho X^{i\dagger} X^i) = \bbrakett{X^i \sqrt{\rho}}{X^i \sqrt{\rho}}
\end{equation}
as the probability of measurement outcome $i$. By properly normalizing in $SP$, we see the state $\sigma[\mX]$ is the mixed state corresponding to the projected ensemble
\begin{equation}
    \left\{ \left( p_i,\, \tfrac{1}{\sqrt{p_i}}\kett{X^i \sqrt{\rho}} \right) \;\middle|\; i=1,\dots,K \right\}.
\end{equation}
Higher moments of such an ensemble are the study of deep thermalization and take the form
\begin{equation}
    \sigma^{(n)}[\mX] = \sum_i p_i \left( \tfrac{1}{p_i} \projj{X^i \sqrt{\rho}} \right)^{\otimes n}
\end{equation}
for the $n^\text{th}$ moment. This notion immediately generalizes to multitime channels $(U\mX)^t$ and has seen some initial study in many-body systems~\cite{odonovan-2025-DiagnosingChaos}, where the higher moments allows for quantification of chaotic dynamics on a more fine-grained level.

One can also define a quantum dynamical entropy as the Shannon entropy production rate of the multitime outcome distribution
\begin{equation}
    p_{\bm{i}} = \Tr \left(\MTK{i} \rho \MTKd{i} \right).
\end{equation}
We refer to this as the post-selected quantum dynamical entropy, which is defined as
\begin{equation}
    h_\text{PS}(\rho, U) = {\adjustlimits \sup_\mX \lim_{t\rightarrow\infty}} \frac{1}{t} \shannon{p_{\bm{i}}}.
\end{equation}
See Ref.~\cite{slomczynski-2017-QuantumDynamical} for a recent work. Observe that $p_{\bm{i}}$ are the diagonal entries of the state of the environment~\eqref{eq:AFL-state}. Thus, we recognize $h_\text{PS}$ as the von Neumann entropy of $\AFLstate{(U\mX)^t}$ after it has undergone a complete dephasing. This dephasing removes the dimensional bound on the cumulative entropy and $h_\text{PS}$ may indeed be nonzero in finite-dimensional systems.

\section{The Cat Map} \label{sect:cat-appendix}

\subsection{Classical Cat Map}

\begin{figure*}[t]
    \centering
    \includegraphics[width=0.8\textwidth]{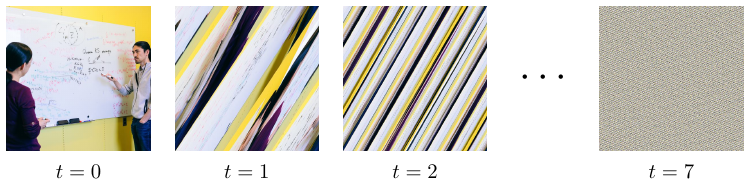}
    \caption{The action of the unperturbed classical cat map $\bfA = \smallcatmat{2}{1}{3}{2}$ for $t$ iterations.}
    \label{fig:classical-cat}
\end{figure*}

The classical cat map, first introduced by Arnol'd~\cite{arnold-1968-ErgodicProblems}, is a map $\bfA$ on the torus $\dsT^2$ (with periodicity set to unity for convenience) given by
\begin{equation}
    \bfA\torvec{q}{p} =
    \catmat{A_{11}}{A_{12}}{A_{21}}{A_{22}} \torvec{q}{p} \mod 1.
\end{equation}
With a slight abuse of notation, we will keep the modulo 1 implicit and use $\bfA$ to refer to the matrix. Given $\bfA \in \operatorname{SL}(2,\dsZ)$ and $\Tr\bfA > 2$, the cat map is continuous, area (Lebesgue measure) preserving, and hyperbolic. This makes the map chaotic (in fact, Anosov) with Lyapunov exponents given by $\log \lambda_+ > 0$ and $\log \lambda_- < 0$, where $\lambda_\pm$ are the eigenvalues of $\bfA$. The KS entropy of the cat map is equal to the positive Lyapunov exponent $\log \lambda_+$~\cite{arnold-1968-ErgodicProblems}. The action of an example classical cat map is shown in Fig.~\ref{fig:classical-cat}. 

The cat map is structurally stable and can retain its chaotic (Anosov) properties in the presence of weak perturbations~\cite{arnold-1968-ErgodicProblems, matos-1993-QuantizationAnosov, boasman-1995-SemiclassicalAsymptotics}. Specifically, the cat map is often studied with nonlinear shears in $q$ or $p$, meaning the dynamical map takes the form $\bfA \circ \kappa_p G_p \circ \kappa_q F_q$, where
\begin{align}
    F_q \torvec{q}{p} &= \torvec{q + F(p)}{p} \\[5pt]
    G_p \torvec{q}{p} &= \torvec{q}{p + G(q)}.
\end{align}
There is not a unique choice of shear, but the form of $F$ and $G$ is constrained by the periodicity of the torus and symmetries one wishes to retain or break, namely parity and time-reversal~\cite{keating-2000-PseudosymmetriesAnosov}. The case of no shears is referred to as the unperturbed cat map.

Our work uses a common choice of perturbation $G(q) = \tfrac{1}{2\pi} \cos (2\pi q)$ with $\kappa \coloneq \kappa_p$. For weakly breaking the $R$ and $W$ symmetries, as shown in Appendix~\ref{sect:cat-aprx-sym}, we add the perturbation $F(p) = \tfrac{1}{4\pi}\cos(4\pi p)$.

\subsection{Quantizing the Cat Map}

The quantization of the unperturbed cat map was first derived by Hannay and Berry~\cite{hannay-1980-QuantizationLinear} based on earlier work on quantum maps~\cite{berry-1979-QuantumMaps}. Following work by Matos and Almeida \cite{matos-1993-QuantizationAnosov} extended the derivation to the perturbed maps. The quantization follows by identifying $(q,p) \in \dsT^2$ as conjugate position and momentum, then requiring wavefunctions to be periodic in both position and momentum space, possibly up to a phase. This produces a Hilbert space of finite dimension $N$ with an effective, rational Planck's constant of $h = 1/N$. A basis for the Hilbert space $\mH$ is given by delta-localized position eigenstates $\ket{q_j}$ for $q_j = j/N$ with $j$ understood modulo $N$. The unitary map $U$ on $\mH$ corresponding to the classical cat map is defined by the semiclassical formula for the propagator (also known as the stationary phase approximation or the Van Vleck formula) in position space. For the unperturbed map, the matrix elements are given by
\begin{align}
    \braket{q_k | U | q_j}
    ={} &\sqrt{\frac{A_{12}}{N}} \nonumber\\
    & \times \exp \Biggl[
        \frac{i\pi}{A_{12}N} (A_{11}j^2 - 2jk + A_{22}k^2)
    \Biggr] \nonumber\\
    & \times G\left( N'A_{11}, A'_{12}, \frac{2(A_{11}j-k)}{\gcd(N,A_{12})} \right)
\end{align}
where $N' = N / \gcd(N,A_{12})$, $A'_{12} = A_{12} / \gcd(N,A_{12})$, and $G$ is a Gauss average function
\begin{multline}
    G(a,b,c) = \lim_{M\rightarrow\infty} \frac{1}{2M} \\
    \times \sum_{m=-M}^{M} \exp \left\{ \frac{i\pi}{b} (am^2 + cm) \right\}
\end{multline}
for coprime integers $a$ and $b$. The $G$ function is evaluated in Ref.~\cite{hannay-1980-QuantizationLinear}. Perturbing shears may be included by composing with additional simple matrices, as detailed in Ref.~\cite{matos-1993-QuantizationAnosov, esposti-2005-QuantumPerturbed}.

For further detail on the quantization of the cat map, see Ref.~\cite{esposti-1993-QuantizationOrientation, esposti-1995-ClassicalLimit, debievre-1996-QuantizationClass}. The toral wavefunctions need only be periodic up to a phase, the effects of which are discussed in Ref.~\cite{keating-1999-QuantumBoundary}. The cat map unitary also has quantum symmetries with no classical counterpart, as constructed in Ref.~\cite{kurlberg-2000-HeckeTheory, keating-2000-PseudosymmetriesAnosov}. The semiclassical limit of quantum maps is well-studied~\cite{eckhardt-1986-ExactEigenfunctions, keating-1991-CatMaps, boasman-1995-SemiclassicalAsymptotics, esposti-1995-ClassicalLimit, bouzouina-1996-EquipartitionEigenfunctions, kurlberg-2001-QuantumErgodicity, horvat-2006-QuantumClassical, horvat-2007-EgorovProperty}. Basics of quantization and semiclassics for quantum maps on the torus are reviewed in Ref.~\cite{esposti-2003-MathematicalAspects}. There there alternative quantization schemes as well, such as a quadratic kick~\cite{ford-1991-ArnoldCat}. One may also remove the constraints on wavefunctions and the Hilbert space, and instead perform a noncommutative deformation at the algebraic level. This allows for any Plank's constant $h \in \mathds{R}$, but the resulting Hilbert space is infinite dimensional for irrational $h$~\cite{benatti-1991-NoncommutativeVersion}. The AFL entropy $\AFL$ has been computed for this algebraic quantum cat map and matches the classical KS value~\cite{andries-1995-DynamicalEntropy}.

\subsection{Proof of Anticommutation} \label{sect:cat-AC-proof}

The following is adapted from Appendix~B of Esposti and Winn (EW)~\cite{esposti-2005-QuantumPerturbed}, which proves that when the Hilbert space dimension $N$ is divisible by 4,
\begin{equation}
    U_{kj} = (-1)^{j-k}  U_{N/2-k,N/2-j}
\end{equation}
where $U$ is the unperturbed quantum cat map unitary. This property proves the unitary $W$ defined in~\eqref{eq:cat-W} commutes with $U$ by way of the relation
\begin{equation}
    ( W^\dagger U W )_{kj} = (-1)^{j-k}  U_{N/2-k,N/2-j}
\end{equation}
What we show here is that when $N$ is even but not divisible by 4, it is possible to instead have the relation
\begin{equation} \label{eq:cat-AC-indices}
    U_{kj} = (-1)^{j-k+1} U_{N/2-k,N/2-j}
\end{equation}
which implies $W$ and $U$ \textit{anticommute}. For particular choices of the perturbation, such as \eqref{eq:pert-cat-map}, $W$ commutes with the quantized perturbation irrespective of $N$, and thus the commutation/anticommutation results extend to the perturbed map. The calculation differs from that of EW by occasional factors of $-1$, which we tally below. Before we begin, note the cat map matrix is always of a chessboard form
\begin{equation}
    A =
    \begin{pmatrix}
        \text{odd} & \text{even} \\
        \text{even} & \text{odd}
    \end{pmatrix}
    \eqtext{or}
    A =
    \begin{pmatrix}
        \text{even} & \text{odd} \\
        \text{odd} & \text{even}
    \end{pmatrix}
\end{equation}
and we use the notation $N' = N / \gcd(N, A_{12})$ and $A_{12}' = A_{12} / \gcd(N, A_{12})$. $N$ is even but not divisible by 4 throughout the calculation.

\subsubsection{Case $N'A_{11}A'_{12}$ Even}

The calculations of EW follow unchanged until the $A_{11}$ even case of equations (EW.B.14--15). With $A_{11}$ even, $A_{12}$ is odd and $N'$ is even. The factor under consideration is
\begin{equation} \label{eq:esposti-B14}
    \exp \left( -\pi \frac{N A_{11}}{4 A_{12}}\ell A'_{12} \right) = \exp \left( -\frac{\pi}{4} N' A_{11} \ell \right)
\end{equation}
which is $+1$ only if $4|A_{11}$ or $2 | \ell$, and is $-1$ otherwise. The integer $\ell$ is defined by
\begin{align}
    \ell A'_{12}
    &= [1 - A_{22} + A_{12}A_{21} \nonumber \\
        &\qquad + mA'_{12}(A_{11}-1)]^2
        - (1 - A_{22})^2 \nonumber \\
    &= m^2 {A'}_{12}^2 (A_{11} - 1)^2 + A_{12}^2 A_{21}^2 + 2 q A'_{12} \nonumber \\
    &= [m^2 A'_{12} (A_{11} - 1)^2 \nonumber \\
        &\qquad + A_{12}A_{21}^2\gcd(N,A_{12}) + 2 q]A'_{12}
\end{align}
where
\begin{multline}
    q = (1 - A_{22})A_{21}\gcd(N,A_{12}) \\
    + m(A_{11} - 1)(1 - A_{22} + A_{12}A_{21})
\end{multline}
Note $A'_{12}(A_{11} - 1)^2$ and $A_{12}A_{21}^2\gcd(N, A_{12})$ are odd, so the parity of $\ell$ is opposite that of $m$. The integer $m$ is itself defined by
\begin{equation}
    N'(N'A_{11} \backslash A'_{12}) = A_{22} + m A'_{12}
\end{equation}
where $x \backslash y$ is the inverse of $x$ modulo $y$ ($N'A_{11}$ and $A'_{12}$ are coprime). $N'$ and $A_{22}$ are even and $A'_{12}$ is odd, so $m$ must be even. Thus, $\ell$ is always odd. The exponential factor is then $-1$ if $A_{11}$ is even but not divisible by 4. Otherwise (including the $A_{11}$ odd case from EW), the factor is $+1$.

The next change is at equation (EW.B.17), with the factor
\begin{equation} \label{eq:esposti-B17}
    \exp \left( -\frac{\pi}{4} N(2-A_{22})A_{21} \right).
\end{equation}
Observe $N$ is even but not divisible by 4 and $(2-A_{22})A_{21}$ is even, so the expression simplifies to $+1$ if $4|(2-A_{22})A_{21}$ and $-1$ otherwise.

\subsubsection{Case $N'A_{11}A'_{12}$ Odd}

Here, each factor in $N'A_{11}A'_{12}$ is odd so we are in the $\bfA = \smallcatmat{\text{odd}}{\text{even}}{\text{even}}{\text{odd}}$ case. The computation of EW follows unchanged until equation (EW.B.25), where the factor~\eqref{eq:esposti-B17} appears again.%
\footnote{To be clear, expression~\eqref{eq:esposti-B17} appears in the expansion of $\exp( -\pi N A_{11} (1 - A_{22})^2 / 4 A_{12})$, which also appears in the algebra preceding (EW.B.25).}
This time, $A_{21}$ is even and $(2-A_{22})$ is odd, so the factor is $+1$ if $4 | A_{21}$ and $-1$ otherwise.

The final change from EW occurs at equation (EW.B.26) in the expression
\begin{equation} \label{eq:esposti-B26}
    \exp \left( \frac{\pi}{2} \gcd(N,A_{12}) \right).
\end{equation}
Note $N'$ is odd while $N = N'\gcd(N, A_{12})$ is even but not divisible by 4. Thus, $\gcd(N, A_{12})$ is also even but not divisible by 4 and this factor is always $-1$.

\subsubsection{Summary}

Let us simplify the above conditions by tallying the possible $-1$ factors. Taking $A_{11}$ as even places us in the $N'A_{11}A'_{12}$ even case with $A_{22}$ even as well. We get a factor of $-1$ from~\eqref{eq:esposti-B14} if $4 \nmid A_{11}$ and another from~\eqref{eq:esposti-B17} if $4 | A_{22}$.

Now we take $A_{11}$ to be odd. We get a factor of $-1$ from~\eqref{eq:esposti-B17} if $4 \nmid A_{21}$. If $4 \nmid A_{12}$, we are in the $N'A_{11}A'_{12}$ odd case and get a $-1$ factor from~\eqref{eq:esposti-B26}. If $4|A_{12}$, we are in the $N'A_{11}A'_{12}$ even case where no such factor appears.

In conclusion, if $A_{11}$ is even, then $W$ and $U$ commute if and only if 4 divides exactly one of $A_{11}$ or $A_{22}$. If $A_{11}$ is odd, then $W$ and $U$ commute if and only if 4 divides both or neither of $A_{12}$ and $A_{21}$. In all other cases, we are left with an overall factor of $-1$ which yields equation~\eqref{eq:cat-AC-indices}, making $W$ and $U$ and anticommute. These results are shown as a table in Fig.~\ref{fig:cat-AC-conditions}. The maps we chose for numerics, $\bfA = \smallcatmat{6}{5}{7}{6}$ and $\bfA = \smallcatmat{2}{1}{3}{2}$, satisfy the anticommutation conditions.

\begin{figure}[h]
    \hspace*{\fill}
    \begin{NiceTabular}{ccc}[hvlines, cell-space-limits=5pt, columns-width=1cm]
        \Block[fill=gray!10]{1-3}{$A_{11}$ even} \\
        & $4|A_{11}$ & $4 \nmid A_{11}$ \\
        $4|A_{22}$ & \Block[fill=red!10]{}{AC} & \Block[fill=green!10]{}{C} \\
        $4 \nmid A_{22}$ & \Block[fill=green!10]{}{C} & \Block[fill=red!10]{}{AC}
    \end{NiceTabular}
    \hspace*{\fill}
    \begin{NiceTabular}{ccc}[hvlines,cell-space-limits=5pt, columns-width=1cm]
        \Block[fill=gray!10]{1-3}{$A_{11}$ odd} \\
        & $4|A_{12}$ & $4 \nmid A_{12}$ \\
        $4|A_{21}$ & \Block[fill=green!10]{}{C} & \Block[fill=red!10]{}{AC} \\
        $4 \nmid A_{21}$ & \Block[fill=red!10]{}{AC} & \Block[fill=green!10]{}{C}
    \end{NiceTabular}
    \hspace*{\fill}
    \caption{\justifying The conditions on the classical cat matrix $\bfA$ for the corresponding quantum cat map unitary $U$ to commute or anticommute with $W$ as defined by equation~\eqref{eq:cat-W}, given the dimension of the Hilbert space $N$ is even but not divisible by 4. Here, ``C'' means commuting ($[W,U]=0$) and ``AC'' means anticommuting ($\{W,U\}=0$).}
    \label{fig:cat-AC-conditions}
\end{figure}

\subsection[The \textit{R} and \textit{W} Algebra]{The $R$ and $W$ Algebra}
\label{sect:cat-nonabelian-algebra}

\newcommand{\qket}[1]{\ket{q_{#1}}}
\newcommand{\ssum}[1]{\sum_{{#1}=0}^{s-1}}
\newcommand{\wsign}{(-1)^{\frac{M}{4}}}

Here, we compute the representation described in Sec.~\ref{sect:nonabelian} of the non-Abelian algebra $\mA$ generated by the cat $R$ and $W$ unitaries, defined as
\begin{align}
    R\qket{j} &= \omega_s^j \qket{j} \\
    W\qket{j} &= (-1)^j \qket{\frac{N}{2}-j}
\end{align}
where $\omega_s = \exp ( i 2\pi / s)$ is a root of unity and $N$ is the dimension of the Hilbert space. For simplicity, we take $s$ to be an odd prime, as it is for the cat map~\eqref{eq:cat-6576}, so that $\omega_s^n$ is a primitive $s$th root of unity for $n \not\equiv 0 \bmod s$. For both $R$ and $W$ to be symmetries, we require $N = sM$ with $M$ divisible by 4.

First, note that $R$ only distinguishes the indices $j$ modulo $s$. In other words, the $n$th power $R^n$ can be understood modulo $s$, with $R^s = R^0 = \id$. The $W$ operator has $W^2 = \id$ and has a nontrivial relation to $R$ due to
\begin{equation}
    \frac{N}{2} - j \equiv -j \bmod s
\end{equation}
which leads directly to the relation
\begin{equation} \label{eq:RW-relation}
    W R = R^{-1} W.
\end{equation}
Therefore, a general operator $A \in \mA$ admits the form
\begin{equation} \label{eq:RW-expansion}
    A = \ssum{n} \left(x_n \id + y_n W \right) R^n
\end{equation}
for some complex coefficients $x_n$ and $y_n$.

For completeness, we compute the center of the algebra, although this is not strictly necessary to find the representation~\eqref{eq:algebra-reps}. The identity $\ssum{n} \omega_s^n = 0$ will prove useful here. Given an operator $Z$
\begin{equation}
    Z = \ssum{m} (\alpha_m \id + \beta_m W) R^m
\end{equation}
in the center $\mZ$ of $\mA$, the commutator with $A$ must vanish. We directly compute
\begin{align}
    [Z, A] = \smash{\sum_{n,m}}\,
    \bigl[
        &\beta_m y_n (R^{n-m} - R^{m-n}) \nonumber \\
        &+ \beta_m x_n W (R^{m+n} - R^{m-n}) \nonumber \\
        &+ \alpha_m y_n W (R^{n-m} - R^{m+n})
    \bigr]
\end{align}
which gives us the conditions
\begin{align}
    0 &= \sum_{n,m} \beta_m y_n (\omega_s^{n-m} - \omega_s^{m-n}) \label{eq:center-cond-1} \\
    \begin{split}
        0 &= \smash{\sum_{n,m}}\, \bigl[ \beta_m x_n (\omega_s^{m+n} - \omega_s^{m-n}) \\
            &\qquad\quad\; + \alpha_m y_n (\omega_s^{n-m} - \omega_s^{m+n}) \bigr]
    \end{split} \label{eq:center-cond-2}
\end{align}
for arbitrary $x_n$ and $y_n$. Relation~\eqref{eq:center-cond-1} constrains $\beta_m$ to be an $m$-independent (possibly zero) constant, so that the sum over $m$ vanishes for each term independently. This also means the first term in~\eqref{eq:center-cond-2} is zero. The second term constrains $\alpha_m$ to be symmetric: $\alpha_m = \alpha_{-m}$. The center is then spanned by $R^n + R^{-n}$ for $n=0,\dots,s-1$ and $W\Delta$ where $\Delta$ is the projector
\begin{align}
    \Delta &= \frac{1}{s} \ssum{n} R^n \\
    \Rightarrow \Delta \qket{j} &= \delta(j \equiv 0 \bmod s) \qket{j}.
\end{align}

The irreps of the center provide the direct sum decomposition~\eqref{eq:algebra-reps}. Observe $\mZ$ is only sensitive to the index $j$ modulo $s$. The $j \equiv 0 \bmod s$ subspace ($R + R^{-1} = 2$) contains two irreps: the eigenspaces $W\Delta=+1$ and $W\Delta=-1$. The $j\not\equiv 0 \bmod s$ subspace ($W\Delta = 0$) splits into $\frac{s-1}{2}$ irreps corresponding to the remaining distinct eigenvalues of $R + R^{-1}$, with $j$ and $-j \bmod s$ in the same representation.

\begin{figure}[t]
    \includegraphics[width=0.75\columnwidth]{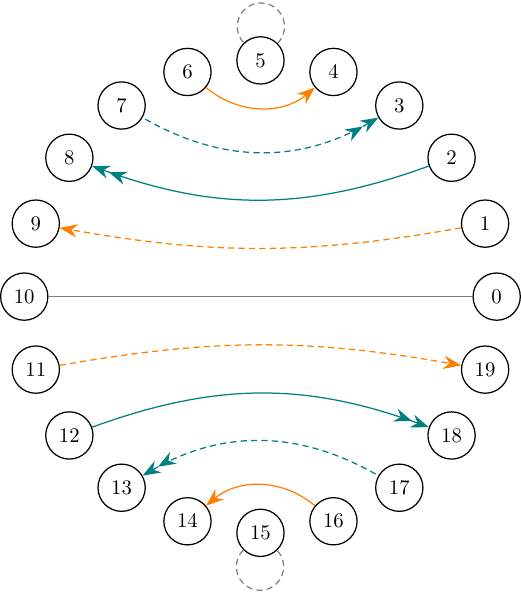}
    \caption{\justifying Graphical representation of the action of $R$ and $W$ on the quantum cat map Hilbert space of dimension $N=20$ with $s=5$. The lines notate the action of $W$ and the arrow heads notate the action of $R$ as described in the text.}
    \label{fig:RW-action}
\end{figure}

To better understand the representation, consider the action of the generators for $N=20$ and $s=5$, as shown in Fig.~\ref{fig:RW-action}. The lines show the indices that swap under $W$, with solid and dashed lines corresponding to phases of $+1$ and $-1$ upon swap respectively. The gray lines with no heads connect to the $R=+1$ indices, which are 0 modulo 5. The orange, single-headed arrows point from 1 to $-1$ modulo 5, and the teal, double-headed arrows point from 2 to $-2$ modulo 5. Thus, the algebra acts on $(\qket{1}, -\qket{9})$ the same as it does on $(\qket{6}, \qket{4})$, $(\qket{11}, -\qket{19})$, and $(\qket{16}, \qket{14})$.

This motivates the following tensor product structure:
\begin{align}
\begin{split} \label{eq:RW-basis}
    \ket{j \uparrow} \otimes \ket{k}_s &\coloneqq \qket{j + ks} \\
    \ket{j \downarrow} \otimes \ket{k}_s &\coloneqq (-1)^{j+ks}\qket{\frac{N}{2} - j - ks}
\end{split}
\end{align}
for $j = 0, \dots, \frac{s-1}{2}$ and $k$ is an index taken modulo $M = N/s$. These vectors span the Hilbert space but double count the following states:
\begin{align}
\begin{split}\label{eq:double-counting}
    \Ket{0 \uparrow}\Ket{\tfrac{M}{4}}_s
        &= \wsign\Ket{0 \downarrow}\Ket{\tfrac{M}{4}}_s \\
    \Ket{0 \uparrow}\Ket{-\tfrac{M}{4}}_s
        &= \wsign\Ket{0 \downarrow}\Ket{-\tfrac{M}{4}}_s
\end{split}
\end{align}
The generators of $\mA$ act as
\begin{align}
\begin{split}
    R \ket{j \uparrow} \ket{k}_s &= \omega_s^j \ket{j \uparrow} \ket{k}_s \\
    R \ket{j \downarrow} \ket{k}_s &= \omega_s^{-j} \ket{j \downarrow} \ket{k}_s
\end{split}
\end{align}
and
\begin{align}
\begin{split}
    W \ket{j \uparrow} \ket{k}_s &= \ket{j \downarrow} \ket{k}_s \\
    W \ket{j \downarrow} \ket{k}_s &= \ket{j \uparrow} \ket{k}_s
\end{split}
\end{align}
The algebra acts trivially on the $\ket{k}_s$ space, but nontrivially on the $\ket{j,\uparrow\!/\!\downarrow}$ space. The double-counted states~\eqref{eq:double-counting} are eigenstates of $W$ with eigenvalue $\wsign$.

Now the subspaces from decomposition~\eqref{eq:algebra-reps} are clear. There is a one-dimensional irrep with $W\Delta = \wsign$ and degeneracy $M/2+1$ given by
\begin{multline}
    \mH_0^+ = \Span \bigl\{ \ket{0 \uparrow}\ket{k}_s + \wsign\ket{0 \downarrow}\ket{k}_s \\
    |\; k=-\tfrac{M}{4}, \dots, \tfrac{M}{4} \bigr\},
\end{multline}
and another one-dimensional irrep with $W\Delta = (-1)^{\frac{M}{4}+1}$ and degeneracy $M/2-1$ given by
\begin{multline}
    \mH_0^- = \Span \bigl\{ \ket{0 \uparrow}\ket{k}_s - \wsign\ket{0 \downarrow}\ket{k}_s \\
    |\; k=1-\tfrac{M}{4}, \dots, \tfrac{M}{4}-1 \bigr\}.
\end{multline}
The subspace with $W\Delta = 0$ splits into $\mH_j = \mK_j \otimes \mK'$ for $j=1,\dots,(s-1)/2$ where
\begin{equation}
    \mK_j = \Span\bigl\{ \ket{j \uparrow}, \ket{j \downarrow} \bigr\}
\end{equation}
is a two-dimensional irrep of $\mA$ and
\begin{equation}
    \mK' = \Span\bigl\{ \ket{k}_s \ |\  k=0, \dots, M-1 \bigr\}
\end{equation}
is the $M$-dimensional space $\mA$ acts trivially on.
We can now decompose the Hilbert space as
\begin{equation}
    \mH = \mH_0^+ \oplus \mH_0^- \oplus \, \bigoplus_{j=1}^{\frac{s-1}{2}} \mK_j \otimes \mK'
\end{equation}
with the algebra taking the form
\begin{equation} \label{eq:RW-algebra}
    \mA = \dsC\, \id_{\frac{M}{2}+1} \oplus \dsC\, \id_{\frac{M}{2}-1} \oplus \,
        \bigoplus_{j=1}^{\frac{s-1}{2}} \mL(\dsC^2) \otimes \id_M.
\end{equation}
Picking $(\ket{j\uparrow}, \ket{j\downarrow})$ as the basis for the two-dimensional irreps $\mK_j$, the generators are matrices
\begin{align}
    R &= \id_{\frac{M}{2}+1} \oplus \id_{\frac{M}{2}-1} \oplus\, \bigoplus_{j=1}^{\frac{s-1}{2}}
    \begin{pmatrix}
        \omega_s^j & 0 \\
        0 & \omega_s^{-j}
    \end{pmatrix}
    \otimes \id_M \\[10pt]
    \begin{split}
        W &= \wsign \id_{\frac{M}{2}+1} \oplus (-1)^{\frac{M}{4}+1} \id_{\frac{M}{2}-1} 
        \\ & \phantom{{}= \id_{\frac{M}{2}+1} \oplus \id_{\frac{M}{2}-1}} \oplus \,
        \bigoplus_{j=1}^{\frac{s-1}{2}}
        \begin{pmatrix}
            0 & 1 \\
            1 & 0
        \end{pmatrix}
        \otimes \id_M
    \end{split}
\end{align}
As a sanity check, the algebra has $1 + 1 + \frac{s-1}{2} \times 2^2 = 2s$ complex degrees of freedom from the decomposition into irreps~\eqref{eq:RW-algebra}, which matches the number of complex parameters in the expansion of a general operator~\eqref{eq:RW-expansion}.

\subsection{AFL Entropy for Varying Perturbation}
\label{sect:cat-perts}

\begin{figure}[t]
    \includegraphics[width=0.9\columnwidth]{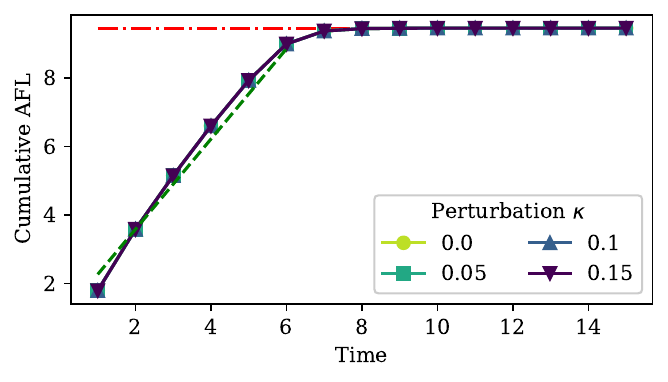}
    \caption{\justifying Plotted is the cumulative AFL entropy of the quantum cat map for varying perturbation strength. The classical KS growth is denoted by the dashed green line. We note the initial growth rate of the cumulative AFL entropy is slightly above the classical rate, as expected. Observe the cumulative AFL entropy is independent of the perturbation strength, despite the dramatic change in spectral statistics.}
    \label{fig:cat-perts}
\end{figure}

The AFL entropy of the quantum cat map~\eqref{eq:pert-cat-map} with $\bf{A} = \smallcatmat{2}{1}{3}{2}$ is plotted in Fig.~\ref{fig:cat-perts} for varying perturbation strength. The partition is 6 projectors in the $q$ (position) basis. As the quantum map is semiclassical, a partition with a smooth semiclassical limit, such as $q$ or $p$ projectors and notably \textit{not} random projectors, is expected to grow at approximately the classical Kolmogorov-Sinai rate (unless dimension-limited by the partition, which is not the case here)~\cite{alicki-2004-DecoherenceRate}. For the unperturbed map, the KS rate is $\log\lambda_+$ where $\lambda_+=2+\sqrt{3}$ is the larger eigenvalue of $\bf{A}$. To be more specific, the growth rate is expected to be slightly above the KS rate due to the extra uncertainty from the quantum measurements, as seen in the figure. This observation is studied in further detail for the quantum baker's map in Ref.~\cite{alicki-2004-DecoherenceRate}. In the classical limit of the quantum cat map, taking the supremum over partition choices, this deviation vanishes and the rate is exactly the KS rate~\cite{benatti-2003-ClassicalLimit}.

Of note here is that the AFL entropies for all the perturbation strengths are essentially identical. This may be surprising, as the unperturbed map does not display random matrix statistics and so is typically not considered quantum chaotic, while the perturbed map does~\cite{matos-1993-QuantizationAnosov, keating-2000-PseudosymmetriesAnosov, backer-2003-NumericalAspects}. However, the map still displays many aspects of chaos in the semiclassical limit~\cite{esposti-2003-MathematicalAspects} and in correlation functions~\cite{chen-2018-OperatorScrambling, garcia-mata-2018-ChaosSignatures, axenides-2018-QuantumCat}. As mentioned in Sec.~\ref{sect:discussion}, this is an example of the separation of basis and spectral chaos as detailed by Magan and Wu~\cite{magan-2024-TwoTypes}, evidencing that AFL entropy is primarily a measure of basis chaos.

\section{Approximate Symmetries} \label{sect:cat-aprx-sym}
\begin{figure}[t]
        \includegraphics[width=\columnwidth]{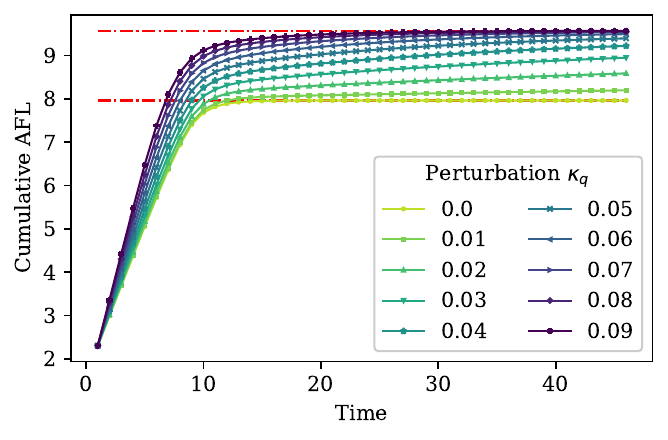}
        \caption{\justifying Approximate symmetry. Plotted is the cumulative AFL entropy of the quantum cat map under $R$-symmetric partitions for various strengths of the symmetry-breaking perturbation strength $\kappa_q$. The horizontal dash-dotted lines show the expected bounds of $2\log N$ without symmetry and $2\log M + \log s$ with $R$ symmetry.}
        \label{fig:cat-aprxsym}
    \end{figure}

    Consider a quantum cat map with nontrivial $R$ and $W$ symmetry. Introducing an additional perturbation of
    \begin{equation}
        \frac{\kappa_q}{4\pi}\cos(4\pi p) \torvec{A_{11}}{A_{21}}
    \end{equation}
    to the classical map weakly breaks the $R$ and $W$ symmetries of the corresponding quantum unitary. We can then explore the behavior of AFL entropy with approximate symmetry. For numerics, we choose $\bfA = \smallcatmat{6}{5}{7}{6}$ with $\kappa=0.001$ and $N=120$. Fig.~\ref{fig:cat-aprxsym} plots the cumulative AFL entropy under an $R$-symmetric partitions of size 10 for various values of $\kappa_q$.
    
    When the $R$ symmetry is weakly broken, the coupling between charge sectors of $R$ is small and the growth of $\cAFL$ is initially dominated by dynamics \textit{within} a charge sector since the measurements do not couple charge sectors. Thus $\cAFL$ grows at roughly the same linear rate as the symmetric case until it nears the charge sector dimensional bound of $2\log M + \log s$. After this point, the cumulative AFL entropy growth is dominated by the symmetry-breaking perturbation. The growth transitions to a slower linear growth proportional to $\kappa_q$ (it is numerically very close to $\kappa_q/2$). Eventually, the AFL entropy nears the full dimensional bound of $2\log N$ and transitions to an exponential approach to the steady state.

    We expect similar results to hold if the symmetry is present but the partition or state weakly breaks it. Similarly, we can imagine weakly breaking tensor product dynamics by having a small coupling between the two Hilbert spaces, either in the dynamics directly or due to measurement of the state.

\section{Proof of Theorem~\ref{thm:abelian} in \textit{SP} Space} \label{sect:abelian-proof2-appendix}

\begin{proof}
After $t$ time steps of the measurement channel, the state of the system and purifier is
\begin{align}
    \mathrlap{\sigAFL{(U\mX)^t}} \nonumber \\
    &\quad= \smash{\sum_{\bm{i}}}\,
        \kett{\MTK{i} \sqrt{\rho}} \nonumber \\
        &\qquad\qquad \otimes \bbra{\MTK{i} \sqrt{\rho}} \nonumber \\
    &\quad= \sum_{\bm{i}}\,
        \Bigl|\,
            \bigoplus_\lambda \MTK[\lambda]{i} \sqrt{p_\lambda \rho_\lambda}
        \,\Bigl\rangle\!\Bigr\rangle
        \nonumber \\ &\qquad\qquad \otimes
        \Bigl\langle\!\Bigr\langle
            \bigoplus_\mu \MTK[\mu]{i} \sqrt{p_\mu \rho_\mu}
        \,\Bigr|
\end{align}
Abusing notation slightly, we can take the direct sum out of the kets to write
\begin{align}
    \mathrlap{\sigAFL{(U\mX)^t}} \nonumber \\
    &\quad= \sum_{\bm{i}}
        \Bigl(
            \bigoplus_{\lambda} \sqrt{p_\lambda} \bigkett{\MTK[\lambda]{i} \sqrt{\rho_\lambda}}
        \Bigr)
        \nonumber \\ &\qquad\qquad \otimes
        \Bigl(
            \bigoplus_{\mu} \sqrt{p_\mu} \bigbbra{\MTK[\mu]{i} \sqrt{\rho_\mu}}
        \Bigr).
\end{align}
To make this expression more transparent, we define
\begin{multline}
    \mS_{\lambda\mu} = \smash{\sum_{\bm{i}}}\, \sqrt{p_\lambda p_\mu}
    \bigkett{\MTK[\lambda]{i} \sqrt{\rho_\lambda}} \\
    \otimes \bigbbra{\MTK[\mu]{i} \sqrt{\rho_\mu}}
\end{multline}
which is a map $\mH_\mu^{\otimes 2} \mapsto \mH_\lambda^{\otimes 2}$. If we abuse notation and use $\mS_{\lambda\mu}$ to refer additionally to the respective map from $\mH^{\otimes 2}$ to itself (with support on $\mH_\mu^{\otimes 2}$ and image in $\mH_\lambda^{\otimes 2}$), we have $\sigma = \sum_{\lambda\mu} \mS_{\lambda\mu}$. This means $\sigma$ is a density matrix on $\mK \coloneqq \bigoplus_\lambda \mH_\lambda^{\otimes 2} \subseteq \mH^{\otimes 2}$ with $\mS_{\lambda\mu}$ as blocks:
\begin{equation} \label{eq:sigmablocks}
    \sigAFL{(U\mX)^t}\Big|_{\mK} =
    \begin{pNiceMatrix}[margin][columns-width=12pt]
        \NiceBlock[b11]{2-2}{\mS_{\lambda_1 \lambda_1}} & & & & \NiceBlock[b12]{2-2}{\mS_{\lambda_1 \lambda_n}} \\
        & & & & & \\
        & & & & & \\
        & & & & & \\
        \NiceBlock[b21]{2-2}{\mS_{\lambda_n \lambda_1}} & & & & \NiceBlock[b22]{2-2}{\mS_{\lambda_n \lambda_n}} \\
        & & & & &
    \CodeAfter
        \line[radius=0.7pt,inter=0.6em,shorten=0.5em]{b11}{b12}
        \line[radius=0.7pt,inter=0.6em,shorten=0.5em]{b11}{b21}
        \line[radius=0.7pt,inter=0.6em,shorten=0.5em]{b12}{b22}
        \line[radius=0.7pt,inter=0.6em,shorten=0.5em]{b21}{b22}
        \line[radius=0.7pt,inter=0.6em,shorten=0.5em]{b11}{b22}
    \end{pNiceMatrix}
\end{equation}
The diagonal blocks can be written as
\begin{align}
    \mS_{\lambda\lambda}
    &= p_\lambda \smash{\sum_{\bm{i}}}\,
    \begin{multlined}[t]
        \bigkett{\MTK[\lambda]{i} \sqrt{\rho_\lambda}} \\
        \otimes \bigbbra{\MTK[\lambda]{i} \sqrt{\rho_\lambda}}
    \end{multlined} \nonumber \\
    &= p_\lambda \sigAFL[\lambda]{(U_\lambda \mX_\lambda)^t}
\end{align}
where $\sigAFL[\lambda]{(U_\lambda \mX_\lambda)^t}$ is the state of the restricted density matrix $\rho_\lambda$ and its purifier after $t$ time steps. Note the block-diagonal part of $\sigma$ is itself a density matrix of the form
\begin{equation}
    \varrho = \bigoplus_\lambda \mS_{\lambda\lambda} = \bigoplus_\lambda p_\lambda \sigAFL[\lambda]{(U_\lambda \mX_\lambda)^t}
\end{equation}
with von Neumann entropy
\begin{equation}
    S (\varrho) =
    \sum_\lambda p_\lambda S \left( \sigAFL[\lambda]{(U_\lambda \mX_\lambda)^t} \right)
    + \shannon{p_\lambda}
\end{equation}
since the $\sigma_\lambda$ have orthogonal support. Now we recall that a Hermitian matrix majorizes the matrix given by its block-diagonal part~\cite{marshall-2011-InequalitiesTheory}, so $\sigma \succ \varrho$. This implies $S(\sigma) \leq S(\varrho)$~\cite{nielsen-2002-IntroductionMajorization}, and so we have
\begin{align}
    \begin{split} \label{eq:symm-bound-sys}
        S \left( \sigAFL{(U\mX)^t} \right)
        \leq \smash{\sum_\lambda}\, p_\lambda 
        S \left( \sigAFL[\lambda]{(U_\lambda \mX_\lambda)^t} \right)\quad& \\
        + \shannon{p_\lambda}&
    \end{split}
\end{align}
as desired.

For the equality case, we now take the Kraus operators in $\mX$ to have support on exactly one charge sector each. Due to equation~\eqref{eq:single-charge}, the nonzero terms of $\mS_{\lambda\mu}$ require $\bm{i} \sim \lambda$ and $\bm{i} \sim \mu$ simultaneously, meaning $\mS_{\lambda\mu}$ vanishes for $\lambda \neq \mu$. Thus, $\sigma$ is equal to its block-diagonal part $\varrho$, and in particular $S(\sigma) = S(\varrho)$ as desired.

\end{proof}

\bibliography{eric_zotero, other_refs}

\begin{thebibliography}{149}%
\makeatletter
\providecommand \@ifxundefined [1]{%
 \@ifx{#1\undefined}
}%
\providecommand \@ifnum [1]{%
 \ifnum #1\expandafter \@firstoftwo
 \else \expandafter \@secondoftwo
 \fi
}%
\providecommand \@ifx [1]{%
 \ifx #1\expandafter \@firstoftwo
 \else \expandafter \@secondoftwo
 \fi
}%
\providecommand \natexlab [1]{#1}%
\providecommand \enquote  [1]{``#1''}%
\providecommand \bibnamefont  [1]{#1}%
\providecommand \bibfnamefont [1]{#1}%
\providecommand \citenamefont [1]{#1}%
\providecommand \href@noop [0]{\@secondoftwo}%
\providecommand \href [0]{\begingroup \@sanitize@url \@href}%
\providecommand \@href[1]{\@@startlink{#1}\@@href}%
\providecommand \@@href[1]{\endgroup#1\@@endlink}%
\providecommand \@sanitize@url [0]{\catcode `\\12\catcode `\$12\catcode `\&12\catcode `\#12\catcode `\^12\catcode `\_12\catcode `\%12\relax}%
\providecommand \@@startlink[1]{}%
\providecommand \@@endlink[0]{}%
\providecommand \url  [0]{\begingroup\@sanitize@url \@url }%
\providecommand \@url [1]{\endgroup\@href {#1}{\urlprefix }}%
\providecommand \urlprefix  [0]{URL }%
\providecommand \Eprint [0]{\href }%
\providecommand \doibase [0]{https://doi.org/}%
\providecommand \selectlanguage [0]{\@gobble}%
\providecommand \bibinfo  [0]{\@secondoftwo}%
\providecommand \bibfield  [0]{\@secondoftwo}%
\providecommand \translation [1]{[#1]}%
\providecommand \BibitemOpen [0]{}%
\providecommand \bibitemStop [0]{}%
\providecommand \bibitemNoStop [0]{.\EOS\space}%
\providecommand \EOS [0]{\spacefactor3000\relax}%
\providecommand \BibitemShut  [1]{\csname bibitem#1\endcsname}%
\let\auto@bib@innerbib\@empty
\bibitem [{\citenamefont {D'Alessio}\ \emph {et~al.}(2016)\citenamefont {D'Alessio}, \citenamefont {Kafri}, \citenamefont {Polkovnikov},\ and\ \citenamefont {Rigol}}]{dalessio-2016-QuantumChaos}%
  \BibitemOpen
  \bibfield  {author} {\bibinfo {author} {\bibfnamefont {L.}~\bibnamefont {D'Alessio}}, \bibinfo {author} {\bibfnamefont {Y.}~\bibnamefont {Kafri}}, \bibinfo {author} {\bibfnamefont {A.}~\bibnamefont {Polkovnikov}},\ and\ \bibinfo {author} {\bibfnamefont {M.}~\bibnamefont {Rigol}},\ }\bibfield  {title} {\bibinfo {title} {From quantum chaos and eigenstate thermalization to statistical mechanics and thermodynamics},\ }\href {https://doi.org/10.1080/00018732.2016.1198134} {\bibfield  {journal} {\bibinfo  {journal} {Advances in Physics}\ }\textbf {\bibinfo {volume} {65}},\ \bibinfo {pages} {239} (\bibinfo {year} {2016})}\BibitemShut {NoStop}%
\bibitem [{\citenamefont {Abanin}\ \emph {et~al.}(2019)\citenamefont {Abanin}, \citenamefont {Altman}, \citenamefont {Bloch},\ and\ \citenamefont {Serbyn}}]{abanin-2019-ColloquiumManybody}%
  \BibitemOpen
  \bibfield  {author} {\bibinfo {author} {\bibfnamefont {D.~A.}\ \bibnamefont {Abanin}}, \bibinfo {author} {\bibfnamefont {E.}~\bibnamefont {Altman}}, \bibinfo {author} {\bibfnamefont {I.}~\bibnamefont {Bloch}},\ and\ \bibinfo {author} {\bibfnamefont {M.}~\bibnamefont {Serbyn}},\ }\bibfield  {title} {\bibinfo {title} {Colloquium: {{Many-body}} localization, thermalization, and entanglement},\ }\href {https://doi.org/10.1103/RevModPhys.91.021001} {\bibfield  {journal} {\bibinfo  {journal} {Reviews of Modern Physics}\ }\textbf {\bibinfo {volume} {91}},\ \bibinfo {pages} {021001} (\bibinfo {year} {2019})}\BibitemShut {NoStop}%
\bibitem [{\citenamefont {Xu}\ and\ \citenamefont {Swingle}(2024)}]{xu-2024-ScramblingDynamics}%
  \BibitemOpen
  \bibfield  {author} {\bibinfo {author} {\bibfnamefont {S.}~\bibnamefont {Xu}}\ and\ \bibinfo {author} {\bibfnamefont {B.}~\bibnamefont {Swingle}},\ }\bibfield  {title} {\bibinfo {title} {Scrambling {{Dynamics}} and {{Out-of-Time-Ordered Correlators}} in {{Quantum Many-Body Systems}}},\ }\href {https://doi.org/10.1103/PRXQuantum.5.010201} {\bibfield  {journal} {\bibinfo  {journal} {PRX Quantum}\ }\textbf {\bibinfo {volume} {5}},\ \bibinfo {pages} {010201} (\bibinfo {year} {2024})}\BibitemShut {NoStop}%
\bibitem [{\citenamefont {{Garc{\'i}a-Mata}}\ \emph {et~al.}(2018)\citenamefont {{Garc{\'i}a-Mata}}, \citenamefont {Saraceno}, \citenamefont {Jalabert}, \citenamefont {Roncaglia},\ and\ \citenamefont {Wisniacki}}]{garcia-mata-2018-ChaosSignatures}%
  \BibitemOpen
  \bibfield  {author} {\bibinfo {author} {\bibfnamefont {I.}~\bibnamefont {{Garc{\'i}a-Mata}}}, \bibinfo {author} {\bibfnamefont {M.}~\bibnamefont {Saraceno}}, \bibinfo {author} {\bibfnamefont {R.~A.}\ \bibnamefont {Jalabert}}, \bibinfo {author} {\bibfnamefont {A.~J.}\ \bibnamefont {Roncaglia}},\ and\ \bibinfo {author} {\bibfnamefont {D.~A.}\ \bibnamefont {Wisniacki}},\ }\bibfield  {title} {\bibinfo {title} {Chaos {{Signatures}} in the {{Short}} and {{Long Time Behavior}} of the {{Out-of-Time Ordered Correlator}}},\ }\href {https://doi.org/10.1103/PhysRevLett.121.210601} {\bibfield  {journal} {\bibinfo  {journal} {Physical Review Letters}\ }\textbf {\bibinfo {volume} {121}},\ \bibinfo {pages} {210601} (\bibinfo {year} {2018})}\BibitemShut {NoStop}%
\bibitem [{\citenamefont {Fortes}\ \emph {et~al.}(2019)\citenamefont {Fortes}, \citenamefont {{Garc{\'i}a-Mata}}, \citenamefont {Jalabert},\ and\ \citenamefont {Wisniacki}}]{fortes-2019-GaugingClassical}%
  \BibitemOpen
  \bibfield  {author} {\bibinfo {author} {\bibfnamefont {E.~M.}\ \bibnamefont {Fortes}}, \bibinfo {author} {\bibfnamefont {I.}~\bibnamefont {{Garc{\'i}a-Mata}}}, \bibinfo {author} {\bibfnamefont {R.~A.}\ \bibnamefont {Jalabert}},\ and\ \bibinfo {author} {\bibfnamefont {D.~A.}\ \bibnamefont {Wisniacki}},\ }\bibfield  {title} {\bibinfo {title} {Gauging classical and quantum integrability through out-of-time-ordered correlators},\ }\href {https://doi.org/10.1103/PhysRevE.100.042201} {\bibfield  {journal} {\bibinfo  {journal} {Physical Review E}\ }\textbf {\bibinfo {volume} {100}},\ \bibinfo {pages} {042201} (\bibinfo {year} {2019})}\BibitemShut {NoStop}%
\bibitem [{\citenamefont {Foini}\ and\ \citenamefont {Kurchan}(2019)}]{foini-2019-OTOCandETH}%
  \BibitemOpen
  \bibfield  {author} {\bibinfo {author} {\bibfnamefont {L.}~\bibnamefont {Foini}}\ and\ \bibinfo {author} {\bibfnamefont {J.}~\bibnamefont {Kurchan}},\ }\bibfield  {title} {\bibinfo {title} {Eigenstate thermalization hypothesis and out of time order correlators},\ }\href {https://doi.org/10.1103/PhysRevE.99.042139} {\bibfield  {journal} {\bibinfo  {journal} {Phys. Rev. E}\ }\textbf {\bibinfo {volume} {99}},\ \bibinfo {pages} {042139} (\bibinfo {year} {2019})}\BibitemShut {NoStop}%
\bibitem [{\citenamefont {Chan}\ \emph {et~al.}(2019)\citenamefont {Chan}, \citenamefont {De~Luca},\ and\ \citenamefont {Chalker}}]{chan-2019-ETH}%
  \BibitemOpen
  \bibfield  {author} {\bibinfo {author} {\bibfnamefont {A.}~\bibnamefont {Chan}}, \bibinfo {author} {\bibfnamefont {A.}~\bibnamefont {De~Luca}},\ and\ \bibinfo {author} {\bibfnamefont {J.~T.}\ \bibnamefont {Chalker}},\ }\bibfield  {title} {\bibinfo {title} {Eigenstate correlations, thermalization, and the butterfly effect},\ }\href {https://doi.org/10.1103/PhysRevLett.122.220601} {\bibfield  {journal} {\bibinfo  {journal} {Phys. Rev. Lett.}\ }\textbf {\bibinfo {volume} {122}},\ \bibinfo {pages} {220601} (\bibinfo {year} {2019})}\BibitemShut {NoStop}%
\bibitem [{\citenamefont {Brenes}\ \emph {et~al.}(2021)\citenamefont {Brenes}, \citenamefont {Pappalardi}, \citenamefont {Mitchison}, \citenamefont {Goold},\ and\ \citenamefont {Silva}}]{brenes-2021-OTOCandETH}%
  \BibitemOpen
  \bibfield  {author} {\bibinfo {author} {\bibfnamefont {M.}~\bibnamefont {Brenes}}, \bibinfo {author} {\bibfnamefont {S.}~\bibnamefont {Pappalardi}}, \bibinfo {author} {\bibfnamefont {M.~T.}\ \bibnamefont {Mitchison}}, \bibinfo {author} {\bibfnamefont {J.}~\bibnamefont {Goold}},\ and\ \bibinfo {author} {\bibfnamefont {A.}~\bibnamefont {Silva}},\ }\bibfield  {title} {\bibinfo {title} {Out-of-time-order correlations and the fine structure of eigenstate thermalization},\ }\href {https://doi.org/10.1103/PhysRevE.104.034120} {\bibfield  {journal} {\bibinfo  {journal} {Phys. Rev. E}\ }\textbf {\bibinfo {volume} {104}},\ \bibinfo {pages} {034120} (\bibinfo {year} {2021})}\BibitemShut {NoStop}%
\bibitem [{\citenamefont {Bohigas}\ \emph {et~al.}(1984)\citenamefont {Bohigas}, \citenamefont {Giannoni},\ and\ \citenamefont {Schmit}}]{bohigas-1984-CharacterizationChaotic}%
  \BibitemOpen
  \bibfield  {author} {\bibinfo {author} {\bibfnamefont {O.}~\bibnamefont {Bohigas}}, \bibinfo {author} {\bibfnamefont {M.~J.}\ \bibnamefont {Giannoni}},\ and\ \bibinfo {author} {\bibfnamefont {C.}~\bibnamefont {Schmit}},\ }\bibfield  {title} {\bibinfo {title} {Characterization of {{Chaotic Quantum Spectra}} and {{Universality}} of {{Level Fluctuation Laws}}},\ }\href {https://doi.org/10.1103/PhysRevLett.52.1} {\bibfield  {journal} {\bibinfo  {journal} {Physical Review Letters}\ }\textbf {\bibinfo {volume} {52}},\ \bibinfo {pages} {1} (\bibinfo {year} {1984})}\BibitemShut {NoStop}%
\bibitem [{\citenamefont {Vikram}\ and\ \citenamefont {Galitski}(2023)}]{vikram-2023-DynamicalQuantum}%
  \BibitemOpen
  \bibfield  {author} {\bibinfo {author} {\bibfnamefont {A.}~\bibnamefont {Vikram}}\ and\ \bibinfo {author} {\bibfnamefont {V.}~\bibnamefont {Galitski}},\ }\bibfield  {title} {\bibinfo {title} {Dynamical quantum ergodicity from energy level statistics},\ }\href {https://doi.org/10.1103/PhysRevResearch.5.033126} {\bibfield  {journal} {\bibinfo  {journal} {Physical Review Research}\ }\textbf {\bibinfo {volume} {5}},\ \bibinfo {pages} {033126} (\bibinfo {year} {2023})},\ \Eprint {https://arxiv.org/abs/2205.05704} {arXiv:2205.05704 [cond-mat, physics:math-ph, physics:nlin, physics:quant-ph]} \BibitemShut {NoStop}%
\bibitem [{\citenamefont {Cotler}\ \emph {et~al.}(2017)\citenamefont {Cotler}, \citenamefont {Hunter-Jones}, \citenamefont {Liu},\ and\ \citenamefont {Yoshida}}]{Cotler-2017-chaoscomplexity}%
  \BibitemOpen
  \bibfield  {author} {\bibinfo {author} {\bibfnamefont {J.}~\bibnamefont {Cotler}}, \bibinfo {author} {\bibfnamefont {N.}~\bibnamefont {Hunter-Jones}}, \bibinfo {author} {\bibfnamefont {J.}~\bibnamefont {Liu}},\ and\ \bibinfo {author} {\bibfnamefont {B.}~\bibnamefont {Yoshida}},\ }\bibfield  {title} {\bibinfo {title} {Chaos, complexity, and random matrices},\ }\href {https://doi.org/10.1007/JHEP11(2017)048} {\bibfield  {journal} {\bibinfo  {journal} {Journal of High Energy Physics}\ }\textbf {\bibinfo {volume} {2017}},\ \bibinfo {pages} {48} (\bibinfo {year} {2017})}\BibitemShut {NoStop}%
\bibitem [{\citenamefont {Kos}\ \emph {et~al.}(2018)\citenamefont {Kos}, \citenamefont {Ljubotina},\ and\ \citenamefont {Prosen}}]{Kos-2018-analyticconnection}%
  \BibitemOpen
  \bibfield  {author} {\bibinfo {author} {\bibfnamefont {P.}~\bibnamefont {Kos}}, \bibinfo {author} {\bibfnamefont {M.}~\bibnamefont {Ljubotina}},\ and\ \bibinfo {author} {\bibfnamefont {T.~c.~v.}\ \bibnamefont {Prosen}},\ }\bibfield  {title} {\bibinfo {title} {Many-body quantum chaos: Analytic connection to random matrix theory},\ }\href {https://doi.org/10.1103/PhysRevX.8.021062} {\bibfield  {journal} {\bibinfo  {journal} {Phys. Rev. X}\ }\textbf {\bibinfo {volume} {8}},\ \bibinfo {pages} {021062} (\bibinfo {year} {2018})}\BibitemShut {NoStop}%
\bibitem [{\citenamefont {Hosur}\ \emph {et~al.}(2016)\citenamefont {Hosur}, \citenamefont {Qi}, \citenamefont {Roberts},\ and\ \citenamefont {Yoshida}}]{hosur-2016-ChaosQuantum}%
  \BibitemOpen
  \bibfield  {author} {\bibinfo {author} {\bibfnamefont {P.}~\bibnamefont {Hosur}}, \bibinfo {author} {\bibfnamefont {X.-L.}\ \bibnamefont {Qi}}, \bibinfo {author} {\bibfnamefont {D.~A.}\ \bibnamefont {Roberts}},\ and\ \bibinfo {author} {\bibfnamefont {B.}~\bibnamefont {Yoshida}},\ }\bibfield  {title} {\bibinfo {title} {Chaos in quantum channels},\ }\href {https://doi.org/10.1007/JHEP02(2016)004} {\bibfield  {journal} {\bibinfo  {journal} {Journal of High Energy Physics}\ }\textbf {\bibinfo {volume} {2016}},\ \bibinfo {pages} {4} (\bibinfo {year} {2016})}\BibitemShut {NoStop}%
\bibitem [{\citenamefont {Nie}\ \emph {et~al.}(2019)\citenamefont {Nie}, \citenamefont {Nozaki}, \citenamefont {Ryu},\ and\ \citenamefont {Tan}}]{nie-2019-SignatureQuantum}%
  \BibitemOpen
  \bibfield  {author} {\bibinfo {author} {\bibfnamefont {L.}~\bibnamefont {Nie}}, \bibinfo {author} {\bibfnamefont {M.}~\bibnamefont {Nozaki}}, \bibinfo {author} {\bibfnamefont {S.}~\bibnamefont {Ryu}},\ and\ \bibinfo {author} {\bibfnamefont {M.~T.}\ \bibnamefont {Tan}},\ }\bibfield  {title} {\bibinfo {title} {Signature of quantum chaos in operator entanglement in 2d {{CFTs}}},\ }\href {https://doi.org/10.1088/1742-5468/ab3a29} {\bibfield  {journal} {\bibinfo  {journal} {Journal of Statistical Mechanics: Theory and Experiment}\ }\textbf {\bibinfo {volume} {2019}},\ \bibinfo {pages} {093107} (\bibinfo {year} {2019})}\BibitemShut {NoStop}%
\bibitem [{\citenamefont {Bianchi}\ \emph {et~al.}(2022)\citenamefont {Bianchi}, \citenamefont {Hackl}, \citenamefont {Kieburg}, \citenamefont {Rigol},\ and\ \citenamefont {Vidmar}}]{bianchi-2022-VolumeLawEntanglement}%
  \BibitemOpen
  \bibfield  {author} {\bibinfo {author} {\bibfnamefont {E.}~\bibnamefont {Bianchi}}, \bibinfo {author} {\bibfnamefont {L.}~\bibnamefont {Hackl}}, \bibinfo {author} {\bibfnamefont {M.}~\bibnamefont {Kieburg}}, \bibinfo {author} {\bibfnamefont {M.}~\bibnamefont {Rigol}},\ and\ \bibinfo {author} {\bibfnamefont {L.}~\bibnamefont {Vidmar}},\ }\bibfield  {title} {\bibinfo {title} {Volume-{{Law Entanglement Entropy}} of {{Typical Pure Quantum States}}},\ }\href {https://doi.org/10.1103/PRXQuantum.3.030201} {\bibfield  {journal} {\bibinfo  {journal} {PRX Quantum}\ }\textbf {\bibinfo {volume} {3}},\ \bibinfo {pages} {030201} (\bibinfo {year} {2022})}\BibitemShut {NoStop}%
\bibitem [{\citenamefont {Li}\ \emph {et~al.}(2018)\citenamefont {Li}, \citenamefont {Chen},\ and\ \citenamefont {Fisher}}]{Li-2018-Zeno}%
  \BibitemOpen
  \bibfield  {author} {\bibinfo {author} {\bibfnamefont {Y.}~\bibnamefont {Li}}, \bibinfo {author} {\bibfnamefont {X.}~\bibnamefont {Chen}},\ and\ \bibinfo {author} {\bibfnamefont {M.~P.~A.}\ \bibnamefont {Fisher}},\ }\bibfield  {title} {\bibinfo {title} {Quantum zeno effect and the many-body entanglement transition},\ }\href {https://doi.org/10.1103/PhysRevB.98.205136} {\bibfield  {journal} {\bibinfo  {journal} {Phys. Rev. B}\ }\textbf {\bibinfo {volume} {98}},\ \bibinfo {pages} {205136} (\bibinfo {year} {2018})}\BibitemShut {NoStop}%
\bibitem [{\citenamefont {Skinner}\ \emph {et~al.}(2019)\citenamefont {Skinner}, \citenamefont {Ruhman},\ and\ \citenamefont {Nahum}}]{Skinner-2019-mipt}%
  \BibitemOpen
  \bibfield  {author} {\bibinfo {author} {\bibfnamefont {B.}~\bibnamefont {Skinner}}, \bibinfo {author} {\bibfnamefont {J.}~\bibnamefont {Ruhman}},\ and\ \bibinfo {author} {\bibfnamefont {A.}~\bibnamefont {Nahum}},\ }\bibfield  {title} {\bibinfo {title} {Measurement-induced phase transitions in the dynamics of entanglement},\ }\href {https://doi.org/10.1103/PhysRevX.9.031009} {\bibfield  {journal} {\bibinfo  {journal} {Phys. Rev. X}\ }\textbf {\bibinfo {volume} {9}},\ \bibinfo {pages} {031009} (\bibinfo {year} {2019})}\BibitemShut {NoStop}%
\bibitem [{\citenamefont {Potter}\ and\ \citenamefont {Vasseur}(2022)}]{potter-2022-MIPT}%
  \BibitemOpen
  \bibfield  {author} {\bibinfo {author} {\bibfnamefont {A.~C.}\ \bibnamefont {Potter}}\ and\ \bibinfo {author} {\bibfnamefont {R.}~\bibnamefont {Vasseur}},\ }\bibinfo {title} {Entanglement dynamics in hybrid quantum circuits},\ in\ \href {https://doi.org/10.1007/978-3-031-03998-0_9} {\emph {\bibinfo {booktitle} {Entanglement in Spin Chains}}}\ (\bibinfo  {publisher} {Springer International Publishing},\ \bibinfo {year} {2022})\ p.\ \bibinfo {pages} {211–249}\BibitemShut {NoStop}%
\bibitem [{\citenamefont {Fisher}\ \emph {et~al.}(2023)\citenamefont {Fisher}, \citenamefont {Khemani}, \citenamefont {Nahum},\ and\ \citenamefont {Vijay}}]{fisher-2023-MIPT}%
  \BibitemOpen
  \bibfield  {author} {\bibinfo {author} {\bibfnamefont {M.~P.}\ \bibnamefont {Fisher}}, \bibinfo {author} {\bibfnamefont {V.}~\bibnamefont {Khemani}}, \bibinfo {author} {\bibfnamefont {A.}~\bibnamefont {Nahum}},\ and\ \bibinfo {author} {\bibfnamefont {S.}~\bibnamefont {Vijay}},\ }\bibfield  {title} {\bibinfo {title} {Random quantum circuits},\ }\href {https://doi.org/10.1146/annurev-conmatphys-031720-030658} {\bibfield  {journal} {\bibinfo  {journal} {Annual Review of Condensed Matter Physics}\ }\textbf {\bibinfo {volume} {14}},\ \bibinfo {pages} {335–379} (\bibinfo {year} {2023})}\BibitemShut {NoStop}%
\bibitem [{\citenamefont {Skinner}(2023)}]{skinner-2023-MIPT}%
  \BibitemOpen
  \bibfield  {author} {\bibinfo {author} {\bibfnamefont {B.}~\bibnamefont {Skinner}},\ }\href {https://arxiv.org/abs/2307.02986} {\bibinfo {title} {Lecture notes: Introduction to random unitary circuits and the measurement-induced entanglement phase transition}} (\bibinfo {year} {2023}),\ \Eprint {https://arxiv.org/abs/2307.02986} {arXiv:2307.02986 [cond-mat.stat-mech]} \BibitemShut {NoStop}%
\bibitem [{\citenamefont {Nonnenmacher}(2003)}]{nonnenmacher-2003-SpectralProperties}%
  \BibitemOpen
  \bibfield  {author} {\bibinfo {author} {\bibfnamefont {S.}~\bibnamefont {Nonnenmacher}},\ }\bibfield  {title} {\bibinfo {title} {Spectral properties of noisy classical and quantum propagators},\ }\href {https://doi.org/10.1088/0951-7715/16/5/309} {\bibfield  {journal} {\bibinfo  {journal} {Nonlinearity}\ }\textbf {\bibinfo {volume} {16}},\ \bibinfo {pages} {1685} (\bibinfo {year} {2003})}\BibitemShut {NoStop}%
\bibitem [{\citenamefont {Prosen}(2004)}]{prosen-2004-RuelleResonances}%
  \BibitemOpen
  \bibfield  {author} {\bibinfo {author} {\bibfnamefont {T.}~\bibnamefont {Prosen}},\ }\bibfield  {title} {\bibinfo {title} {Ruelle resonances in kicked quantum spin chain},\ }\href {https://doi.org/10.1016/j.physd.2003.09.017} {\bibfield  {journal} {\bibinfo  {journal} {Physica D: Nonlinear Phenomena}\ }\bibinfo {series} {Microscopic {{Chaos}} and {{Transport}} in {{Many-Particle Systems}}},\ \textbf {\bibinfo {volume} {187}},\ \bibinfo {pages} {244} (\bibinfo {year} {2004})}\BibitemShut {NoStop}%
\bibitem [{\citenamefont {{Garc{\'i}a-Mata}}\ and\ \citenamefont {Saraceno}(2005)}]{garcia-mata-2005-SpectralApproach}%
  \BibitemOpen
  \bibfield  {author} {\bibinfo {author} {\bibfnamefont {I.}~\bibnamefont {{Garc{\'i}a-Mata}}}\ and\ \bibinfo {author} {\bibfnamefont {M.}~\bibnamefont {Saraceno}},\ }\bibfield  {title} {\bibinfo {title} {Spectral approach to chaos and quantum-classical correspondence in quantum maps},\ }\href {https://doi.org/10.1142/S0217984905008402} {\bibfield  {journal} {\bibinfo  {journal} {Modern Physics Letters B}\ }\textbf {\bibinfo {volume} {19}},\ \bibinfo {pages} {341} (\bibinfo {year} {2005})}\BibitemShut {NoStop}%
\bibitem [{\citenamefont {Yoshimura}\ and\ \citenamefont {S{\'a}}(2024)}]{yoshimura-2024-RobustnessQuantum}%
  \BibitemOpen
  \bibfield  {author} {\bibinfo {author} {\bibfnamefont {T.}~\bibnamefont {Yoshimura}}\ and\ \bibinfo {author} {\bibfnamefont {L.}~\bibnamefont {S{\'a}}},\ }\bibfield  {title} {\bibinfo {title} {Robustness of quantum chaos and anomalous relaxation in open quantum circuits},\ }\href {https://doi.org/10.1038/s41467-024-54164-7} {\bibfield  {journal} {\bibinfo  {journal} {Nature Communications}\ }\textbf {\bibinfo {volume} {15}},\ \bibinfo {pages} {9808} (\bibinfo {year} {2024})}\BibitemShut {NoStop}%
\bibitem [{\citenamefont {Mori}(2024)}]{mori-2024-LiouvilliangapAnalysis}%
  \BibitemOpen
  \bibfield  {author} {\bibinfo {author} {\bibfnamefont {T.}~\bibnamefont {Mori}},\ }\href {http://arxiv.org/abs/2311.10304} {\bibinfo {title} {Liouvillian-gap analysis of open quantum many-body systems in the weak dissipation limit}} (\bibinfo {year} {2024}),\ \Eprint {https://arxiv.org/abs/2311.10304} {arXiv:2311.10304 [cond-mat, physics:quant-ph]} \BibitemShut {NoStop}%
\bibitem [{\citenamefont {Jacoby}\ \emph {et~al.}(2024)\citenamefont {Jacoby}, \citenamefont {Huse},\ and\ \citenamefont {Gopalakrishnan}}]{jacoby-2024-SpectralGaps}%
  \BibitemOpen
  \bibfield  {author} {\bibinfo {author} {\bibfnamefont {J.~A.}\ \bibnamefont {Jacoby}}, \bibinfo {author} {\bibfnamefont {D.~A.}\ \bibnamefont {Huse}},\ and\ \bibinfo {author} {\bibfnamefont {S.}~\bibnamefont {Gopalakrishnan}},\ }\href {https://doi.org/10.48550/arXiv.2409.17238} {\bibinfo {title} {Spectral gaps of local quantum channels in the weak-dissipation limit}} (\bibinfo {year} {2024}),\ \Eprint {https://arxiv.org/abs/2409.17238} {arXiv:2409.17238} \BibitemShut {NoStop}%
\bibitem [{\citenamefont {{\v Z}nidari{\v c}}(2024)}]{znidaric-2024-MomentumdependentQuantum}%
  \BibitemOpen
  \bibfield  {author} {\bibinfo {author} {\bibfnamefont {M.}~\bibnamefont {{\v Z}nidari{\v c}}},\ }\bibfield  {title} {\bibinfo {title} {Momentum-dependent quantum {{Ruelle-Pollicott}} resonances in translationally invariant many-body systems},\ }\href {https://doi.org/10.1103/PhysRevE.110.054204} {\bibfield  {journal} {\bibinfo  {journal} {Physical Review E}\ }\textbf {\bibinfo {volume} {110}},\ \bibinfo {pages} {054204} (\bibinfo {year} {2024})}\BibitemShut {NoStop}%
\bibitem [{\citenamefont {Zhang}\ \emph {et~al.}(2024)\citenamefont {Zhang}, \citenamefont {Nie},\ and\ \citenamefont {{von Keyserlingk}}}]{zhang-2024-ThermalizationRates}%
  \BibitemOpen
  \bibfield  {author} {\bibinfo {author} {\bibfnamefont {C.}~\bibnamefont {Zhang}}, \bibinfo {author} {\bibfnamefont {L.}~\bibnamefont {Nie}},\ and\ \bibinfo {author} {\bibfnamefont {C.}~\bibnamefont {{von Keyserlingk}}},\ }\href {https://doi.org/10.48550/arXiv.2409.17251} {\bibinfo {title} {Thermalization rates and quantum {{Ruelle-Pollicott}} resonances: Insights from operator hydrodynamics}} (\bibinfo {year} {2024}),\ \Eprint {https://arxiv.org/abs/2409.17251} {arXiv:2409.17251} \BibitemShut {NoStop}%
\bibitem [{\citenamefont {Yoshimura}\ and\ \citenamefont {Sá}(2025)}]{yoshimura2025irreversibility}%
  \BibitemOpen
  \bibfield  {author} {\bibinfo {author} {\bibfnamefont {T.}~\bibnamefont {Yoshimura}}\ and\ \bibinfo {author} {\bibfnamefont {L.}~\bibnamefont {Sá}},\ }\href {https://arxiv.org/abs/2501.06183} {\bibinfo {title} {Theory of irreversibility in quantum many-body systems}} (\bibinfo {year} {2025}),\ \Eprint {https://arxiv.org/abs/2501.06183} {arXiv:2501.06183 [cond-mat.stat-mech]} \BibitemShut {NoStop}%
\bibitem [{\citenamefont {Walters}(1982)}]{WaltersBook}%
  \BibitemOpen
  \bibfield  {author} {\bibinfo {author} {\bibfnamefont {P.}~\bibnamefont {Walters}},\ }\href@noop {} {\emph {\bibinfo {title} {An Introduction to Ergodic Theory}}}\ (\bibinfo  {publisher} {Springer New York},\ \bibinfo {year} {1982})\BibitemShut {NoStop}%
\bibitem [{\citenamefont {Connes}\ and\ \citenamefont {St{\o}rmer}(1975)}]{connes-1975-EntropyAutomorphisms}%
  \BibitemOpen
  \bibfield  {author} {\bibinfo {author} {\bibfnamefont {A.}~\bibnamefont {Connes}}\ and\ \bibinfo {author} {\bibfnamefont {E.}~\bibnamefont {St{\o}rmer}},\ }\bibfield  {title} {\bibinfo {title} {Entropy for automorphisms of {{II}}{$_1$} von {{Neumann}} algebras},\ }\href {https://doi.org/10.1007/BF02392105} {\bibfield  {journal} {\bibinfo  {journal} {Acta Mathematica}\ }\textbf {\bibinfo {volume} {134}},\ \bibinfo {pages} {289} (\bibinfo {year} {1975})}\BibitemShut {NoStop}%
\bibitem [{\citenamefont {Connes}\ \emph {et~al.}(1987)\citenamefont {Connes}, \citenamefont {Narnhofer},\ and\ \citenamefont {Thirring}}]{connes-1987-DynamicalEntropy}%
  \BibitemOpen
  \bibfield  {author} {\bibinfo {author} {\bibfnamefont {A.}~\bibnamefont {Connes}}, \bibinfo {author} {\bibfnamefont {H.}~\bibnamefont {Narnhofer}},\ and\ \bibinfo {author} {\bibfnamefont {W.}~\bibnamefont {Thirring}},\ }\bibfield  {title} {\bibinfo {title} {Dynamical entropy of {{C}}* algebras and von {{Neumann}} algebras},\ }\href {https://doi.org/10.1007/BF01225381} {\bibfield  {journal} {\bibinfo  {journal} {Communications in Mathematical Physics}\ }\textbf {\bibinfo {volume} {112}},\ \bibinfo {pages} {691} (\bibinfo {year} {1987})}\BibitemShut {NoStop}%
\bibitem [{\citenamefont {S{\l}omczy{\'n}ski}\ and\ \citenamefont {{\.Z}yczkowski}(1994)}]{slomczynski-1994-QuantumChaos}%
  \BibitemOpen
  \bibfield  {author} {\bibinfo {author} {\bibfnamefont {W.}~\bibnamefont {S{\l}omczy{\'n}ski}}\ and\ \bibinfo {author} {\bibfnamefont {K.}~\bibnamefont {{\.Z}yczkowski}},\ }\bibfield  {title} {\bibinfo {title} {Quantum chaos: {{An}} entropy approach},\ }\href {https://doi.org/10.1063/1.530704} {\bibfield  {journal} {\bibinfo  {journal} {Journal of Mathematical Physics}\ }\textbf {\bibinfo {volume} {35}},\ \bibinfo {pages} {5674} (\bibinfo {year} {1994})}\BibitemShut {NoStop}%
\bibitem [{\citenamefont {Voiculescu}(1995)}]{voiculescu-1995-DynamicalApproximation}%
  \BibitemOpen
  \bibfield  {author} {\bibinfo {author} {\bibfnamefont {D.}~\bibnamefont {Voiculescu}},\ }\bibfield  {title} {\bibinfo {title} {Dynamical approximation entropies and topological entropy in operator algebras},\ }\href {https://doi.org/10.1007/BF02108329} {\bibfield  {journal} {\bibinfo  {journal} {Communications in Mathematical Physics}\ }\textbf {\bibinfo {volume} {170}},\ \bibinfo {pages} {249} (\bibinfo {year} {1995})}\BibitemShut {NoStop}%
\bibitem [{\citenamefont {Alicki}\ and\ \citenamefont {Fannes}(1994)}]{alicki-1994-DefiningQuantum}%
  \BibitemOpen
  \bibfield  {author} {\bibinfo {author} {\bibfnamefont {R.}~\bibnamefont {Alicki}}\ and\ \bibinfo {author} {\bibfnamefont {M.}~\bibnamefont {Fannes}},\ }\bibfield  {title} {\bibinfo {title} {Defining quantum dynamical entropy},\ }\href {https://doi.org/10.1007/BF00761125} {\bibfield  {journal} {\bibinfo  {journal} {Letters in Mathematical Physics}\ }\textbf {\bibinfo {volume} {32}},\ \bibinfo {pages} {75} (\bibinfo {year} {1994})}\BibitemShut {NoStop}%
\bibitem [{\citenamefont {S{\l}omczy{\'n}ski}\ and\ \citenamefont {Szczepanek}(2017)}]{slomczynski-2017-QuantumDynamical}%
  \BibitemOpen
  \bibfield  {author} {\bibinfo {author} {\bibfnamefont {W.}~\bibnamefont {S{\l}omczy{\'n}ski}}\ and\ \bibinfo {author} {\bibfnamefont {A.}~\bibnamefont {Szczepanek}},\ }\bibfield  {title} {\bibinfo {title} {Quantum {{Dynamical Entropy}}, {{Chaotic Unitaries}} and {{Complex Hadamard Matrices}}},\ }\href {https://doi.org/10.1109/TIT.2017.2751507} {\bibfield  {journal} {\bibinfo  {journal} {IEEE Transactions on Information Theory}\ }\textbf {\bibinfo {volume} {63}},\ \bibinfo {pages} {7821} (\bibinfo {year} {2017})}\BibitemShut {NoStop}%
\bibitem [{\citenamefont {Gharibyan}\ \emph {et~al.}(2019)\citenamefont {Gharibyan}, \citenamefont {Hanada}, \citenamefont {Swingle},\ and\ \citenamefont {Tezuka}}]{Swingle2019LyapunovSpectrum}%
  \BibitemOpen
  \bibfield  {author} {\bibinfo {author} {\bibfnamefont {H.}~\bibnamefont {Gharibyan}}, \bibinfo {author} {\bibfnamefont {M.}~\bibnamefont {Hanada}}, \bibinfo {author} {\bibfnamefont {B.}~\bibnamefont {Swingle}},\ and\ \bibinfo {author} {\bibfnamefont {M.}~\bibnamefont {Tezuka}},\ }\bibfield  {title} {\bibinfo {title} {Quantum lyapunov spectrum},\ }\href {https://doi.org/10.1007/JHEP04(2019)082} {\bibfield  {journal} {\bibinfo  {journal} {Journal of High Energy Physics}\ }\textbf {\bibinfo {volume} {2019}},\ \bibinfo {pages} {82} (\bibinfo {year} {2019})}\BibitemShut {NoStop}%
\bibitem [{\citenamefont {Goldfriend}\ and\ \citenamefont {Kurchan}(2021)}]{Goldfriend2021KSPesin}%
  \BibitemOpen
  \bibfield  {author} {\bibinfo {author} {\bibfnamefont {T.}~\bibnamefont {Goldfriend}}\ and\ \bibinfo {author} {\bibfnamefont {J.}~\bibnamefont {Kurchan}},\ }\bibfield  {title} {\bibinfo {title} {Quantum kolmogorov-sinai entropy and pesin relation},\ }\href {https://doi.org/10.1103/PhysRevResearch.3.023234} {\bibfield  {journal} {\bibinfo  {journal} {Phys. Rev. Res.}\ }\textbf {\bibinfo {volume} {3}},\ \bibinfo {pages} {023234} (\bibinfo {year} {2021})}\BibitemShut {NoStop}%
\bibitem [{\citenamefont {Maier}\ \emph {et~al.}(2022)\citenamefont {Maier}, \citenamefont {Sch{\"a}fer},\ and\ \citenamefont {Waeber}}]{Maier2022holographicKS}%
  \BibitemOpen
  \bibfield  {author} {\bibinfo {author} {\bibfnamefont {G.}~\bibnamefont {Maier}}, \bibinfo {author} {\bibfnamefont {A.}~\bibnamefont {Sch{\"a}fer}},\ and\ \bibinfo {author} {\bibfnamefont {S.}~\bibnamefont {Waeber}},\ }\bibfield  {title} {\bibinfo {title} {Holographic kolmogorov-sinai entropy and the quantum lyapunov spectrum},\ }\href {https://doi.org/10.1007/JHEP01(2022)165} {\bibfield  {journal} {\bibinfo  {journal} {Journal of High Energy Physics}\ }\textbf {\bibinfo {volume} {2022}},\ \bibinfo {pages} {165} (\bibinfo {year} {2022})}\BibitemShut {NoStop}%
\bibitem [{\citenamefont {Alicki}\ \emph {et~al.}(1996{\natexlab{a}})\citenamefont {Alicki}, \citenamefont {Makowiec},\ and\ \citenamefont {Miklaszewski}}]{alicki-1996-KickedTop}%
  \BibitemOpen
  \bibfield  {author} {\bibinfo {author} {\bibfnamefont {R.}~\bibnamefont {Alicki}}, \bibinfo {author} {\bibfnamefont {D.}~\bibnamefont {Makowiec}},\ and\ \bibinfo {author} {\bibfnamefont {W.}~\bibnamefont {Miklaszewski}},\ }\bibfield  {title} {\bibinfo {title} {Quantum {{Chaos}} in {{Terms}} of {{Entropy}} for a {{Periodically Kicked Top}}},\ }\href {https://doi.org/10.1103/PhysRevLett.77.838} {\bibfield  {journal} {\bibinfo  {journal} {Physical Review Letters}\ }\textbf {\bibinfo {volume} {77}},\ \bibinfo {pages} {838} (\bibinfo {year} {1996}{\natexlab{a}})}\BibitemShut {NoStop}%
\bibitem [{\citenamefont {Alicki}\ \emph {et~al.}(2004)\citenamefont {Alicki}, \citenamefont {{\L}ozi{\'n}ski}, \citenamefont {Pako{\'n}ski},\ and\ \citenamefont {{\.Z}yczkowski}}]{alicki-2004-DecoherenceRate}%
  \BibitemOpen
  \bibfield  {author} {\bibinfo {author} {\bibfnamefont {R.}~\bibnamefont {Alicki}}, \bibinfo {author} {\bibfnamefont {A.}~\bibnamefont {{\L}ozi{\'n}ski}}, \bibinfo {author} {\bibfnamefont {P.}~\bibnamefont {Pako{\'n}ski}},\ and\ \bibinfo {author} {\bibfnamefont {K.}~\bibnamefont {{\.Z}yczkowski}},\ }\bibfield  {title} {\bibinfo {title} {Quantum dynamical entropy and decoherence rate},\ }\href {https://doi.org/10.1088/0305-4470/37/19/004} {\bibfield  {journal} {\bibinfo  {journal} {Journal of Physics A: Mathematical and General}\ }\textbf {\bibinfo {volume} {37}},\ \bibinfo {pages} {5157} (\bibinfo {year} {2004})}\BibitemShut {NoStop}%
\bibitem [{\citenamefont {Alicki}(1997)}]{alicki-1997-QuantumErgodic}%
  \BibitemOpen
  \bibfield  {author} {\bibinfo {author} {\bibfnamefont {R.}~\bibnamefont {Alicki}},\ }\bibfield  {title} {\bibinfo {title} {Quantum {{Ergodic Theory}} and {{Communication Channels}}},\ }\href {https://doi.org/10.1023/A:1009657518056} {\bibfield  {journal} {\bibinfo  {journal} {Open Systems \& Information Dynamics}\ }\textbf {\bibinfo {volume} {4}},\ \bibinfo {pages} {53} (\bibinfo {year} {1997})}\BibitemShut {NoStop}%
\bibitem [{\citenamefont {Alicki}(2002)}]{alicki-2002-InformationtheoreticalMeaning}%
  \BibitemOpen
  \bibfield  {author} {\bibinfo {author} {\bibfnamefont {R.}~\bibnamefont {Alicki}},\ }\bibfield  {title} {\bibinfo {title} {Information-theoretical meaning of quantum-dynamical entropy},\ }\href {https://doi.org/10.1103/PhysRevA.66.052302} {\bibfield  {journal} {\bibinfo  {journal} {Physical Review A}\ }\textbf {\bibinfo {volume} {66}},\ \bibinfo {pages} {052302} (\bibinfo {year} {2002})}\BibitemShut {NoStop}%
\bibitem [{\citenamefont {Cotler}\ \emph {et~al.}(2018)\citenamefont {Cotler}, \citenamefont {Jian}, \citenamefont {Qi},\ and\ \citenamefont {Wilczek}}]{cotler-2018-SuperdensityOperators}%
  \BibitemOpen
  \bibfield  {author} {\bibinfo {author} {\bibfnamefont {J.}~\bibnamefont {Cotler}}, \bibinfo {author} {\bibfnamefont {C.-M.}\ \bibnamefont {Jian}}, \bibinfo {author} {\bibfnamefont {X.-L.}\ \bibnamefont {Qi}},\ and\ \bibinfo {author} {\bibfnamefont {F.}~\bibnamefont {Wilczek}},\ }\bibfield  {title} {\bibinfo {title} {Superdensity operators for spacetime quantum mechanics},\ }\href {https://doi.org/10.1007/JHEP09(2018)093} {\bibfield  {journal} {\bibinfo  {journal} {Journal of High Energy Physics}\ }\textbf {\bibinfo {volume} {2018}},\ \bibinfo {pages} {93} (\bibinfo {year} {2018})}\BibitemShut {NoStop}%
\bibitem [{\citenamefont {O'Donovan}\ \emph {et~al.}(2025)\citenamefont {O'Donovan}, \citenamefont {Dowling}, \citenamefont {Modi},\ and\ \citenamefont {Mitchison}}]{odonovan-2025-DiagnosingChaos}%
  \BibitemOpen
  \bibfield  {author} {\bibinfo {author} {\bibfnamefont {P.}~\bibnamefont {O'Donovan}}, \bibinfo {author} {\bibfnamefont {N.}~\bibnamefont {Dowling}}, \bibinfo {author} {\bibfnamefont {K.}~\bibnamefont {Modi}},\ and\ \bibinfo {author} {\bibfnamefont {M.~T.}\ \bibnamefont {Mitchison}},\ }\href {https://doi.org/10.48550/arXiv.2502.13930} {\bibinfo {title} {Diagnosing chaos with projected ensembles of process tensors}} (\bibinfo {year} {2025}),\ \Eprint {https://arxiv.org/abs/2502.13930} {arXiv:2502.13930 [quant-ph]} \BibitemShut {NoStop}%
\bibitem [{\citenamefont {Yoshimura}\ and\ \citenamefont {S{\'a}}(2025)}]{yoshimura-2025-TheoryIrreversibility}%
  \BibitemOpen
  \bibfield  {author} {\bibinfo {author} {\bibfnamefont {T.}~\bibnamefont {Yoshimura}}\ and\ \bibinfo {author} {\bibfnamefont {L.}~\bibnamefont {S{\'a}}},\ }\href {https://doi.org/10.48550/arXiv.2501.06183} {\bibinfo {title} {Theory of {{Irreversibility}} in {{Quantum Many-Body Systems}}}} (\bibinfo {year} {2025}),\ \Eprint {https://arxiv.org/abs/2501.06183} {arXiv:2501.06183 [cond-mat]} \BibitemShut {NoStop}%
\bibitem [{\citenamefont {Dowling}\ \emph {et~al.}(2023{\natexlab{a}})\citenamefont {Dowling}, \citenamefont {{Figueroa-Romero}}, \citenamefont {Pollock}, \citenamefont {Strasberg},\ and\ \citenamefont {Modi}}]{dowling-2023-RelaxationMultitime}%
  \BibitemOpen
  \bibfield  {author} {\bibinfo {author} {\bibfnamefont {N.}~\bibnamefont {Dowling}}, \bibinfo {author} {\bibfnamefont {P.}~\bibnamefont {{Figueroa-Romero}}}, \bibinfo {author} {\bibfnamefont {F.~A.}\ \bibnamefont {Pollock}}, \bibinfo {author} {\bibfnamefont {P.}~\bibnamefont {Strasberg}},\ and\ \bibinfo {author} {\bibfnamefont {K.}~\bibnamefont {Modi}},\ }\bibfield  {title} {\bibinfo {title} {Relaxation of {{Multitime Statistics}} in {{Quantum Systems}}},\ }\href {https://doi.org/10.22331/q-2023-06-01-1027} {\bibfield  {journal} {\bibinfo  {journal} {Quantum}\ }\textbf {\bibinfo {volume} {7}},\ \bibinfo {pages} {1027} (\bibinfo {year} {2023}{\natexlab{a}})},\ \Eprint {https://arxiv.org/abs/2108.07420} {arXiv:2108.07420 [quant-ph]} \BibitemShut {NoStop}%
\bibitem [{\citenamefont {Dowling}\ and\ \citenamefont {Modi}(2024)}]{dowling-2024-OperationalMetric}%
  \BibitemOpen
  \bibfield  {author} {\bibinfo {author} {\bibfnamefont {N.}~\bibnamefont {Dowling}}\ and\ \bibinfo {author} {\bibfnamefont {K.}~\bibnamefont {Modi}},\ }\bibfield  {title} {\bibinfo {title} {Operational {{Metric}} for {{Quantum Chaos}} and the {{Corresponding Spatiotemporal-Entanglement Structure}}},\ }\href {https://doi.org/10.1103/PRXQuantum.5.010314} {\bibfield  {journal} {\bibinfo  {journal} {PRX Quantum}\ }\textbf {\bibinfo {volume} {5}},\ \bibinfo {pages} {010314} (\bibinfo {year} {2024})}\BibitemShut {NoStop}%
\bibitem [{\citenamefont {Kaneko}\ \emph {et~al.}(2020)\citenamefont {Kaneko}, \citenamefont {Iyoda},\ and\ \citenamefont {Sagawa}}]{kaneko-2020-DeepTherm}%
  \BibitemOpen
  \bibfield  {author} {\bibinfo {author} {\bibfnamefont {K.}~\bibnamefont {Kaneko}}, \bibinfo {author} {\bibfnamefont {E.}~\bibnamefont {Iyoda}},\ and\ \bibinfo {author} {\bibfnamefont {T.}~\bibnamefont {Sagawa}},\ }\bibfield  {title} {\bibinfo {title} {Characterizing complexity of many-body quantum dynamics by higher-order eigenstate thermalization},\ }\href {https://doi.org/10.1103/PhysRevA.101.042126} {\bibfield  {journal} {\bibinfo  {journal} {Phys. Rev. A}\ }\textbf {\bibinfo {volume} {101}},\ \bibinfo {pages} {042126} (\bibinfo {year} {2020})}\BibitemShut {NoStop}%
\bibitem [{\citenamefont {Ho}\ and\ \citenamefont {Choi}(2022)}]{ho-2022-DeepTherm}%
  \BibitemOpen
  \bibfield  {author} {\bibinfo {author} {\bibfnamefont {W.~W.}\ \bibnamefont {Ho}}\ and\ \bibinfo {author} {\bibfnamefont {S.}~\bibnamefont {Choi}},\ }\bibfield  {title} {\bibinfo {title} {Exact emergent quantum state designs from quantum chaotic dynamics},\ }\href {https://doi.org/10.1103/PhysRevLett.128.060601} {\bibfield  {journal} {\bibinfo  {journal} {Phys. Rev. Lett.}\ }\textbf {\bibinfo {volume} {128}},\ \bibinfo {pages} {060601} (\bibinfo {year} {2022})}\BibitemShut {NoStop}%
\bibitem [{\citenamefont {Ippoliti}\ and\ \citenamefont {Ho}(2022)}]{ippoliti-2022-DeepTherm}%
  \BibitemOpen
  \bibfield  {author} {\bibinfo {author} {\bibfnamefont {M.}~\bibnamefont {Ippoliti}}\ and\ \bibinfo {author} {\bibfnamefont {W.~W.}\ \bibnamefont {Ho}},\ }\bibfield  {title} {\bibinfo {title} {Solvable model of deep thermalization with distinct design times},\ }\href {https://doi.org/10.22331/q-2022-12-29-886} {\bibfield  {journal} {\bibinfo  {journal} {{Quantum}}\ }\textbf {\bibinfo {volume} {6}},\ \bibinfo {pages} {886} (\bibinfo {year} {2022})}\BibitemShut {NoStop}%
\bibitem [{\citenamefont {Cotler}\ \emph {et~al.}(2023)\citenamefont {Cotler}, \citenamefont {Mark}, \citenamefont {Huang}, \citenamefont {Hern\'andez}, \citenamefont {Choi}, \citenamefont {Shaw}, \citenamefont {Endres},\ and\ \citenamefont {Choi}}]{cotler-2023-DeepTherm}%
  \BibitemOpen
  \bibfield  {author} {\bibinfo {author} {\bibfnamefont {J.~S.}\ \bibnamefont {Cotler}}, \bibinfo {author} {\bibfnamefont {D.~K.}\ \bibnamefont {Mark}}, \bibinfo {author} {\bibfnamefont {H.-Y.}\ \bibnamefont {Huang}}, \bibinfo {author} {\bibfnamefont {F.}~\bibnamefont {Hern\'andez}}, \bibinfo {author} {\bibfnamefont {J.}~\bibnamefont {Choi}}, \bibinfo {author} {\bibfnamefont {A.~L.}\ \bibnamefont {Shaw}}, \bibinfo {author} {\bibfnamefont {M.}~\bibnamefont {Endres}},\ and\ \bibinfo {author} {\bibfnamefont {S.}~\bibnamefont {Choi}},\ }\bibfield  {title} {\bibinfo {title} {Emergent quantum state designs from individual many-body wave functions},\ }\href {https://doi.org/10.1103/PRXQuantum.4.010311} {\bibfield  {journal} {\bibinfo  {journal} {PRX Quantum}\ }\textbf {\bibinfo {volume} {4}},\ \bibinfo {pages} {010311} (\bibinfo {year} {2023})}\BibitemShut {NoStop}%
\bibitem [{\citenamefont {Ippoliti}\ and\ \citenamefont {Ho}(2023)}]{ippoliti-2023-DeepTherm}%
  \BibitemOpen
  \bibfield  {author} {\bibinfo {author} {\bibfnamefont {M.}~\bibnamefont {Ippoliti}}\ and\ \bibinfo {author} {\bibfnamefont {W.~W.}\ \bibnamefont {Ho}},\ }\bibfield  {title} {\bibinfo {title} {Dynamical purification and the emergence of quantum state designs from the projected ensemble},\ }\href {https://doi.org/10.1103/PRXQuantum.4.030322} {\bibfield  {journal} {\bibinfo  {journal} {PRX Quantum}\ }\textbf {\bibinfo {volume} {4}},\ \bibinfo {pages} {030322} (\bibinfo {year} {2023})}\BibitemShut {NoStop}%
\bibitem [{\citenamefont {Bhore}\ \emph {et~al.}(2023)\citenamefont {Bhore}, \citenamefont {Desaules},\ and\ \citenamefont {Papi\ifmmode~\acute{c}\else \'{c}\fi{}}}]{bhore-2023-DeepTherm}%
  \BibitemOpen
  \bibfield  {author} {\bibinfo {author} {\bibfnamefont {T.}~\bibnamefont {Bhore}}, \bibinfo {author} {\bibfnamefont {J.-Y.}\ \bibnamefont {Desaules}},\ and\ \bibinfo {author} {\bibfnamefont {Z.}~\bibnamefont {Papi\ifmmode~\acute{c}\else \'{c}\fi{}}},\ }\bibfield  {title} {\bibinfo {title} {Deep thermalization in constrained quantum systems},\ }\href {https://doi.org/10.1103/PhysRevB.108.104317} {\bibfield  {journal} {\bibinfo  {journal} {Phys. Rev. B}\ }\textbf {\bibinfo {volume} {108}},\ \bibinfo {pages} {104317} (\bibinfo {year} {2023})}\BibitemShut {NoStop}%
\bibitem [{\citenamefont {Mark}\ \emph {et~al.}(2024)\citenamefont {Mark}, \citenamefont {Surace}, \citenamefont {Elben}, \citenamefont {Shaw}, \citenamefont {Choi}, \citenamefont {Refael}, \citenamefont {Endres},\ and\ \citenamefont {Choi}}]{daniel-2024-DeepTherm}%
  \BibitemOpen
  \bibfield  {author} {\bibinfo {author} {\bibfnamefont {D.~K.}\ \bibnamefont {Mark}}, \bibinfo {author} {\bibfnamefont {F.}~\bibnamefont {Surace}}, \bibinfo {author} {\bibfnamefont {A.}~\bibnamefont {Elben}}, \bibinfo {author} {\bibfnamefont {A.~L.}\ \bibnamefont {Shaw}}, \bibinfo {author} {\bibfnamefont {J.}~\bibnamefont {Choi}}, \bibinfo {author} {\bibfnamefont {G.}~\bibnamefont {Refael}}, \bibinfo {author} {\bibfnamefont {M.}~\bibnamefont {Endres}},\ and\ \bibinfo {author} {\bibfnamefont {S.}~\bibnamefont {Choi}},\ }\bibfield  {title} {\bibinfo {title} {Maximum entropy principle in deep thermalization and in hilbert-space ergodicity},\ }\href {https://doi.org/10.1103/PhysRevX.14.041051} {\bibfield  {journal} {\bibinfo  {journal} {Phys. Rev. X}\ }\textbf {\bibinfo {volume} {14}},\ \bibinfo {pages} {041051} (\bibinfo {year} {2024})}\BibitemShut {NoStop}%
\bibitem [{\citenamefont {Srinivas}(1978)}]{srinivas-1978-QuantumGeneralization}%
  \BibitemOpen
  \bibfield  {author} {\bibinfo {author} {\bibfnamefont {M.~D.}\ \bibnamefont {Srinivas}},\ }\bibfield  {title} {\bibinfo {title} {Quantum generalization of {{Kolmogorov}} entropy},\ }\href {https://doi.org/10.1063/1.523916} {\bibfield  {journal} {\bibinfo  {journal} {Journal of Mathematical Physics}\ }\textbf {\bibinfo {volume} {19}},\ \bibinfo {pages} {1952} (\bibinfo {year} {1978})}\BibitemShut {NoStop}%
\bibitem [{\citenamefont {Pechukas}(1982)}]{pechukas-1982-KolmogorovEntropy}%
  \BibitemOpen
  \bibfield  {author} {\bibinfo {author} {\bibfnamefont {P.}~\bibnamefont {Pechukas}},\ }\bibfield  {title} {\bibinfo {title} {Kolmogorov entropy and "quantum chaos"},\ }\href {https://doi.org/10.1021/j100209a019} {\bibfield  {journal} {\bibinfo  {journal} {The Journal of Physical Chemistry}\ }\textbf {\bibinfo {volume} {86}},\ \bibinfo {pages} {2239} (\bibinfo {year} {1982})}\BibitemShut {NoStop}%
\bibitem [{\citenamefont {Beck}\ and\ \citenamefont {Graudenz}(1992)}]{beck-1992-SymbolicDynamics}%
  \BibitemOpen
  \bibfield  {author} {\bibinfo {author} {\bibfnamefont {C.}~\bibnamefont {Beck}}\ and\ \bibinfo {author} {\bibfnamefont {D.}~\bibnamefont {Graudenz}},\ }\bibfield  {title} {\bibinfo {title} {Symbolic dynamics of successive quantum-mechanical measurements},\ }\href {https://doi.org/10.1103/PhysRevA.46.6265} {\bibfield  {journal} {\bibinfo  {journal} {Physical Review A}\ }\textbf {\bibinfo {volume} {46}},\ \bibinfo {pages} {6265} (\bibinfo {year} {1992})}\BibitemShut {NoStop}%
\bibitem [{\citenamefont {Koll{\'a}r}\ and\ \citenamefont {Koniorczyk}(2014)}]{kollar-2014-EntropyRate}%
  \BibitemOpen
  \bibfield  {author} {\bibinfo {author} {\bibfnamefont {B.}~\bibnamefont {Koll{\'a}r}}\ and\ \bibinfo {author} {\bibfnamefont {M.}~\bibnamefont {Koniorczyk}},\ }\bibfield  {title} {\bibinfo {title} {Entropy rate of message sources driven by quantum walks},\ }\href {https://doi.org/10.1103/PhysRevA.89.022338} {\bibfield  {journal} {\bibinfo  {journal} {Physical Review A}\ }\textbf {\bibinfo {volume} {89}},\ \bibinfo {pages} {022338} (\bibinfo {year} {2014})},\ \Eprint {https://arxiv.org/abs/1402.6731} {arXiv:1402.6731 [quant-ph]} \BibitemShut {NoStop}%
\bibitem [{\citenamefont {Pandey}\ \emph {et~al.}(1993)\citenamefont {Pandey}, \citenamefont {Ramaswamy},\ and\ \citenamefont {Shukla}}]{Pandey-1993-symmetry-breaking}%
  \BibitemOpen
  \bibfield  {author} {\bibinfo {author} {\bibfnamefont {A.}~\bibnamefont {Pandey}}, \bibinfo {author} {\bibfnamefont {R.}~\bibnamefont {Ramaswamy}},\ and\ \bibinfo {author} {\bibfnamefont {P.}~\bibnamefont {Shukla}},\ }\bibfield  {title} {\bibinfo {title} {Symmetry breaking in quantum chaotic systems},\ }\href {https://doi.org/10.1007/BF02847320} {\bibfield  {journal} {\bibinfo  {journal} {Pramana}\ }\textbf {\bibinfo {volume} {41}},\ \bibinfo {pages} {L75} (\bibinfo {year} {1993})}\BibitemShut {NoStop}%
\bibitem [{\citenamefont {de~la Cruz}\ \emph {et~al.}(2020)\citenamefont {de~la Cruz}, \citenamefont {Lerma-Hern\'andez},\ and\ \citenamefont {Hirsch}}]{Javier-2020-symmetry-chaos}%
  \BibitemOpen
  \bibfield  {author} {\bibinfo {author} {\bibfnamefont {J.}~\bibnamefont {de~la Cruz}}, \bibinfo {author} {\bibfnamefont {S.}~\bibnamefont {Lerma-Hern\'andez}},\ and\ \bibinfo {author} {\bibfnamefont {J.~G.}\ \bibnamefont {Hirsch}},\ }\bibfield  {title} {\bibinfo {title} {Quantum chaos in a system with high degree of symmetries},\ }\href {https://doi.org/10.1103/PhysRevE.102.032208} {\bibfield  {journal} {\bibinfo  {journal} {Phys. Rev. E}\ }\textbf {\bibinfo {volume} {102}},\ \bibinfo {pages} {032208} (\bibinfo {year} {2020})}\BibitemShut {NoStop}%
\bibitem [{\citenamefont {Murthy}\ \emph {et~al.}(2023)\citenamefont {Murthy}, \citenamefont {Babakhani}, \citenamefont {Iniguez}, \citenamefont {Srednicki},\ and\ \citenamefont {Yunger~Halpern}}]{murthy-2023-NonAbelianEigenstate}%
  \BibitemOpen
  \bibfield  {author} {\bibinfo {author} {\bibfnamefont {C.}~\bibnamefont {Murthy}}, \bibinfo {author} {\bibfnamefont {A.}~\bibnamefont {Babakhani}}, \bibinfo {author} {\bibfnamefont {F.}~\bibnamefont {Iniguez}}, \bibinfo {author} {\bibfnamefont {M.}~\bibnamefont {Srednicki}},\ and\ \bibinfo {author} {\bibfnamefont {N.}~\bibnamefont {Yunger~Halpern}},\ }\bibfield  {title} {\bibinfo {title} {Non-{{Abelian Eigenstate Thermalization Hypothesis}}},\ }\href {https://doi.org/10.1103/PhysRevLett.130.140402} {\bibfield  {journal} {\bibinfo  {journal} {Physical Review Letters}\ }\textbf {\bibinfo {volume} {130}},\ \bibinfo {pages} {140402} (\bibinfo {year} {2023})}\BibitemShut {NoStop}%
\bibitem [{\citenamefont {Chang}\ \emph {et~al.}(2024)\citenamefont {Chang}, \citenamefont {Shrotriya}, \citenamefont {Ho},\ and\ \citenamefont {Ippoliti}}]{chang-2024-DeepThermCharges}%
  \BibitemOpen
  \bibfield  {author} {\bibinfo {author} {\bibfnamefont {R.-A.}\ \bibnamefont {Chang}}, \bibinfo {author} {\bibfnamefont {H.}~\bibnamefont {Shrotriya}}, \bibinfo {author} {\bibfnamefont {W.~W.}\ \bibnamefont {Ho}},\ and\ \bibinfo {author} {\bibfnamefont {M.}~\bibnamefont {Ippoliti}},\ }\href {https://arxiv.org/abs/2408.15325} {\bibinfo {title} {Deep thermalization under charge-conserving quantum dynamics}} (\bibinfo {year} {2024}),\ \Eprint {https://arxiv.org/abs/2408.15325} {arXiv:2408.15325 [quant-ph]} \BibitemShut {NoStop}%
\bibitem [{\citenamefont {Chen}\ and\ \citenamefont {Fang}(2024)}]{Chen:2024pxm}%
  \BibitemOpen
  \bibfield  {author} {\bibinfo {author} {\bibfnamefont {F.}~\bibnamefont {Chen}}\ and\ \bibinfo {author} {\bibfnamefont {P.}~\bibnamefont {Fang}},\ }\href@noop {} {\bibinfo {title} {{System Symmetry and the Classification of Out-of-Time-Ordered Correlator Dynamics in Quantum Chaos}}} (\bibinfo {year} {2024}),\ \Eprint {https://arxiv.org/abs/2410.04712} {arXiv:2410.04712 [cond-mat.stat-mech]} \BibitemShut {NoStop}%
\bibitem [{\citenamefont {Sharma}\ \emph {et~al.}(2024)\citenamefont {Sharma}, \citenamefont {Sahu},\ and\ \citenamefont {Mukerjee}}]{Sharma:2024fqc}%
  \BibitemOpen
  \bibfield  {author} {\bibinfo {author} {\bibfnamefont {K.}~\bibnamefont {Sharma}}, \bibinfo {author} {\bibfnamefont {H.}~\bibnamefont {Sahu}},\ and\ \bibinfo {author} {\bibfnamefont {S.}~\bibnamefont {Mukerjee}},\ }\href@noop {} {\bibinfo {title} {{Quantum chaos in PT symmetric quantum systems}}} (\bibinfo {year} {2024}),\ \Eprint {https://arxiv.org/abs/2401.07215} {arXiv:2401.07215 [quant-ph]} \BibitemShut {NoStop}%
\bibitem [{\citenamefont {Bracken}(2020)}]{Bracken20}%
  \BibitemOpen
  \bibfield  {author} {\bibinfo {author} {\bibfnamefont {P.}~\bibnamefont {Bracken}},\ }\bibfield  {title} {\bibinfo {title} {Introductory chapter: Dynamical symmetries and quantum chaos},\ }in\ \href {https://doi.org/10.5772/intechopen.88551} {\emph {\bibinfo {booktitle} {Research Advances in Chaos Theory}}},\ \bibinfo {editor} {edited by\ \bibinfo {editor} {\bibfnamefont {P.}~\bibnamefont {Bracken}}}\ (\bibinfo  {publisher} {IntechOpen},\ \bibinfo {address} {Rijeka},\ \bibinfo {year} {2020})\ Chap.~\bibinfo {chapter} {1}\BibitemShut {NoStop}%
\bibitem [{\citenamefont {{Kudler-Flam}}\ \emph {et~al.}(2022)\citenamefont {{Kudler-Flam}}, \citenamefont {Sohal},\ and\ \citenamefont {Nie}}]{kudler-flam-2022-InformationScrambling}%
  \BibitemOpen
  \bibfield  {author} {\bibinfo {author} {\bibfnamefont {J.}~\bibnamefont {{Kudler-Flam}}}, \bibinfo {author} {\bibfnamefont {R.}~\bibnamefont {Sohal}},\ and\ \bibinfo {author} {\bibfnamefont {L.}~\bibnamefont {Nie}},\ }\bibfield  {title} {\bibinfo {title} {Information {{Scrambling}} with {{Conservation Laws}}},\ }\href {https://doi.org/10.21468/SciPostPhys.12.4.117} {\bibfield  {journal} {\bibinfo  {journal} {SciPost Physics}\ }\textbf {\bibinfo {volume} {12}},\ \bibinfo {pages} {117} (\bibinfo {year} {2022})}\BibitemShut {NoStop}%
\bibitem [{\citenamefont {Balasubramanian}\ \emph {et~al.}(2024)\citenamefont {Balasubramanian}, \citenamefont {Das}, \citenamefont {Erdmenger},\ and\ \citenamefont {Xian}}]{Balasubramanian:2024ghv}%
  \BibitemOpen
  \bibfield  {author} {\bibinfo {author} {\bibfnamefont {V.}~\bibnamefont {Balasubramanian}}, \bibinfo {author} {\bibfnamefont {R.~N.}\ \bibnamefont {Das}}, \bibinfo {author} {\bibfnamefont {J.}~\bibnamefont {Erdmenger}},\ and\ \bibinfo {author} {\bibfnamefont {Z.-Y.}\ \bibnamefont {Xian}},\ }\href@noop {} {\bibinfo {title} {{Chaos and integrability in triangular billiards}}} (\bibinfo {year} {2024}),\ \Eprint {https://arxiv.org/abs/2407.11114} {arXiv:2407.11114 [hep-th]} \BibitemShut {NoStop}%
\bibitem [{\citenamefont {Abreu}\ and\ \citenamefont {Vallejos}(2006)}]{abreu-2006-EntanglingPower}%
  \BibitemOpen
  \bibfield  {author} {\bibinfo {author} {\bibfnamefont {R.~F.}\ \bibnamefont {Abreu}}\ and\ \bibinfo {author} {\bibfnamefont {R.~O.}\ \bibnamefont {Vallejos}},\ }\bibfield  {title} {\bibinfo {title} {Entangling power of the baker's map: {{Role}} of symmetries},\ }\href {https://doi.org/10.1103/PhysRevA.73.052327} {\bibfield  {journal} {\bibinfo  {journal} {Physical Review A}\ }\textbf {\bibinfo {volume} {73}},\ \bibinfo {pages} {052327} (\bibinfo {year} {2006})}\BibitemShut {NoStop}%
\bibitem [{\citenamefont {Santhanam}\ \emph {et~al.}(2008)\citenamefont {Santhanam}, \citenamefont {Sheorey},\ and\ \citenamefont {Lakshminarayan}}]{santhanam-2008-EffectClassicalBifurcations}%
  \BibitemOpen
  \bibfield  {author} {\bibinfo {author} {\bibfnamefont {M.~S.}\ \bibnamefont {Santhanam}}, \bibinfo {author} {\bibfnamefont {V.~B.}\ \bibnamefont {Sheorey}},\ and\ \bibinfo {author} {\bibfnamefont {A.}~\bibnamefont {Lakshminarayan}},\ }\bibfield  {title} {\bibinfo {title} {Effect of classical bifurcations on the quantum entanglement of two coupled quartic oscillators},\ }\href {https://doi.org/10.1103/PhysRevE.77.026213} {\bibfield  {journal} {\bibinfo  {journal} {Phys. Rev. E}\ }\textbf {\bibinfo {volume} {77}},\ \bibinfo {pages} {026213} (\bibinfo {year} {2008})}\BibitemShut {NoStop}%
\bibitem [{\citenamefont {Goldstein}\ and\ \citenamefont {Sela}(2018)}]{Goldstein-2018-SymmetryResolvedEntanglement}%
  \BibitemOpen
  \bibfield  {author} {\bibinfo {author} {\bibfnamefont {M.}~\bibnamefont {Goldstein}}\ and\ \bibinfo {author} {\bibfnamefont {E.}~\bibnamefont {Sela}},\ }\bibfield  {title} {\bibinfo {title} {Symmetry-resolved entanglement in many-body systems},\ }\href {https://doi.org/10.1103/PhysRevLett.120.200602} {\bibfield  {journal} {\bibinfo  {journal} {Phys. Rev. Lett.}\ }\textbf {\bibinfo {volume} {120}},\ \bibinfo {pages} {200602} (\bibinfo {year} {2018})}\BibitemShut {NoStop}%
\bibitem [{\citenamefont {Bianchi}\ \emph {et~al.}(2024)\citenamefont {Bianchi}, \citenamefont {Dona},\ and\ \citenamefont {Kumar}}]{bianchi-2024-NonAbelianSymmetryResolved}%
  \BibitemOpen
  \bibfield  {author} {\bibinfo {author} {\bibfnamefont {E.}~\bibnamefont {Bianchi}}, \bibinfo {author} {\bibfnamefont {P.}~\bibnamefont {Dona}},\ and\ \bibinfo {author} {\bibfnamefont {R.}~\bibnamefont {Kumar}},\ }\bibfield  {title} {\bibinfo {title} {Non-{{Abelian}} symmetry-resolved entanglement entropy},\ }\href {https://doi.org/10.21468/SciPostPhys.17.5.127} {\bibfield  {journal} {\bibinfo  {journal} {SciPost Physics}\ }\textbf {\bibinfo {volume} {17}},\ \bibinfo {pages} {127} (\bibinfo {year} {2024})}\BibitemShut {NoStop}%
\bibitem [{\citenamefont {Moharramipour}\ \emph {et~al.}(2024)\citenamefont {Moharramipour}, \citenamefont {Lessa}, \citenamefont {Wang}, \citenamefont {Hsieh},\ and\ \citenamefont {Sahu}}]{moharramipour-2024-SymmetryEnforcedEntanglement}%
  \BibitemOpen
  \bibfield  {author} {\bibinfo {author} {\bibfnamefont {A.}~\bibnamefont {Moharramipour}}, \bibinfo {author} {\bibfnamefont {L.~A.}\ \bibnamefont {Lessa}}, \bibinfo {author} {\bibfnamefont {C.}~\bibnamefont {Wang}}, \bibinfo {author} {\bibfnamefont {T.~H.}\ \bibnamefont {Hsieh}},\ and\ \bibinfo {author} {\bibfnamefont {S.}~\bibnamefont {Sahu}},\ }\bibfield  {title} {\bibinfo {title} {Symmetry-{{Enforced Entanglement}} in {{Maximally Mixed States}}},\ }\href {https://doi.org/10.1103/PRXQuantum.5.040336} {\bibfield  {journal} {\bibinfo  {journal} {PRX Quantum}\ }\textbf {\bibinfo {volume} {5}},\ \bibinfo {pages} {040336} (\bibinfo {year} {2024})}\BibitemShut {NoStop}%
\bibitem [{\citenamefont {Li}\ \emph {et~al.}(2025)\citenamefont {Li}, \citenamefont {Pollmann}, \citenamefont {Read},\ and\ \citenamefont {Sala}}]{li-2025-HighlyEntangled}%
  \BibitemOpen
  \bibfield  {author} {\bibinfo {author} {\bibfnamefont {Y.}~\bibnamefont {Li}}, \bibinfo {author} {\bibfnamefont {F.}~\bibnamefont {Pollmann}}, \bibinfo {author} {\bibfnamefont {N.}~\bibnamefont {Read}},\ and\ \bibinfo {author} {\bibfnamefont {P.}~\bibnamefont {Sala}},\ }\bibfield  {title} {\bibinfo {title} {Highly {{Entangled Stationary States}} from {{Strong Symmetries}}},\ }\href {https://doi.org/10.1103/PhysRevX.15.011068} {\bibfield  {journal} {\bibinfo  {journal} {Physical Review X}\ }\textbf {\bibinfo {volume} {15}},\ \bibinfo {pages} {011068} (\bibinfo {year} {2025})}\BibitemShut {NoStop}%
\bibitem [{\citenamefont {Chen}\ and\ \citenamefont {Zhou}(2018)}]{chen-2018-OperatorScrambling}%
  \BibitemOpen
  \bibfield  {author} {\bibinfo {author} {\bibfnamefont {X.}~\bibnamefont {Chen}}\ and\ \bibinfo {author} {\bibfnamefont {T.}~\bibnamefont {Zhou}},\ }\href {http://arxiv.org/abs/1804.08655} {\bibinfo {title} {Operator scrambling and quantum chaos}} (\bibinfo {year} {2018}),\ \Eprint {https://arxiv.org/abs/1804.08655} {arXiv:1804.08655 [cond-mat, physics:hep-th, physics:quant-ph]} \BibitemShut {NoStop}%
\bibitem [{\citenamefont {Moudgalya}\ \emph {et~al.}(2019)\citenamefont {Moudgalya}, \citenamefont {Devakul}, \citenamefont {Von~Keyserlingk},\ and\ \citenamefont {Sondhi}}]{moudgalya-2019-OperatorSpreading}%
  \BibitemOpen
  \bibfield  {author} {\bibinfo {author} {\bibfnamefont {S.}~\bibnamefont {Moudgalya}}, \bibinfo {author} {\bibfnamefont {T.}~\bibnamefont {Devakul}}, \bibinfo {author} {\bibfnamefont {C.~W.}\ \bibnamefont {Von~Keyserlingk}},\ and\ \bibinfo {author} {\bibfnamefont {S.~L.}\ \bibnamefont {Sondhi}},\ }\bibfield  {title} {\bibinfo {title} {Operator spreading in quantum maps},\ }\href {https://doi.org/10.1103/PhysRevB.99.094312} {\bibfield  {journal} {\bibinfo  {journal} {Physical Review B}\ }\textbf {\bibinfo {volume} {99}},\ \bibinfo {pages} {094312} (\bibinfo {year} {2019})}\BibitemShut {NoStop}%
\bibitem [{\citenamefont {Nie}(2021)}]{nie-2021-OperatorGrowth}%
  \BibitemOpen
  \bibfield  {author} {\bibinfo {author} {\bibfnamefont {L.}~\bibnamefont {Nie}},\ }\href {http://arxiv.org/abs/2111.08729} {\bibinfo {title} {Operator {{Growth}} and {{Symmetry-Resolved Coefficient Entropy}} in {{Quantum Maps}}}} (\bibinfo {year} {2021}),\ \Eprint {https://arxiv.org/abs/2111.08729} {arXiv:2111.08729 [cond-mat, physics:quant-ph]} \BibitemShut {NoStop}%
\bibitem [{\citenamefont {Lindblad}(1979)}]{lindblad-1979-NonMarkovianQuantum}%
  \BibitemOpen
  \bibfield  {author} {\bibinfo {author} {\bibfnamefont {G.}~\bibnamefont {Lindblad}},\ }\bibfield  {title} {\bibinfo {title} {Non-{{Markovian}} quantum stochastic processes and their entropy},\ }\href {https://doi.org/10.1007/BF01197883} {\bibfield  {journal} {\bibinfo  {journal} {Communications in Mathematical Physics}\ }\textbf {\bibinfo {volume} {65}},\ \bibinfo {pages} {281} (\bibinfo {year} {1979})}\BibitemShut {NoStop}%
\bibitem [{\citenamefont {Lindblad}(1991)}]{lindblad-1991-QuantumEntropy}%
  \BibitemOpen
  \bibfield  {author} {\bibinfo {author} {\bibfnamefont {G.}~\bibnamefont {Lindblad}},\ }\bibfield  {title} {\bibinfo {title} {Quantum entropy and quantum measurements},\ }in\ \href {https://doi.org/10.1007/3-540-53862-3_168} {\emph {\bibinfo {booktitle} {Quantum {{Aspects}} of {{Optical Communications}}}}},\ \bibinfo {editor} {edited by\ \bibinfo {editor} {\bibfnamefont {C.}~\bibnamefont {Bendjaballah}}, \bibinfo {editor} {\bibfnamefont {O.}~\bibnamefont {Hirota}},\ and\ \bibinfo {editor} {\bibfnamefont {S.}~\bibnamefont {Reynaud}}}\ (\bibinfo  {publisher} {Springer},\ \bibinfo {address} {Berlin, Heidelberg},\ \bibinfo {year} {1991})\ pp.\ \bibinfo {pages} {79--80}\BibitemShut {NoStop}%
\bibitem [{\citenamefont {Nielsen}\ and\ \citenamefont {Chuang}(2010)}]{nielsen-2010-QuantumComputation}%
  \BibitemOpen
  \bibfield  {author} {\bibinfo {author} {\bibfnamefont {M.~A.}\ \bibnamefont {Nielsen}}\ and\ \bibinfo {author} {\bibfnamefont {I.~L.}\ \bibnamefont {Chuang}},\ }\href@noop {} {\emph {\bibinfo {title} {Quantum Computation and Quantum Information: 10th Anniversary Edition}}}\ (\bibinfo  {publisher} {Cambridge University Press},\ \bibinfo {year} {2010})\BibitemShut {NoStop}%
\bibitem [{\citenamefont {Wilde}(2021)}]{wilde-2021-ClassicalQuantum}%
  \BibitemOpen
  \bibfield  {author} {\bibinfo {author} {\bibfnamefont {M.~M.}\ \bibnamefont {Wilde}},\ }\href {https://doi.org/10.1017/9781316809976.001} {\bibinfo {title} {From {{Classical}} to {{Quantum Shannon Theory}}}} (\bibinfo {year} {2021}),\ \Eprint {https://arxiv.org/abs/1106.1445} {arXiv:1106.1445 [quant-ph]} \BibitemShut {NoStop}%
\bibitem [{\citenamefont {Schumacher}(1996)}]{schumacher-1996-SendingEntanglement}%
  \BibitemOpen
  \bibfield  {author} {\bibinfo {author} {\bibfnamefont {B.}~\bibnamefont {Schumacher}},\ }\bibfield  {title} {\bibinfo {title} {Sending entanglement through noisy quantum channels},\ }\href {https://doi.org/10.1103/PhysRevA.54.2614} {\bibfield  {journal} {\bibinfo  {journal} {Physical Review A}\ }\textbf {\bibinfo {volume} {54}},\ \bibinfo {pages} {2614} (\bibinfo {year} {1996})}\BibitemShut {NoStop}%
\bibitem [{\citenamefont {Bridgeman}\ and\ \citenamefont {Chubb}(2017)}]{bridgeman-2017-HandwavingInterpretive}%
  \BibitemOpen
  \bibfield  {author} {\bibinfo {author} {\bibfnamefont {J.~C.}\ \bibnamefont {Bridgeman}}\ and\ \bibinfo {author} {\bibfnamefont {C.~T.}\ \bibnamefont {Chubb}},\ }\bibfield  {title} {\bibinfo {title} {Hand-waving and interpretive dance: An introductory course on tensor networks},\ }\href {https://doi.org/10.1088/1751-8121/aa6dc3} {\bibfield  {journal} {\bibinfo  {journal} {Journal of Physics A: Mathematical and Theoretical}\ }\textbf {\bibinfo {volume} {50}},\ \bibinfo {pages} {223001} (\bibinfo {year} {2017})}\BibitemShut {NoStop}%
\bibitem [{\citenamefont {Biamonte}(2020)}]{biamonte-2020-LecturesQuantum}%
  \BibitemOpen
  \bibfield  {author} {\bibinfo {author} {\bibfnamefont {J.}~\bibnamefont {Biamonte}},\ }\href {https://doi.org/10.48550/arXiv.1912.10049} {\bibinfo {title} {Lectures on {{Quantum Tensor Networks}}}} (\bibinfo {year} {2020}),\ \Eprint {https://arxiv.org/abs/1912.10049} {arXiv:1912.10049} \BibitemShut {NoStop}%
\bibitem [{\citenamefont {Alicki}\ \emph {et~al.}(1996{\natexlab{b}})\citenamefont {Alicki}, \citenamefont {Andries}, \citenamefont {Fannes},\ and\ \citenamefont {Tuyls}}]{alicki-1996-AlgebraicApproach}%
  \BibitemOpen
  \bibfield  {author} {\bibinfo {author} {\bibfnamefont {R.}~\bibnamefont {Alicki}}, \bibinfo {author} {\bibfnamefont {J.}~\bibnamefont {Andries}}, \bibinfo {author} {\bibfnamefont {M.}~\bibnamefont {Fannes}},\ and\ \bibinfo {author} {\bibfnamefont {P.}~\bibnamefont {Tuyls}},\ }\bibfield  {title} {\bibinfo {title} {An algebraic approach to the kolmogorov-sinai entropy},\ }\href {https://doi.org/10.1142/S0129055X96000068} {\bibfield  {journal} {\bibinfo  {journal} {Reviews in Mathematical Physics}\ }\textbf {\bibinfo {volume} {08}},\ \bibinfo {pages} {167} (\bibinfo {year} {1996}{\natexlab{b}})}\BibitemShut {NoStop}%
\bibitem [{\citenamefont {Alicki}(1998)}]{alicki-1998-QuantumMechanical}%
  \BibitemOpen
  \bibfield  {author} {\bibinfo {author} {\bibfnamefont {R.}~\bibnamefont {Alicki}},\ }\bibfield  {title} {\bibinfo {title} {Quantum {{Mechanical Tools}} in {{Applications}} to {{Classical Dynamical Systems}}},\ }\href {https://doi.org/10.1103/PhysRevLett.81.2040} {\bibfield  {journal} {\bibinfo  {journal} {Physical Review Letters}\ }\textbf {\bibinfo {volume} {81}},\ \bibinfo {pages} {2040} (\bibinfo {year} {1998})}\BibitemShut {NoStop}%
\bibitem [{\citenamefont {Benatti}\ \emph {et~al.}(2003)\citenamefont {Benatti}, \citenamefont {Cappellini}, \citenamefont {De~Cock}, \citenamefont {Fannes},\ and\ \citenamefont {Vanpeteghem}}]{benatti-2003-ClassicalLimit}%
  \BibitemOpen
  \bibfield  {author} {\bibinfo {author} {\bibfnamefont {F.}~\bibnamefont {Benatti}}, \bibinfo {author} {\bibfnamefont {V.}~\bibnamefont {Cappellini}}, \bibinfo {author} {\bibfnamefont {M.}~\bibnamefont {De~Cock}}, \bibinfo {author} {\bibfnamefont {M.}~\bibnamefont {Fannes}},\ and\ \bibinfo {author} {\bibfnamefont {D.}~\bibnamefont {Vanpeteghem}},\ }\bibfield  {title} {\bibinfo {title} {Classical {{Limit}} of {{Quantum Dynamical Entropies}}},\ }\href {https://doi.org/10.1142/S0129055X03001837} {\bibfield  {journal} {\bibinfo  {journal} {Reviews in Mathematical Physics}\ }\textbf {\bibinfo {volume} {15}},\ \bibinfo {pages} {847} (\bibinfo {year} {2003})},\ \Eprint {https://arxiv.org/abs/quant-ph/0308069} {arXiv:quant-ph/0308069} \BibitemShut {NoStop}%
\bibitem [{\citenamefont {Shepelyansky}(2020)}]{shepelyansky-2020-EhrenfestTime}%
  \BibitemOpen
  \bibfield  {author} {\bibinfo {author} {\bibfnamefont {D.}~\bibnamefont {Shepelyansky}},\ }\bibfield  {title} {\bibinfo {title} {Ehrenfest time and chaos},\ }\href {https://doi.org/10.4249/scholarpedia.55031} {\bibfield  {journal} {\bibinfo  {journal} {Scholarpedia}\ }\textbf {\bibinfo {volume} {15}},\ \bibinfo {pages} {55031} (\bibinfo {year} {2020})}\BibitemShut {NoStop}%
\bibitem [{\citenamefont {Schultz}(2025)}]{github-link}%
  \BibitemOpen
  \bibfield  {author} {\bibinfo {author} {\bibfnamefont {E.~D.}\ \bibnamefont {Schultz}},\ }\href@noop {} {\bibinfo {title} {Numerics for {A}licki-{F}annes-{L}indblad {E}ntropy}} (\bibinfo {year} {2025}),\ \bibinfo {note} {\url{https://github.com/ericdschultz/AFL_numerics}}\BibitemShut {NoStop}%
\bibitem [{\citenamefont {Arnol'd}(1968)}]{arnold-1968-ErgodicProblems}%
  \BibitemOpen
  \bibfield  {author} {\bibinfo {author} {\bibfnamefont {V.~I.}\ \bibnamefont {Arnol'd}},\ }\href@noop {} {\emph {\bibinfo {title} {Ergodic Problems of Classical Mechanics}}},\ The {{Mathematical}} Physics Monograph Series\ (\bibinfo  {publisher} {Benjamin},\ \bibinfo {address} {New York},\ \bibinfo {year} {1968})\BibitemShut {NoStop}%
\bibitem [{\citenamefont {Hannay}\ and\ \citenamefont {Berry}(1980)}]{hannay-1980-QuantizationLinear}%
  \BibitemOpen
  \bibfield  {author} {\bibinfo {author} {\bibfnamefont {J.}~\bibnamefont {Hannay}}\ and\ \bibinfo {author} {\bibfnamefont {M.}~\bibnamefont {Berry}},\ }\bibfield  {title} {\bibinfo {title} {Quantization of linear maps on a torus-fresnel diffraction by a periodic grating},\ }\href {https://doi.org/10.1016/0167-2789(80)90026-3} {\bibfield  {journal} {\bibinfo  {journal} {Physica D: Nonlinear Phenomena}\ }\textbf {\bibinfo {volume} {1}},\ \bibinfo {pages} {267} (\bibinfo {year} {1980})}\BibitemShut {NoStop}%
\bibitem [{\citenamefont {Esposti}(1993)}]{esposti-1993-QuantizationOrientation}%
  \BibitemOpen
  \bibfield  {author} {\bibinfo {author} {\bibfnamefont {M.~D.}\ \bibnamefont {Esposti}},\ }\bibfield  {title} {\bibinfo {title} {Quantization of the orientation preserving automorphisms of the torus},\ }\href {https://eudml.org/doc/76609} {\bibfield  {journal} {\bibinfo  {journal} {Annales de l'I.H.P. Physique th{\'e}orique}\ }\textbf {\bibinfo {volume} {58}},\ \bibinfo {pages} {323} (\bibinfo {year} {1993})}\BibitemShut {NoStop}%
\bibitem [{\citenamefont {de~Matos}\ and\ \citenamefont {de~Almeida}(1993)}]{matos-1993-QuantizationAnosov}%
  \BibitemOpen
  \bibfield  {author} {\bibinfo {author} {\bibfnamefont {M.~B.}\ \bibnamefont {de~Matos}}\ and\ \bibinfo {author} {\bibfnamefont {A.~M.~O.}\ \bibnamefont {de~Almeida}},\ }\bibfield  {title} {\bibinfo {title} {Quantization of {{Anosov Maps}}},\ }\href {https://doi.org/10.1006/aphy.1995.1003} {\bibfield  {journal} {\bibinfo  {journal} {Annals of Physics}\ }\textbf {\bibinfo {volume} {237}},\ \bibinfo {pages} {46} (\bibinfo {year} {1993})}\BibitemShut {NoStop}%
\bibitem [{\citenamefont {Kurlberg}\ and\ \citenamefont {Rudnick}(2000)}]{kurlberg-2000-HeckeTheory}%
  \BibitemOpen
  \bibfield  {author} {\bibinfo {author} {\bibfnamefont {P.}~\bibnamefont {Kurlberg}}\ and\ \bibinfo {author} {\bibfnamefont {Z.}~\bibnamefont {Rudnick}},\ }\bibfield  {title} {\bibinfo {title} {Hecke theory and equidistribution for the quantization of linear maps of the torus},\ }\href {https://doi.org/10.1215/S0012-7094-00-10314-6} {\bibfield  {journal} {\bibinfo  {journal} {Duke Mathematical Journal}\ }\textbf {\bibinfo {volume} {103}},\ \bibinfo {pages} {47} (\bibinfo {year} {2000})}\BibitemShut {NoStop}%
\bibitem [{\citenamefont {Keating}\ and\ \citenamefont {Mezzadri}(2000)}]{keating-2000-PseudosymmetriesAnosov}%
  \BibitemOpen
  \bibfield  {author} {\bibinfo {author} {\bibfnamefont {J.~P.}\ \bibnamefont {Keating}}\ and\ \bibinfo {author} {\bibfnamefont {F.}~\bibnamefont {Mezzadri}},\ }\bibfield  {title} {\bibinfo {title} {Pseudo-symmetries of {{Anosov}} maps and spectral statistics},\ }\href {https://doi.org/10.1088/0951-7715/13/3/313} {\bibfield  {journal} {\bibinfo  {journal} {Nonlinearity}\ }\textbf {\bibinfo {volume} {13}},\ \bibinfo {pages} {747} (\bibinfo {year} {2000})}\BibitemShut {NoStop}%
\bibitem [{\citenamefont {Esposti}\ and\ \citenamefont {Winn}(2005)}]{esposti-2005-QuantumPerturbed}%
  \BibitemOpen
  \bibfield  {author} {\bibinfo {author} {\bibfnamefont {M.~D.}\ \bibnamefont {Esposti}}\ and\ \bibinfo {author} {\bibfnamefont {B.}~\bibnamefont {Winn}},\ }\bibfield  {title} {\bibinfo {title} {The quantum perturbed cat map and symmetry},\ }\href {https://doi.org/10.1088/0305-4470/38/26/005} {\bibfield  {journal} {\bibinfo  {journal} {Journal of Physics A: Mathematical and General}\ }\textbf {\bibinfo {volume} {38}},\ \bibinfo {pages} {5895} (\bibinfo {year} {2005})}\BibitemShut {NoStop}%
\bibitem [{\citenamefont {B{\"a}cker}(2003)}]{backer-2003-NumericalAspects}%
  \BibitemOpen
  \bibfield  {author} {\bibinfo {author} {\bibfnamefont {A.}~\bibnamefont {B{\"a}cker}},\ }\bibfield  {title} {\bibinfo {title} {Numerical {{Aspects}} of {{Eigenvalue}} and {{Eigenfunction Computations}} for {{Chaotic Quantum Systems}}},\ }in\ \href {https://doi.org/10.1007/3-540-37045-5_4} {\emph {\bibinfo {booktitle} {The {{Mathematical Aspects}} of {{Quantum Maps}}}}},\ Vol.\ \bibinfo {volume} {618},\ \bibinfo {editor} {edited by\ \bibinfo {editor} {\bibfnamefont {M.~D.}\ \bibnamefont {Esposti}}\ and\ \bibinfo {editor} {\bibfnamefont {S.}~\bibnamefont {Graffi}}}\ (\bibinfo  {publisher} {Springer},\ \bibinfo {address} {Berlin, Heidelberg},\ \bibinfo {year} {2003})\ pp.\ \bibinfo {pages} {91--144}\BibitemShut {NoStop}%
\bibitem [{\citenamefont {{Garc{\'i}a-Mata}}\ \emph {et~al.}(2003)\citenamefont {{Garc{\'i}a-Mata}}, \citenamefont {Saraceno},\ and\ \citenamefont {Spina}}]{garcia-mata-2003-ClassicalDecays}%
  \BibitemOpen
  \bibfield  {author} {\bibinfo {author} {\bibfnamefont {I.}~\bibnamefont {{Garc{\'i}a-Mata}}}, \bibinfo {author} {\bibfnamefont {M.}~\bibnamefont {Saraceno}},\ and\ \bibinfo {author} {\bibfnamefont {M.~E.}\ \bibnamefont {Spina}},\ }\bibfield  {title} {\bibinfo {title} {Classical {{Decays}} in {{Decoherent Quantum Maps}}},\ }\href {https://doi.org/10.1103/PhysRevLett.91.064101} {\bibfield  {journal} {\bibinfo  {journal} {Physical Review Letters}\ }\textbf {\bibinfo {volume} {91}},\ \bibinfo {pages} {064101} (\bibinfo {year} {2003})}\BibitemShut {NoStop}%
\bibitem [{\citenamefont {{Garc{\'i}a-Mata}}\ and\ \citenamefont {Saraceno}(2004)}]{garcia-mata-2004-SpectralProperties}%
  \BibitemOpen
  \bibfield  {author} {\bibinfo {author} {\bibfnamefont {I.}~\bibnamefont {{Garc{\'i}a-Mata}}}\ and\ \bibinfo {author} {\bibfnamefont {M.}~\bibnamefont {Saraceno}},\ }\bibfield  {title} {\bibinfo {title} {Spectral properties and classical decays in quantum open systems},\ }\href {https://doi.org/10.1103/PhysRevE.69.056211} {\bibfield  {journal} {\bibinfo  {journal} {Physical Review E}\ }\textbf {\bibinfo {volume} {69}},\ \bibinfo {pages} {056211} (\bibinfo {year} {2004})}\BibitemShut {NoStop}%
\bibitem [{\citenamefont {Nielsen}(2002)}]{nielsen-2002-IntroductionMajorization}%
  \BibitemOpen
  \bibfield  {author} {\bibinfo {author} {\bibfnamefont {M.~A.}\ \bibnamefont {Nielsen}},\ }\href@noop {} {\emph {\bibinfo {title} {An Introduction to Majorization and Its Applications to Quantum Mechanics}}}\ (\bibinfo  {publisher} {University of Queensland},\ \bibinfo {year} {2002})\BibitemShut {NoStop}%
\bibitem [{\citenamefont {Buča}\ and\ \citenamefont {Prosen}(2012)}]{buca-2012-SymmetryReductions}%
  \BibitemOpen
  \bibfield  {author} {\bibinfo {author} {\bibfnamefont {B.}~\bibnamefont {Buča}}\ and\ \bibinfo {author} {\bibfnamefont {T.}~\bibnamefont {Prosen}},\ }\bibfield  {title} {\bibinfo {title} {A note on symmetry reductions of the lindblad equation: transport in constrained open spin chains},\ }\href {https://doi.org/10.1088/1367-2630/14/7/073007} {\bibfield  {journal} {\bibinfo  {journal} {New Journal of Physics}\ }\textbf {\bibinfo {volume} {14}},\ \bibinfo {pages} {073007} (\bibinfo {year} {2012})}\BibitemShut {NoStop}%
\bibitem [{\citenamefont {B{\'e}ny}\ and\ \citenamefont {Richter}(2020)}]{beny-2020-AlgebraicApproach}%
  \BibitemOpen
  \bibfield  {author} {\bibinfo {author} {\bibfnamefont {C.}~\bibnamefont {B{\'e}ny}}\ and\ \bibinfo {author} {\bibfnamefont {F.}~\bibnamefont {Richter}},\ }\href {https://doi.org/10.48550/arXiv.1505.03106} {\bibinfo {title} {Algebraic approach to quantum theory: A finite-dimensional guide}} (\bibinfo {year} {2020}),\ \Eprint {https://arxiv.org/abs/1505.03106} {arXiv:1505.03106 [quant-ph]} \BibitemShut {NoStop}%
\bibitem [{\citenamefont {Harlow}(2017)}]{harlow-2017-RyuTakayanagi}%
  \BibitemOpen
  \bibfield  {author} {\bibinfo {author} {\bibfnamefont {D.}~\bibnamefont {Harlow}},\ }\bibfield  {title} {\bibinfo {title} {The {{Ryu}}--{{Takayanagi Formula}} from {{Quantum Error Correction}}},\ }\href {https://doi.org/10.1007/s00220-017-2904-z} {\bibfield  {journal} {\bibinfo  {journal} {Communications in Mathematical Physics}\ }\textbf {\bibinfo {volume} {354}},\ \bibinfo {pages} {865} (\bibinfo {year} {2017})}\BibitemShut {NoStop}%
\bibitem [{\citenamefont {Moudgalya}\ and\ \citenamefont {Motrunich}(2022)}]{moudgalya-2022-HilbertSpace}%
  \BibitemOpen
  \bibfield  {author} {\bibinfo {author} {\bibfnamefont {S.}~\bibnamefont {Moudgalya}}\ and\ \bibinfo {author} {\bibfnamefont {O.~I.}\ \bibnamefont {Motrunich}},\ }\bibfield  {title} {\bibinfo {title} {Hilbert {{Space Fragmentation}} and {{Commutant Algebras}}},\ }\href {https://doi.org/10.1103/PhysRevX.12.011050} {\bibfield  {journal} {\bibinfo  {journal} {Physical Review X}\ }\textbf {\bibinfo {volume} {12}},\ \bibinfo {pages} {011050} (\bibinfo {year} {2022})}\BibitemShut {NoStop}%
\bibitem [{\citenamefont {Moudgalya}\ and\ \citenamefont {Motrunich}(2023)}]{moudgalya-2023-SymmetriesCommutant}%
  \BibitemOpen
  \bibfield  {author} {\bibinfo {author} {\bibfnamefont {S.}~\bibnamefont {Moudgalya}}\ and\ \bibinfo {author} {\bibfnamefont {O.~I.}\ \bibnamefont {Motrunich}},\ }\bibfield  {title} {\bibinfo {title} {From symmetries to commutant algebras in standard {{Hamiltonians}}},\ }\href {https://doi.org/10.1016/j.aop.2023.169384} {\bibfield  {journal} {\bibinfo  {journal} {Annals of Physics}\ }\textbf {\bibinfo {volume} {455}},\ \bibinfo {pages} {169384} (\bibinfo {year} {2023})}\BibitemShut {NoStop}%
\bibitem [{\citenamefont {Fannes}\ and\ \citenamefont {Tuyls}(1999)}]{fannes-1999-ContinuityProperty}%
  \BibitemOpen
  \bibfield  {author} {\bibinfo {author} {\bibfnamefont {M.}~\bibnamefont {Fannes}}\ and\ \bibinfo {author} {\bibfnamefont {P.}~\bibnamefont {Tuyls}},\ }\bibfield  {title} {\bibinfo {title} {A continuity property of quantum dynamical entropy},\ }\href {https://doi.org/10.1142/S0219025799000308} {\bibfield  {journal} {\bibinfo  {journal} {Infinite Dimensional Analysis, Quantum Probability and Related Topics}\ }\textbf {\bibinfo {volume} {02}},\ \bibinfo {pages} {511} (\bibinfo {year} {1999})}\BibitemShut {NoStop}%
\bibitem [{\citenamefont {de~Groot}\ \emph {et~al.}(2022)\citenamefont {de~Groot}, \citenamefont {Turzillo},\ and\ \citenamefont {Schuch}}]{deGroot-2022-SymmetryProtected}%
  \BibitemOpen
  \bibfield  {author} {\bibinfo {author} {\bibfnamefont {C.}~\bibnamefont {de~Groot}}, \bibinfo {author} {\bibfnamefont {A.}~\bibnamefont {Turzillo}},\ and\ \bibinfo {author} {\bibfnamefont {N.}~\bibnamefont {Schuch}},\ }\bibfield  {title} {\bibinfo {title} {Symmetry {P}rotected {T}opological {O}rder in {O}pen {Q}uantum {S}ystems},\ }\href {https://doi.org/10.22331/q-2022-11-10-856} {\bibfield  {journal} {\bibinfo  {journal} {{Quantum}}\ }\textbf {\bibinfo {volume} {6}},\ \bibinfo {pages} {856} (\bibinfo {year} {2022})}\BibitemShut {NoStop}%
\bibitem [{\citenamefont {Sohal}\ and\ \citenamefont {Prem}(2025)}]{ramanjit-2025-NoisyApproach}%
  \BibitemOpen
  \bibfield  {author} {\bibinfo {author} {\bibfnamefont {R.}~\bibnamefont {Sohal}}\ and\ \bibinfo {author} {\bibfnamefont {A.}~\bibnamefont {Prem}},\ }\bibfield  {title} {\bibinfo {title} {Noisy approach to intrinsically mixed-state topological order},\ }\href {https://doi.org/10.1103/PRXQuantum.6.010313} {\bibfield  {journal} {\bibinfo  {journal} {PRX Quantum}\ }\textbf {\bibinfo {volume} {6}},\ \bibinfo {pages} {010313} (\bibinfo {year} {2025})}\BibitemShut {NoStop}%
\bibitem [{\citenamefont {Esposti}\ and\ \citenamefont {Graffi}(2003)}]{esposti-2003-MathematicalAspects}%
  \BibitemOpen
  \bibfield  {author} {\bibinfo {author} {\bibfnamefont {M.~D.}\ \bibnamefont {Esposti}}\ and\ \bibinfo {author} {\bibfnamefont {S.}~\bibnamefont {Graffi}},\ }\bibfield  {title} {\bibinfo {title} {Mathematical {{Aspects}} of {{Quantum Maps}}},\ }in\ \href@noop {} {\emph {\bibinfo {booktitle} {The {{Mathematical Aspects}} of {{Quantum Maps}}}}},\ \bibinfo {editor} {edited by\ \bibinfo {editor} {\bibfnamefont {M.~D.}\ \bibnamefont {Esposti}}\ and\ \bibinfo {editor} {\bibfnamefont {S.}~\bibnamefont {Graffi}}}\ (\bibinfo  {publisher} {Springer},\ \bibinfo {address} {Berlin, Heidelberg},\ \bibinfo {year} {2003})\ pp.\ \bibinfo {pages} {49--90}\BibitemShut {NoStop}%
\bibitem [{\citenamefont {Axenides}\ \emph {et~al.}(2018)\citenamefont {Axenides}, \citenamefont {Floratos},\ and\ \citenamefont {Nicolis}}]{axenides-2018-QuantumCat}%
  \BibitemOpen
  \bibfield  {author} {\bibinfo {author} {\bibfnamefont {M.}~\bibnamefont {Axenides}}, \bibinfo {author} {\bibfnamefont {E.}~\bibnamefont {Floratos}},\ and\ \bibinfo {author} {\bibfnamefont {S.}~\bibnamefont {Nicolis}},\ }\bibfield  {title} {\bibinfo {title} {The quantum cat map on the modular discretization of extremal black hole horizons},\ }\href {https://doi.org/10.1140/epjc/s10052-018-5850-9} {\bibfield  {journal} {\bibinfo  {journal} {The European Physical Journal C}\ }\textbf {\bibinfo {volume} {78}},\ \bibinfo {pages} {412} (\bibinfo {year} {2018})}\BibitemShut {NoStop}%
\bibitem [{\citenamefont {Magan}\ and\ \citenamefont {Wu}(2024{\natexlab{a}})}]{Magan-2024n-BGS-chaostype}%
  \BibitemOpen
  \bibfield  {author} {\bibinfo {author} {\bibfnamefont {J.~M.}\ \bibnamefont {Magan}}\ and\ \bibinfo {author} {\bibfnamefont {Q.}~\bibnamefont {Wu}},\ }\href@noop {} {\bibinfo {title} {{Two types of quantum chaos: testing the limits of the Bohigas-Giannoni-Schmit conjecture}}} (\bibinfo {year} {2024}{\natexlab{a}}),\ \Eprint {https://arxiv.org/abs/2411.08186} {arXiv:2411.08186 [quant-ph]} \BibitemShut {NoStop}%
\bibitem [{\citenamefont {Hudetz}(1998)}]{hudetz-1998-QuantumDynamical}%
  \BibitemOpen
  \bibfield  {author} {\bibinfo {author} {\bibfnamefont {T.}~\bibnamefont {Hudetz}},\ }\bibfield  {title} {\bibinfo {title} {Quantum dynamical entropy revisited},\ }\href {https://doi.org/10.4064/-43-1-241-251} {\bibfield  {journal} {\bibinfo  {journal} {Banach Center Publications}\ }\textbf {\bibinfo {volume} {43}},\ \bibinfo {pages} {241} (\bibinfo {year} {1998})}\BibitemShut {NoStop}%
\bibitem [{\citenamefont {Benatti}\ \emph {et~al.}(1998)\citenamefont {Benatti}, \citenamefont {Hudetz},\ and\ \citenamefont {Knauf}}]{benatti-1998-QuantumChaos}%
  \BibitemOpen
  \bibfield  {author} {\bibinfo {author} {\bibfnamefont {F.}~\bibnamefont {Benatti}}, \bibinfo {author} {\bibfnamefont {T.}~\bibnamefont {Hudetz}},\ and\ \bibinfo {author} {\bibfnamefont {A.}~\bibnamefont {Knauf}},\ }\bibfield  {title} {\bibinfo {title} {Quantum {{Chaos}} and {{Dynamical Entropy}}},\ }\href {https://doi.org/10.1007/s002200050489} {\bibfield  {journal} {\bibinfo  {journal} {Communications in Mathematical Physics}\ }\textbf {\bibinfo {volume} {198}},\ \bibinfo {pages} {607} (\bibinfo {year} {1998})}\BibitemShut {NoStop}%
\bibitem [{\citenamefont {Benatti}(2023)}]{benatti-2023-DynamicsInformation}%
  \BibitemOpen
  \bibfield  {author} {\bibinfo {author} {\bibfnamefont {F.}~\bibnamefont {Benatti}},\ }\href {https://doi.org/10.1007/978-3-031-34248-6} {\emph {\bibinfo {title} {Dynamics, {{Information}} and {{Complexity}} in {{Quantum Systems}}}}},\ Theoretical and {{Mathematical Physics}}\ (\bibinfo  {publisher} {Springer International Publishing},\ \bibinfo {address} {Cham},\ \bibinfo {year} {2023})\BibitemShut {NoStop}%
\bibitem [{\citenamefont {Alicki}\ and\ \citenamefont {Narnhofer}(1995)}]{alicki-1995-ComparisonDynamical}%
  \BibitemOpen
  \bibfield  {author} {\bibinfo {author} {\bibfnamefont {R.}~\bibnamefont {Alicki}}\ and\ \bibinfo {author} {\bibfnamefont {H.}~\bibnamefont {Narnhofer}},\ }\bibfield  {title} {\bibinfo {title} {Comparison of dynamical entropies for the noncommutative shifts},\ }\href {https://doi.org/10.1007/BF00749625} {\bibfield  {journal} {\bibinfo  {journal} {Letters in Mathematical Physics}\ }\textbf {\bibinfo {volume} {33}},\ \bibinfo {pages} {241} (\bibinfo {year} {1995})}\BibitemShut {NoStop}%
\bibitem [{\citenamefont {Accardi}\ \emph {et~al.}(1996)\citenamefont {Accardi}, \citenamefont {Ohya},\ and\ \citenamefont {Watanabe}}]{accardi-1996-NoteQuantum}%
  \BibitemOpen
  \bibfield  {author} {\bibinfo {author} {\bibfnamefont {L.}~\bibnamefont {Accardi}}, \bibinfo {author} {\bibfnamefont {M.}~\bibnamefont {Ohya}},\ and\ \bibinfo {author} {\bibfnamefont {N.}~\bibnamefont {Watanabe}},\ }\bibfield  {title} {\bibinfo {title} {Note on quantum dynamical entropies},\ }\href {https://doi.org/10.1016/S0034-4877(97)84895-1} {\bibfield  {journal} {\bibinfo  {journal} {Reports on Mathematical Physics}\ }\bibinfo {series} {{{XXVIII}} Symposium on Mathematical Physics},\ \textbf {\bibinfo {volume} {38}},\ \bibinfo {pages} {457} (\bibinfo {year} {1996})}\BibitemShut {NoStop}%
\bibitem [{\citenamefont {Tuyls}(1998)}]{tuyls-1998-ComparingQuantum}%
  \BibitemOpen
  \bibfield  {author} {\bibinfo {author} {\bibfnamefont {P.}~\bibnamefont {Tuyls}},\ }\bibfield  {title} {\bibinfo {title} {Comparing quantum dynamical entropies},\ }\href {https://bibliotekanauki.pl/articles/1342268} {\bibfield  {journal} {\bibinfo  {journal} {Banach Center Publications}\ }\textbf {\bibinfo {volume} {43}},\ \bibinfo {pages} {411} (\bibinfo {year} {1998})}\BibitemShut {NoStop}%
\bibitem [{\citenamefont {Fannes}\ and\ \citenamefont {Haegeman}(2003)}]{fannes-2003-QuantumDynamical}%
  \BibitemOpen
  \bibfield  {author} {\bibinfo {author} {\bibfnamefont {M.}~\bibnamefont {Fannes}}\ and\ \bibinfo {author} {\bibfnamefont {B.}~\bibnamefont {Haegeman}},\ }\bibfield  {title} {\bibinfo {title} {Quantum dynamical entropies for classical stochastic systems},\ }\href {https://doi.org/10.1016/S0034-4877(03)90009-7} {\bibfield  {journal} {\bibinfo  {journal} {Reports on Mathematical Physics}\ }\textbf {\bibinfo {volume} {52}},\ \bibinfo {pages} {151} (\bibinfo {year} {2003})},\ \Eprint {https://arxiv.org/abs/math-ph/0206037} {arXiv:math-ph/0206037} \BibitemShut {NoStop}%
\bibitem [{\citenamefont {Numasawa}\ \emph {et~al.}(2016)\citenamefont {Numasawa}, \citenamefont {Shiba}, \citenamefont {Takayanagi},\ and\ \citenamefont {Watanabe}}]{Numasawa-2016-localmeasurements}%
  \BibitemOpen
  \bibfield  {author} {\bibinfo {author} {\bibfnamefont {T.}~\bibnamefont {Numasawa}}, \bibinfo {author} {\bibfnamefont {N.}~\bibnamefont {Shiba}}, \bibinfo {author} {\bibfnamefont {T.}~\bibnamefont {Takayanagi}},\ and\ \bibinfo {author} {\bibfnamefont {K.}~\bibnamefont {Watanabe}},\ }\bibfield  {title} {\bibinfo {title} {Epr pairs, local projections and quantum teleportation in holography},\ }\href {https://doi.org/10.1007/JHEP08(2016)077} {\bibfield  {journal} {\bibinfo  {journal} {Journal of High Energy Physics}\ }\textbf {\bibinfo {volume} {2016}},\ \bibinfo {pages} {77} (\bibinfo {year} {2016})}\BibitemShut {NoStop}%
\bibitem [{\citenamefont {Antonini}\ \emph {et~al.}(2022)\citenamefont {Antonini}, \citenamefont {Bentsen}, \citenamefont {Cao}, \citenamefont {Harper}, \citenamefont {Jian},\ and\ \citenamefont {Swingle}}]{Antonini-2022-HMbulkteleportation}%
  \BibitemOpen
  \bibfield  {author} {\bibinfo {author} {\bibfnamefont {S.}~\bibnamefont {Antonini}}, \bibinfo {author} {\bibfnamefont {G.}~\bibnamefont {Bentsen}}, \bibinfo {author} {\bibfnamefont {C.}~\bibnamefont {Cao}}, \bibinfo {author} {\bibfnamefont {J.}~\bibnamefont {Harper}}, \bibinfo {author} {\bibfnamefont {S.-K.}\ \bibnamefont {Jian}},\ and\ \bibinfo {author} {\bibfnamefont {B.}~\bibnamefont {Swingle}},\ }\bibfield  {title} {\bibinfo {title} {Holographic measurement and bulk teleportation},\ }\href {https://doi.org/10.1007/JHEP12(2022)124} {\bibfield  {journal} {\bibinfo  {journal} {Journal of High Energy Physics}\ }\textbf {\bibinfo {volume} {2022}},\ \bibinfo {pages} {124} (\bibinfo {year} {2022})}\BibitemShut {NoStop}%
\bibitem [{\citenamefont {Antonini}\ \emph {et~al.}(2023)\citenamefont {Antonini}, \citenamefont {Grado-White}, \citenamefont {Jian},\ and\ \citenamefont {Swingle}}]{Antonini-2023-HMCFT}%
  \BibitemOpen
  \bibfield  {author} {\bibinfo {author} {\bibfnamefont {S.}~\bibnamefont {Antonini}}, \bibinfo {author} {\bibfnamefont {B.}~\bibnamefont {Grado-White}}, \bibinfo {author} {\bibfnamefont {S.-K.}\ \bibnamefont {Jian}},\ and\ \bibinfo {author} {\bibfnamefont {B.}~\bibnamefont {Swingle}},\ }\bibfield  {title} {\bibinfo {title} {Holographic measurement in cft thermofield doubles},\ }\href {https://doi.org/10.1007/JHEP07(2023)014} {\bibfield  {journal} {\bibinfo  {journal} {Journal of High Energy Physics}\ }\textbf {\bibinfo {volume} {2023}},\ \bibinfo {pages} {14} (\bibinfo {year} {2023})}\BibitemShut {NoStop}%
\bibitem [{\citenamefont {Dowling}\ \emph {et~al.}(2023{\natexlab{b}})\citenamefont {Dowling}, \citenamefont {Kos},\ and\ \citenamefont {Modi}}]{dowling-2023-ScramblingNecessary}%
  \BibitemOpen
  \bibfield  {author} {\bibinfo {author} {\bibfnamefont {N.}~\bibnamefont {Dowling}}, \bibinfo {author} {\bibfnamefont {P.}~\bibnamefont {Kos}},\ and\ \bibinfo {author} {\bibfnamefont {K.}~\bibnamefont {Modi}},\ }\bibfield  {title} {\bibinfo {title} {Scrambling is {{Necessary}} but {{Not Sufficient}} for {{Chaos}}},\ }\href {https://doi.org/10.1103/PhysRevLett.131.180403} {\bibfield  {journal} {\bibinfo  {journal} {Physical Review Letters}\ }\textbf {\bibinfo {volume} {131}},\ \bibinfo {pages} {180403} (\bibinfo {year} {2023}{\natexlab{b}})},\ \Eprint {https://arxiv.org/abs/2304.07319} {arXiv:2304.07319 [cond-mat, physics:hep-th, physics:nlin, physics:quant-ph]} \BibitemShut {NoStop}%
\bibitem [{\citenamefont {Doi}\ \emph {et~al.}(2023)\citenamefont {Doi}, \citenamefont {Harper}, \citenamefont {Mollabashi}, \citenamefont {Takayanagi},\ and\ \citenamefont {Taki}}]{Doi-2023-timelikeEE}%
  \BibitemOpen
  \bibfield  {author} {\bibinfo {author} {\bibfnamefont {K.}~\bibnamefont {Doi}}, \bibinfo {author} {\bibfnamefont {J.}~\bibnamefont {Harper}}, \bibinfo {author} {\bibfnamefont {A.}~\bibnamefont {Mollabashi}}, \bibinfo {author} {\bibfnamefont {T.}~\bibnamefont {Takayanagi}},\ and\ \bibinfo {author} {\bibfnamefont {Y.}~\bibnamefont {Taki}},\ }\bibfield  {title} {\bibinfo {title} {Timelike entanglement entropy},\ }\href {https://doi.org/10.1007/JHEP05(2023)052} {\bibfield  {journal} {\bibinfo  {journal} {Journal of High Energy Physics}\ }\textbf {\bibinfo {volume} {2023}},\ \bibinfo {pages} {52} (\bibinfo {year} {2023})}\BibitemShut {NoStop}%
\bibitem [{\citenamefont {Heller}\ \emph {et~al.}(2025)\citenamefont {Heller}, \citenamefont {Ori},\ and\ \citenamefont {Serantes}}]{Heller-2025-GeometrictimelikeEE}%
  \BibitemOpen
  \bibfield  {author} {\bibinfo {author} {\bibfnamefont {M.~P.}\ \bibnamefont {Heller}}, \bibinfo {author} {\bibfnamefont {F.}~\bibnamefont {Ori}},\ and\ \bibinfo {author} {\bibfnamefont {A.}~\bibnamefont {Serantes}},\ }\bibfield  {title} {\bibinfo {title} {Geometric interpretation of timelike entanglement entropy},\ }\href {https://doi.org/10.1103/PhysRevLett.134.131601} {\bibfield  {journal} {\bibinfo  {journal} {Phys. Rev. Lett.}\ }\textbf {\bibinfo {volume} {134}},\ \bibinfo {pages} {131601} (\bibinfo {year} {2025})}\BibitemShut {NoStop}%
\bibitem [{\citenamefont {Li}\ \emph {et~al.}(2023)\citenamefont {Li}, \citenamefont {Xiao},\ and\ \citenamefont {Yang}}]{Li-2023-holographicTEE}%
  \BibitemOpen
  \bibfield  {author} {\bibinfo {author} {\bibfnamefont {Z.}~\bibnamefont {Li}}, \bibinfo {author} {\bibfnamefont {Z.-Q.}\ \bibnamefont {Xiao}},\ and\ \bibinfo {author} {\bibfnamefont {R.-Q.}\ \bibnamefont {Yang}},\ }\bibfield  {title} {\bibinfo {title} {On holographic time-like entanglement entropy},\ }\href {https://doi.org/10.1007/JHEP04(2023)004} {\bibfield  {journal} {\bibinfo  {journal} {Journal of High Energy Physics}\ }\textbf {\bibinfo {volume} {2023}},\ \bibinfo {pages} {4} (\bibinfo {year} {2023})}\BibitemShut {NoStop}%
\bibitem [{\citenamefont {Mollabashi}\ \emph {et~al.}(2021{\natexlab{a}})\citenamefont {Mollabashi}, \citenamefont {Shiba}, \citenamefont {Takayanagi}, \citenamefont {Tamaoka},\ and\ \citenamefont {Wei}}]{Mollabashi-2021-PseudoEntropy}%
  \BibitemOpen
  \bibfield  {author} {\bibinfo {author} {\bibfnamefont {A.}~\bibnamefont {Mollabashi}}, \bibinfo {author} {\bibfnamefont {N.}~\bibnamefont {Shiba}}, \bibinfo {author} {\bibfnamefont {T.}~\bibnamefont {Takayanagi}}, \bibinfo {author} {\bibfnamefont {K.}~\bibnamefont {Tamaoka}},\ and\ \bibinfo {author} {\bibfnamefont {Z.}~\bibnamefont {Wei}},\ }\bibfield  {title} {\bibinfo {title} {Pseudo-entropy in free quantum field theories},\ }\href {https://doi.org/10.1103/PhysRevLett.126.081601} {\bibfield  {journal} {\bibinfo  {journal} {Phys. Rev. Lett.}\ }\textbf {\bibinfo {volume} {126}},\ \bibinfo {pages} {081601} (\bibinfo {year} {2021}{\natexlab{a}})}\BibitemShut {NoStop}%
\bibitem [{\citenamefont {Mollabashi}\ \emph {et~al.}(2021{\natexlab{b}})\citenamefont {Mollabashi}, \citenamefont {Shiba}, \citenamefont {Takayanagi}, \citenamefont {Tamaoka},\ and\ \citenamefont {Wei}}]{Mollabashi-2021-pseudoentropyfreefields}%
  \BibitemOpen
  \bibfield  {author} {\bibinfo {author} {\bibfnamefont {A.}~\bibnamefont {Mollabashi}}, \bibinfo {author} {\bibfnamefont {N.}~\bibnamefont {Shiba}}, \bibinfo {author} {\bibfnamefont {T.}~\bibnamefont {Takayanagi}}, \bibinfo {author} {\bibfnamefont {K.}~\bibnamefont {Tamaoka}},\ and\ \bibinfo {author} {\bibfnamefont {Z.}~\bibnamefont {Wei}},\ }\bibfield  {title} {\bibinfo {title} {Aspects of pseudoentropy in field theories},\ }\href {https://doi.org/10.1103/PhysRevResearch.3.033254} {\bibfield  {journal} {\bibinfo  {journal} {Phys. Rev. Res.}\ }\textbf {\bibinfo {volume} {3}},\ \bibinfo {pages} {033254} (\bibinfo {year} {2021}{\natexlab{b}})}\BibitemShut {NoStop}%
\bibitem [{\citenamefont {Nishioka}\ \emph {et~al.}(2021)\citenamefont {Nishioka}, \citenamefont {Takayanagi},\ and\ \citenamefont {Taki}}]{Nishioka-2021-topologicalPE}%
  \BibitemOpen
  \bibfield  {author} {\bibinfo {author} {\bibfnamefont {T.}~\bibnamefont {Nishioka}}, \bibinfo {author} {\bibfnamefont {T.}~\bibnamefont {Takayanagi}},\ and\ \bibinfo {author} {\bibfnamefont {Y.}~\bibnamefont {Taki}},\ }\bibfield  {title} {\bibinfo {title} {Topological pseudo entropy},\ }\href {https://doi.org/10.1007/JHEP09(2021)015} {\bibfield  {journal} {\bibinfo  {journal} {Journal of High Energy Physics}\ }\textbf {\bibinfo {volume} {2021}},\ \bibinfo {pages} {15} (\bibinfo {year} {2021})}\BibitemShut {NoStop}%
\bibitem [{\citenamefont {Goto}\ \emph {et~al.}(2021)\citenamefont {Goto}, \citenamefont {Nozaki},\ and\ \citenamefont {Tamaoka}}]{Goto-2021-sff}%
  \BibitemOpen
  \bibfield  {author} {\bibinfo {author} {\bibfnamefont {K.}~\bibnamefont {Goto}}, \bibinfo {author} {\bibfnamefont {M.}~\bibnamefont {Nozaki}},\ and\ \bibinfo {author} {\bibfnamefont {K.}~\bibnamefont {Tamaoka}},\ }\bibfield  {title} {\bibinfo {title} {Subregion spectrum form factor via pseudoentropy},\ }\href {https://doi.org/10.1103/PhysRevD.104.L121902} {\bibfield  {journal} {\bibinfo  {journal} {Phys. Rev. D}\ }\textbf {\bibinfo {volume} {104}},\ \bibinfo {pages} {L121902} (\bibinfo {year} {2021})}\BibitemShut {NoStop}%
\bibitem [{\citenamefont {Miyaji}(2021)}]{Miyaji-2021-Island}%
  \BibitemOpen
  \bibfield  {author} {\bibinfo {author} {\bibfnamefont {M.}~\bibnamefont {Miyaji}},\ }\bibfield  {title} {\bibinfo {title} {Island for gravitationally prepared state and pseudo entanglement wedge},\ }\href {https://doi.org/10.1007/JHEP12(2021)013} {\bibfield  {journal} {\bibinfo  {journal} {Journal of High Energy Physics}\ }\textbf {\bibinfo {volume} {2021}},\ \bibinfo {pages} {13} (\bibinfo {year} {2021})}\BibitemShut {NoStop}%
\bibitem [{\citenamefont {Guo}\ \emph {et~al.}(2023)\citenamefont {Guo}, \citenamefont {He},\ and\ \citenamefont {Zhang}}]{Guo-2023-pseudo-Hermiticity}%
  \BibitemOpen
  \bibfield  {author} {\bibinfo {author} {\bibfnamefont {W.-z.}\ \bibnamefont {Guo}}, \bibinfo {author} {\bibfnamefont {S.}~\bibnamefont {He}},\ and\ \bibinfo {author} {\bibfnamefont {Y.-X.}\ \bibnamefont {Zhang}},\ }\bibfield  {title} {\bibinfo {title} {Constructible reality condition of pseudo entropy via pseudo-hermiticity},\ }\href {https://doi.org/10.1007/JHEP05(2023)021} {\bibfield  {journal} {\bibinfo  {journal} {Journal of High Energy Physics}\ }\textbf {\bibinfo {volume} {2023}},\ \bibinfo {pages} {21} (\bibinfo {year} {2023})}\BibitemShut {NoStop}%
\bibitem [{\citenamefont {He}\ \emph {et~al.}(2024)\citenamefont {He}, \citenamefont {Yang}, \citenamefont {Zhang},\ and\ \citenamefont {Zhao}}]{He:2023eap}%
  \BibitemOpen
  \bibfield  {author} {\bibinfo {author} {\bibfnamefont {S.}~\bibnamefont {He}}, \bibinfo {author} {\bibfnamefont {J.}~\bibnamefont {Yang}}, \bibinfo {author} {\bibfnamefont {Y.-X.}\ \bibnamefont {Zhang}},\ and\ \bibinfo {author} {\bibfnamefont {Z.-X.}\ \bibnamefont {Zhao}},\ }\bibfield  {title} {\bibinfo {title} {{Pseudoentropy for descendant operators in two-dimensional conformal field theories}},\ }\href {https://doi.org/10.1103/PhysRevD.109.025014} {\bibfield  {journal} {\bibinfo  {journal} {Phys. Rev. D}\ }\textbf {\bibinfo {volume} {109}},\ \bibinfo {pages} {025014} (\bibinfo {year} {2024})},\ \Eprint {https://arxiv.org/abs/2301.04891} {arXiv:2301.04891 [hep-th]} \BibitemShut {NoStop}%
\bibitem [{\citenamefont {Shi}\ \emph {et~al.}(2023)\citenamefont {Shi}, \citenamefont {Vardhan},\ and\ \citenamefont {Liu}}]{Shi2023localdynamics-ETH}%
  \BibitemOpen
  \bibfield  {author} {\bibinfo {author} {\bibfnamefont {Z.~D.}\ \bibnamefont {Shi}}, \bibinfo {author} {\bibfnamefont {S.}~\bibnamefont {Vardhan}},\ and\ \bibinfo {author} {\bibfnamefont {H.}~\bibnamefont {Liu}},\ }\bibfield  {title} {\bibinfo {title} {{Local dynamics and the structure of chaotic eigenstates}},\ }\href {https://doi.org/10.1103/PhysRevB.108.224305} {\bibfield  {journal} {\bibinfo  {journal} {Phys. Rev. B}\ }\textbf {\bibinfo {volume} {108}},\ \bibinfo {pages} {224305} (\bibinfo {year} {2023})},\ \Eprint {https://arxiv.org/abs/2306.08032} {arXiv:2306.08032 [quant-ph]} \BibitemShut {NoStop}%
\bibitem [{\citenamefont {Boasman}\ and\ \citenamefont {Keating}(1995)}]{boasman-1995-SemiclassicalAsymptotics}%
  \BibitemOpen
  \bibfield  {author} {\bibinfo {author} {\bibfnamefont {P.~A.}\ \bibnamefont {Boasman}}\ and\ \bibinfo {author} {\bibfnamefont {J.~P.}\ \bibnamefont {Keating}},\ }\bibfield  {title} {\bibinfo {title} {Semiclassical {{Asymptotics}} of {{Perturbed Cat Maps}}},\ }\href {https://www.jstor.org/stable/52684} {\bibfield  {journal} {\bibinfo  {journal} {Proceedings: Mathematical and Physical Sciences}\ }\textbf {\bibinfo {volume} {449}},\ \bibinfo {pages} {629} (\bibinfo {year} {1995})},\ \Eprint {https://arxiv.org/abs/52684} {52684} \BibitemShut {NoStop}%
\bibitem [{\citenamefont {Berry}\ \emph {et~al.}(1979)\citenamefont {Berry}, \citenamefont {Balazs}, \citenamefont {Tabor},\ and\ \citenamefont {Voros}}]{berry-1979-QuantumMaps}%
  \BibitemOpen
  \bibfield  {author} {\bibinfo {author} {\bibfnamefont {M.~V.}\ \bibnamefont {Berry}}, \bibinfo {author} {\bibfnamefont {N.~L.}\ \bibnamefont {Balazs}}, \bibinfo {author} {\bibfnamefont {M.}~\bibnamefont {Tabor}},\ and\ \bibinfo {author} {\bibfnamefont {A.}~\bibnamefont {Voros}},\ }\bibfield  {title} {\bibinfo {title} {Quantum maps},\ }\href {https://doi.org/10.1016/0003-4916(79)90296-3} {\bibfield  {journal} {\bibinfo  {journal} {Annals of Physics}\ }\textbf {\bibinfo {volume} {122}},\ \bibinfo {pages} {26} (\bibinfo {year} {1979})}\BibitemShut {NoStop}%
\bibitem [{\citenamefont {Esposti}\ \emph {et~al.}(1995)\citenamefont {Esposti}, \citenamefont {Graffi},\ and\ \citenamefont {Isola}}]{esposti-1995-ClassicalLimit}%
  \BibitemOpen
  \bibfield  {author} {\bibinfo {author} {\bibfnamefont {M.~D.}\ \bibnamefont {Esposti}}, \bibinfo {author} {\bibfnamefont {S.}~\bibnamefont {Graffi}},\ and\ \bibinfo {author} {\bibfnamefont {S.}~\bibnamefont {Isola}},\ }\bibfield  {title} {\bibinfo {title} {Classical limit of the quantized hyperbolic toral automorphisms},\ }\href {https://doi.org/10.1007/BF02101532} {\bibfield  {journal} {\bibinfo  {journal} {Communications in Mathematical Physics}\ }\textbf {\bibinfo {volume} {167}},\ \bibinfo {pages} {471} (\bibinfo {year} {1995})}\BibitemShut {NoStop}%
\bibitem [{\citenamefont {De~Bi{\`e}vre}\ \emph {et~al.}(1996)\citenamefont {De~Bi{\`e}vre}, \citenamefont {Esposti},\ and\ \citenamefont {Giachetti}}]{debievre-1996-QuantizationClass}%
  \BibitemOpen
  \bibfield  {author} {\bibinfo {author} {\bibfnamefont {S.}~\bibnamefont {De~Bi{\`e}vre}}, \bibinfo {author} {\bibfnamefont {M.~D.}\ \bibnamefont {Esposti}},\ and\ \bibinfo {author} {\bibfnamefont {R.}~\bibnamefont {Giachetti}},\ }\bibfield  {title} {\bibinfo {title} {Quantization of a class of piecewise affine transformations on the torus},\ }\href {https://doi.org/10.1007/BF02099363} {\bibfield  {journal} {\bibinfo  {journal} {Communications in Mathematical Physics}\ }\textbf {\bibinfo {volume} {176}},\ \bibinfo {pages} {73} (\bibinfo {year} {1996})}\BibitemShut {NoStop}%
\bibitem [{\citenamefont {Keating}\ \emph {et~al.}(1999)\citenamefont {Keating}, \citenamefont {Mezzadri},\ and\ \citenamefont {Robbins}}]{keating-1999-QuantumBoundary}%
  \BibitemOpen
  \bibfield  {author} {\bibinfo {author} {\bibfnamefont {J.~P.}\ \bibnamefont {Keating}}, \bibinfo {author} {\bibfnamefont {F.}~\bibnamefont {Mezzadri}},\ and\ \bibinfo {author} {\bibfnamefont {J.~M.}\ \bibnamefont {Robbins}},\ }\bibfield  {title} {\bibinfo {title} {Quantum boundary conditions for torus maps},\ }\href {https://doi.org/10.1088/0951-7715/12/3/010} {\bibfield  {journal} {\bibinfo  {journal} {Nonlinearity}\ }\textbf {\bibinfo {volume} {12}},\ \bibinfo {pages} {579} (\bibinfo {year} {1999})}\BibitemShut {NoStop}%
\bibitem [{\citenamefont {Eckhardt}(1986)}]{eckhardt-1986-ExactEigenfunctions}%
  \BibitemOpen
  \bibfield  {author} {\bibinfo {author} {\bibfnamefont {B.}~\bibnamefont {Eckhardt}},\ }\bibfield  {title} {\bibinfo {title} {Exact eigenfunctions for a quantised map},\ }\href {https://doi.org/10.1088/0305-4470/19/10/023} {\bibfield  {journal} {\bibinfo  {journal} {Journal of Physics A: Mathematical and General}\ }\textbf {\bibinfo {volume} {19}},\ \bibinfo {pages} {1823} (\bibinfo {year} {1986})}\BibitemShut {NoStop}%
\bibitem [{\citenamefont {Keating}(1991)}]{keating-1991-CatMaps}%
  \BibitemOpen
  \bibfield  {author} {\bibinfo {author} {\bibfnamefont {J.~P.}\ \bibnamefont {Keating}},\ }\bibfield  {title} {\bibinfo {title} {The cat maps: Quantum mechanics and classical motion},\ }\href {https://doi.org/10.1088/0951-7715/4/2/006} {\bibfield  {journal} {\bibinfo  {journal} {Nonlinearity}\ }\textbf {\bibinfo {volume} {4}},\ \bibinfo {pages} {309} (\bibinfo {year} {1991})}\BibitemShut {NoStop}%
\bibitem [{\citenamefont {Bouzouina}\ and\ \citenamefont {De~Bi{\`e}vre}(1996)}]{bouzouina-1996-EquipartitionEigenfunctions}%
  \BibitemOpen
  \bibfield  {author} {\bibinfo {author} {\bibfnamefont {A.}~\bibnamefont {Bouzouina}}\ and\ \bibinfo {author} {\bibfnamefont {S.}~\bibnamefont {De~Bi{\`e}vre}},\ }\bibfield  {title} {\bibinfo {title} {Equipartition of the eigenfunctions of quantized ergodic maps on the torus},\ }\href {https://doi.org/10.1007/BF02104909} {\bibfield  {journal} {\bibinfo  {journal} {Communications in Mathematical Physics}\ }\textbf {\bibinfo {volume} {178}},\ \bibinfo {pages} {83} (\bibinfo {year} {1996})}\BibitemShut {NoStop}%
\bibitem [{\citenamefont {Kurlberg}\ and\ \citenamefont {Rudnick}(2001)}]{kurlberg-2001-QuantumErgodicity}%
  \BibitemOpen
  \bibfield  {author} {\bibinfo {author} {\bibfnamefont {P.}~\bibnamefont {Kurlberg}}\ and\ \bibinfo {author} {\bibfnamefont {Z.}~\bibnamefont {Rudnick}},\ }\bibfield  {title} {\bibinfo {title} {On {{Quantum Ergodicity}} for {{Linear Maps}} of the {{Torus}}},\ }\href {https://doi.org/10.1007/s002200100501} {\bibfield  {journal} {\bibinfo  {journal} {Communications in Mathematical Physics}\ }\textbf {\bibinfo {volume} {222}},\ \bibinfo {pages} {201} (\bibinfo {year} {2001})}\BibitemShut {NoStop}%
\bibitem [{\citenamefont {Horvat}\ \emph {et~al.}(2006)\citenamefont {Horvat}, \citenamefont {Prosen},\ and\ \citenamefont {Esposti}}]{horvat-2006-QuantumClassical}%
  \BibitemOpen
  \bibfield  {author} {\bibinfo {author} {\bibfnamefont {M.}~\bibnamefont {Horvat}}, \bibinfo {author} {\bibfnamefont {T.}~\bibnamefont {Prosen}},\ and\ \bibinfo {author} {\bibfnamefont {M.~D.}\ \bibnamefont {Esposti}},\ }\bibfield  {title} {\bibinfo {title} {Quantum--classical correspondence on compact phase space},\ }\href {https://doi.org/10.1088/0951-7715/19/6/013} {\bibfield  {journal} {\bibinfo  {journal} {Nonlinearity}\ }\textbf {\bibinfo {volume} {19}},\ \bibinfo {pages} {1471} (\bibinfo {year} {2006})}\BibitemShut {NoStop}%
\bibitem [{\citenamefont {Horvat}\ and\ \citenamefont {Esposti}(2007)}]{horvat-2007-EgorovProperty}%
  \BibitemOpen
  \bibfield  {author} {\bibinfo {author} {\bibfnamefont {M.}~\bibnamefont {Horvat}}\ and\ \bibinfo {author} {\bibfnamefont {M.~D.}\ \bibnamefont {Esposti}},\ }\bibfield  {title} {\bibinfo {title} {The {{Egorov}} property in perturbed cat maps},\ }\href {https://doi.org/10.1088/1751-8113/40/32/004} {\bibfield  {journal} {\bibinfo  {journal} {Journal of Physics A: Mathematical and Theoretical}\ }\textbf {\bibinfo {volume} {40}},\ \bibinfo {pages} {9771} (\bibinfo {year} {2007})}\BibitemShut {NoStop}%
\bibitem [{\citenamefont {Ford}\ \emph {et~al.}(1991)\citenamefont {Ford}, \citenamefont {Mantica},\ and\ \citenamefont {Ristow}}]{ford-1991-ArnoldCat}%
  \BibitemOpen
  \bibfield  {author} {\bibinfo {author} {\bibfnamefont {J.}~\bibnamefont {Ford}}, \bibinfo {author} {\bibfnamefont {G.}~\bibnamefont {Mantica}},\ and\ \bibinfo {author} {\bibfnamefont {G.~H.}\ \bibnamefont {Ristow}},\ }\bibfield  {title} {\bibinfo {title} {The {{Arnol}}'d cat: {{Failure}} of the correspondence principle},\ }\href {https://doi.org/10.1016/0167-2789(91)90012-X} {\bibfield  {journal} {\bibinfo  {journal} {Physica D: Nonlinear Phenomena}\ }\textbf {\bibinfo {volume} {50}},\ \bibinfo {pages} {493} (\bibinfo {year} {1991})}\BibitemShut {NoStop}%
\bibitem [{\citenamefont {Benatti}\ \emph {et~al.}(1991)\citenamefont {Benatti}, \citenamefont {Narnhofer},\ and\ \citenamefont {Sewell}}]{benatti-1991-NoncommutativeVersion}%
  \BibitemOpen
  \bibfield  {author} {\bibinfo {author} {\bibfnamefont {F.}~\bibnamefont {Benatti}}, \bibinfo {author} {\bibfnamefont {H.}~\bibnamefont {Narnhofer}},\ and\ \bibinfo {author} {\bibfnamefont {G.~L.}\ \bibnamefont {Sewell}},\ }\bibfield  {title} {\bibinfo {title} {A non-commutative version of the {{Arnold}} cat map},\ }\href {https://doi.org/10.1007/BF00401650} {\bibfield  {journal} {\bibinfo  {journal} {Letters in Mathematical Physics}\ }\textbf {\bibinfo {volume} {21}},\ \bibinfo {pages} {157} (\bibinfo {year} {1991})}\BibitemShut {NoStop}%
\bibitem [{\citenamefont {Andries}\ \emph {et~al.}(1995)\citenamefont {Andries}, \citenamefont {Fannes}, \citenamefont {Tuyls},\ and\ \citenamefont {Alicki}}]{andries-1995-DynamicalEntropy}%
  \BibitemOpen
  \bibfield  {author} {\bibinfo {author} {\bibfnamefont {J.}~\bibnamefont {Andries}}, \bibinfo {author} {\bibfnamefont {M.}~\bibnamefont {Fannes}}, \bibinfo {author} {\bibfnamefont {P.}~\bibnamefont {Tuyls}},\ and\ \bibinfo {author} {\bibfnamefont {R.}~\bibnamefont {Alicki}},\ }\bibfield  {title} {\bibinfo {title} {The dynamical entropy of the quantum {{Arnold}} cat map},\ }\href {https://doi.org/10.1007/BF00750844} {\bibfield  {journal} {\bibinfo  {journal} {Letters in Mathematical Physics}\ }\textbf {\bibinfo {volume} {35}},\ \bibinfo {pages} {375} (\bibinfo {year} {1995})}\BibitemShut {NoStop}%
\bibitem [{\citenamefont {Magan}\ and\ \citenamefont {Wu}(2024{\natexlab{b}})}]{magan-2024-TwoTypes}%
  \BibitemOpen
  \bibfield  {author} {\bibinfo {author} {\bibfnamefont {J.~M.}\ \bibnamefont {Magan}}\ and\ \bibinfo {author} {\bibfnamefont {Q.}~\bibnamefont {Wu}},\ }\href {http://arxiv.org/abs/2411.08186} {\bibinfo {title} {Two types of quantum chaos: Testing the limits of the {{Bohigas-Giannoni-Schmit}} conjecture}} (\bibinfo {year} {2024}{\natexlab{b}}),\ \Eprint {https://arxiv.org/abs/2411.08186} {arXiv:2411.08186 [quant-ph]} \BibitemShut {NoStop}%
\bibitem [{\citenamefont {Marshall}\ \emph {et~al.}(2011)\citenamefont {Marshall}, \citenamefont {Olkin},\ and\ \citenamefont {Arnold}}]{marshall-2011-InequalitiesTheory}%
  \BibitemOpen
  \bibfield  {author} {\bibinfo {author} {\bibfnamefont {A.~W.}\ \bibnamefont {Marshall}}, \bibinfo {author} {\bibfnamefont {I.}~\bibnamefont {Olkin}},\ and\ \bibinfo {author} {\bibfnamefont {B.~C.}\ \bibnamefont {Arnold}},\ }\href {https://doi.org/10.1007/978-0-387-68276-1} {\emph {\bibinfo {title} {Inequalities: {{Theory}} of {{Majorization}} and {{Its Applications}}}}},\ Springer {{Series}} in {{Statistics}}\ (\bibinfo  {publisher} {Springer},\ \bibinfo {address} {New York, NY},\ \bibinfo {year} {2011})\BibitemShut {NoStop}%
\end{thebibliography}%

\end{document}